\documentclass[pra,twocolumn,pra,groupedaddress]{revtex4-1}

\usepackage{graphicx}
\usepackage{dcolumn}
\usepackage{bm}
\usepackage{epsfig}
\usepackage{amsmath}
\usepackage{amssymb}
\usepackage{color}
\usepackage{bbm}
\usepackage{braket}
\usepackage{float}
\usepackage{dsfont}

\usepackage[colorlinks=true,citecolor=blue,linkcolor=blue,urlcolor=blue]{hyperref}

\begin{document}

\title{Relaxation breakdown and resonant tunneling in ultrastrong-coupling cavity QED}

\author{Daniele De Bernardis$^1$}
\affiliation{$^1$Pitaevskii BEC Center, CNR-INO and Dipartimento di Fisica, Università di Trento, I-38123 Trento, Italy.}

\date{\today}

\begin{abstract} 
We study the open relaxation dynamics of an asymmetric dipole that is ultrastrongly coupled to a single electromagnetic cavity mode. By using a thermalizing master equation for the whole interacting system we derive a phase diagram of the Liouvillian gap. It emerges that the ultrastrong coupling inhibits the system's relaxation toward the equilibrium state due to an exponential suppression of the dipole tunneling rate.
However, we find that polaronic multi-photon resonances restore fast relaxation by a cavity-mediated dipole resonant tunneling process.
Aside of the numerical evidences, we develop a fully analytical description by diagonalizing the Rabi model through a generalized rotating-wave approximation, valid in the so-called polaron frame. 
The relaxation physics of such ultrastrong-coupling systems is then reduced to a multi-photon polaron version of the standard text-book dressed states picture.
At the end we discuss an extension to a multi-well dipole that can set the basis of a cascaded resonant tunnelling setup in the ultrastrong coupling regime.

\end{abstract}
 
\maketitle

\section{Introduction}

Relaxation from a metastable state toward equilibrium is a central problem in many branches of physics, such as chemical reactions, radiaoactive decay and electronic transport, to name a few \cite{landau2013statistical}.
The energy barrier separating a local minimum from the stable equilibrium, i.e. the activation barrier of chemical reactions \cite{Piskulich_2019_activation_energy_ACS}, can be overcome by thermal fluctuations, for which, after an initial absorption of energy from the bath, the system is kicked out the metastable state, rolling down to its absolute equilibrium state and releasing the energy excess.
When the temperature is too small to kick the system over the metastable energy barrier, relaxation is then dominated by the \emph{tunnel effect} (or quantum tunneling), which is one of the first surprising consequences of the quantum theory \cite{Merzbacher2002_History_Quantum_Tunneling_doi:10.1063/1.1510281}.

Following the hand-wavy intuition that quantum fluctuations replace thermal ones in kicking the system out of the metastability, one might speculate that including in these systems a supplemental quantum reservoir could sensibly alter the tunneling dynamics.
The work of Leggett et al. on tunneling-systems coupled to an environment \cite{Leggett_PhysRevLett.46.211, Leggett_RevModPhys.59.1} has shown that this is actually the case, and tunneling can be sensibly changed as a function of the environment parameters.
Since for most systems the natural environment is provided by the electromagnetic radiation, here quantum tunneling is crossing its path with another fundamental concept of quantum physics: the non empty vacuum of quantum electrodynamics (QED) \cite{milonni1994quantum}, rising the question:
can vacuum fluctuations of the electromagnetic field affect tunneling and relaxation in material systems?


Experiments have shown strong suggestions that the answer may be positive, and
that the electromagnetic vacuum of a resonant cavity could have a major role
in chemical reaction and electronic transport where important differences are observed when molecules, atoms or electrons couple strongly or ultrastrongly to
such a extreme resonant electromagnetic environment \cite{Ebbesen_2012_reactionRate-cQED, Faist_magneto_transport_2019NatPhys, Valmorra2021_natCom_Vacuum_field_induced_transport, faist_Science_2022}. 

All these exciting observations have stimulated 
multiple theoretical debates in various communities opening new research lines such as: polaritonic chemistry \cite{Rubio_2017_pnas.1615509114,Feist_2022_ACS_review_polaritonic_chemistry, Rubio_NatCom_2022_ShiningLightCavityChemestry}, cavity QED control of electronic transport in mesoscopic devices or in quantum Hall systems  \cite{ciuti_magnetotransport_nature,Ciuti_transport_2023_PhysRevB.107.045425}, cavity QED modification of ferromagnetism, ferroelectricity and superconductivity \cite{DeBernardis_PhysRevA.97.043820,Schuler_SciPostPhys.9.5.066, Demler_PhysRevX.10.041027, GMAndolina_PhysRevB.102.125137, Zueco_PhysRevLett.127.167201, Cavalleri_PhysRevLett.122.133602}, and their out-of-equilibrium extensions \cite{Hausinger_2008, Savasta_RabiIncoherentPump_PhysRevResearch.4.023048, Savasta_relaxation_generalized_MasterEq_PhysRevA.98.053834,Pistolesi_2019_PhysRevLett.123.246601, Shane_PhysRevLett.126.133603, Grifoni_2021_PhysRevA.104.053711, Shane_PhysRevResearch.4.L042032, Ren_2022_PhysRevResearch.4.013152},
all with the general aim to explore and understand up to which degree the quantum vacuum of cavity QED can be a resource to modify and control properties of matter \cite{Ciuti_USC_original_PhysRevB.72.115303,Sentef_cavity_qmaterial_2022_5.0083825,Hafezi_review_electronPhotons,Ebbesen_review_science.abd0336}.

However in the community there is still not a full consensus about the origin, validity and interpretation of these theories and they relation with the actual experimental evidences \cite{ Andolina_nogo_PhysRevB.100.121109, Feist_CasimirPolderMolecule_PhysRevX.9.021057, andolina2022deep, saezblazquez2022observe}, suggesting that more research and additional examples are needed in order to completely make clear these physical mechanisms.


In this article we explicitly address the problem of how the electromagnetic vacuum of ultrastrong-coupling cavity QED can affect the relaxation toward equilibrium of a polarizable material.
In order to isolate every single different effect we consider a simple paradigmatic setup: an asymmetric double well dipole in a single-mode resonant cavity.
Its low-energy dynamics can be approximated to the \emph{quantum Rabi model}, which is the simplest theoretical framework to study light-matter interactions.
We complete the description of the model including two basic dissipative mechanism: Ohmic cavity dissipation and dipole radiative losses.
Under these circumstances the system's relaxation is described through a thermalizing master equation valid for arbitrary light-matter coupling values, whose steady state is the correct thermal equilibrium state.
From the spectral gap $\lambda$ of its Liouvillian operator we derive a phase diagram describing how relaxation toward equilibrium is modified by the coupling to the cavity.

The intuition arising from all recent works regarding thermalization and transport in cavity QED would suggest that the coupling with the cavity always favours and accelerates the relaxation properties of the system.
However, here
we show that an increasing light-matter coupling strength from the strong to the ultrastrong coupling regime exponentially suppresses the relaxation rate of the dipole, being a prototype for the so-called localization transition in the spin-boson model \cite{Leggett_RevModPhys.59.1}.
The cavity-induced inhibition of the dipole relaxation is only restored thanks to the occurrence of polaronic multi-photon resonant tunneling processes, in very close analogy to Franck-Condon physics describing electron tunneling assisted by vibrational transitions \cite{Andreev_PhysRevB.74.205438, Oppen_NatPhys_FrankCondonExp_2009, Joachim_phonon_electron_tranport_PhysRevLett.114.016804, Eaves_phonon_resonant_tunnel_2016_PhysRevLett.116.186603}.
After showing that this mechanism is already observable in current experimental platforms such as superconducting circuits we comment on the possible consequence for cavity assisted quantum transport and cascaded ultrastrong-coupling setups with multi-well dipoles.

Differently from previous studies \cite{Hausinger_2008, Savasta_relaxation_generalized_MasterEq_PhysRevA.98.053834, Savasta_RabiIncoherentPump_PhysRevResearch.4.023048}, here 
we exploit a generalized rotating-wave approximation of the Rabi model from which we analytically derive the transition rates of the master equation in the ultrastrong coupling regime. 
From this calculation we obtain a complete and simple picture on how relaxation and thermalization work in terms of polaronic dressed states, valid in the ultrastrong coupling regime.

The article is organized as follows.
In Sec. \ref{sec:model} we introduce the physical system and its approximated description in terms of the asymmetric quantum Rabi model.
By considering the Liouvillan gap of its open dynamics, in Sec. \ref{sec:relax_regimes} we study how the ultrastrong coupling regime changes the relaxation and thermalization rate.
By using a generalized rotating-wave approximation to diagonalize the Rabi model we explicitly show an exponential slow-down of the system's relaxation due to the ultrastrong coupling regime.
In Sec. \ref{sec:USC_res_tunnel} we show that the fast relaxation can be restored by a cavity assisted multi-photon resonant tunnelling process. Exploiting again the generalized rotating-wave approximation we develop the discussion in terms of multi-photon polaron dressed states.
In Sec. \ref{sec:discussion_cascaded} we extend this setup to the extended Dicke model leading to a cascaded resonant tunnelling device.
Finally, in Sec. \ref{sec:conclusion} we draw our conclusions.

\section{Model}
\label{sec:model}

\begin{figure}
    \centering
    \includegraphics[width=\columnwidth]{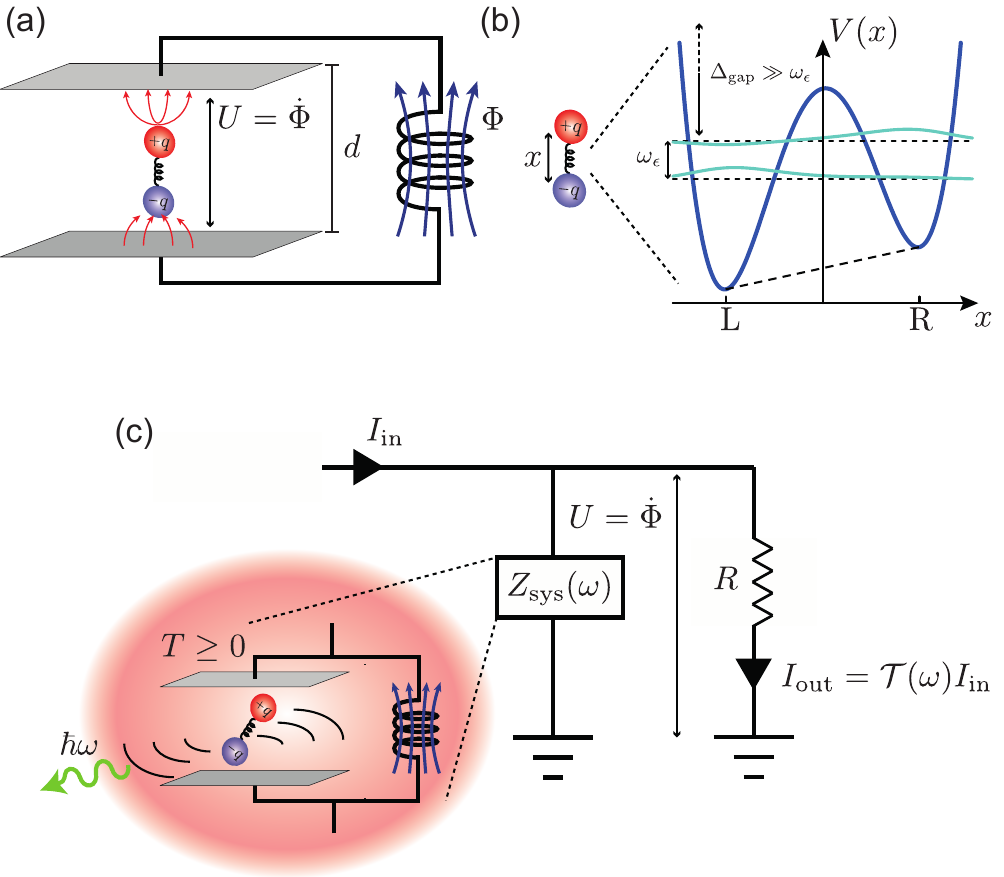}
    \caption{(a) Cavity QED system. The cavity is modelled as an LC-circuit, where the inductor magnetic flux $\Phi$ takes the role of the dynamical variable of the electromagnetic field, usually given by the vector potential $\vec A$. The dipole inside the capacitor couples to the voltage drop $U=\dot \Phi$ between the plates, separated by the distance $d$. (b) The dipole is described as a particle in a tilted double-well potential. The position $x$ represent the displacement between the two charges $q$, $-q$, such that the dipole moment is $qx$}. When the central well of the potential is large enough the system is approximated by only the two lowest levels (two-level approximation, see App. \ref{sec:TLA}).
    (c) Open-system schematic view. The cavity QED system can be interpreted as an element of a dissipative circuit.
    \label{fig:1}
\end{figure}

We consider the paradigmatic cavity quantum electrodynamics (cQED) setup described in Fig. \ref{fig:1}(a), where a single  electrically polarizable object (a dipole) is placed into the planar capacitor of a resonant LC circuit.
This simple toy model is able to reproduce most of the features of cQED in all various coupling regimes, and is particularly important in giving a simple and intuitive description of many solid-state or circuit cQED setups relevant for experiment in the ultrastrong coupling (USC) regime in the GHz or THz range \cite{Tuomas_PhysRevA.94.033850, DeBernardis_PhysRevA.97.043820, Solano_review_RevModPhys.91.025005, USC_Review_Nature_2019, Yoshihara_natcom_2022}.

The system cavity QED Hamiltonian is given by the so-called \emph{asymmetric Rabi model} ($\hbar = 1$)
\begin{equation}\label{eq:ham_Rabi}
\begin{split}
    H_{\rm cQED} \approx H_{\rm Rabi} = \omega_c a^{\dag} a +  \omega_d s_z + \epsilon s_x  + g \left( a + a^{\dag}\right) s_x,
\end{split}
\end{equation}
where $a$ is the annihilation operator of a cavity photon with frequency $\omega_c$. 
The pseudo-spin operators $s_{x,z}$ are linked, respectively, to the dipole moment $x$ and the dipole internal energy through the two-level approximation, $\omega_d$ is the dipole lowest transition frequency, $\epsilon$ is the dipole asymmetry (which breaks the $\mathbb{Z}_2$ symmetry of the Rabi model). The dipole eigenstates are also asymmetric with frequencies $\pm \omega_{\epsilon}/2=\sqrt{\omega_d^2+\epsilon^2}/2$ and this picture holds until the two-level subspace is well separated in energy from the rest of the spectrum, see Fig. \ref{fig:1}(b) for a schematic view.
Finally, $g$ is the light-matter interaction strength due to the dipole coupling to the cavity.
A complete derivation of the model is presented in App. \ref{sec:TLA}-\ref{sec:cQED_ham}.

As schematically shown in Fig. \ref{fig:1}(c), the system dissipates energy mainly in two external environments: a resistive element (or transmission line) for the cavity, and free-space radiative modes for the dipole.
The full system dynamics is thus obtained from the contribution of three Liouvillian super operators
\begin{equation}\label{eq:master_eq_total}
    \partial_t \rho = \mathcal{L}_{H}(\rho) + \mathcal{L}_c (\rho ) + \mathcal{L}_{\rm dip}(\rho ),
\end{equation}
where
\begin{equation}\label{eq:liouvillian_H}
    \mathcal{L}_{H}(\rho ) = - i \left[ H_{\rm Rabi}, \rho \right]
\end{equation}
generates the coherent time evolution, while the cavity and dipole dissipative dynamics are given by
\begin{equation}\label{eq:liouvillian_c_dip}
    \begin{split}
        & \mathcal{L}_{c/{\rm dip}}(\rho ) = \sum_{n < m} \left[ 1 + N_T (\omega_{mn}) \right] \Gamma_{nm}^{c/{\rm dip}} D \left( | n \rangle \langle m |, \rho  \right) +
        \\
        & + \sum_{n < m}  N_T (\omega_{mn}) \Gamma_{nm}^{c/{\rm dip}}  D \left( | m \rangle \langle n |, \rho   \right).
    \end{split}
\end{equation}
Here $D\left( c, \rho  \right) = c\, \rho\, c^{\dag} - \frac{1}{2}\left[c^{\dag} c \, , \, \rho\right]_+$ is the usual dissipator super-operator \cite{zoller_quantum_world_2_doi:10.1142/p983} ($\left[\cdot , \cdot\right]_+$ is the anticommutator), and $N_T (\omega ) = 1/(\exp\left[\omega/(k_B T)\right] - 1)$ is the bosonic thermal population, where $k_B$ is the Boltzmann constant.
The transition rates of the relaxation dynamics are given by
\begin{equation}\label{eq:therm_rates}
    \begin{split}
        &\Gamma^c_{nm} = J_{\rm Ohm}(\omega_{mn})|\braket{n| c^c_{nm}|m}|^2 = \gamma \frac{|\omega_{mn}|}{\omega_c}|\braket{n | a - a^{\dag} | m}|^2,
        \\
        & \Gamma_{nm}^{\rm dip} = J_{\rm rad}(\omega_{mn})|\braket{n| c^{\rm dip}_{nm}|m}|^2 = \kappa \frac{|\omega_{mn}|^3}{\omega_d^3} |\braket{n | s_x | m}|^2,
    \end{split}
\end{equation}
where $\omega_{mn} = \omega_m - \omega_n$ is the difference between the eigenfrequencies of the Rabi Hamiltonian in Eq. \eqref{eq:ham_Rabi}, 
while $J_{\rm Ohm}(\omega ) = \gamma \omega/\omega_c$ is the spectral density of the resistance (cavity bath), which is Ohmic, with photon loss rate $\gamma$, and $J_{\rm rad}(\omega ) = \kappa \omega^3/\omega_{d}^3$ is the spectral density of the radiative modes (dipole bath), which is super Ohmic with dipole decay rate $\kappa$.
Notice that a different choice for these spectral densities does not change our main conclusions, as long as the spectral densities are Ohmic or super Ohmic.
See App. \ref{app:physical_dissipators} for major details regarding the modelling of dissipation.

It is then easy to verify that the steady state of such defined master equation is correctly given by the thermal density matrix $\rho(t=+\infty) = \rho_{T} = e^{- H_{\rm Rabi}/(k_B T)}/\mathcal{Z}$, where $\mathcal{Z} = {\rm Tr}[e^{- H_{\rm Rabi}/(k_B T)}]$.

\section{Relaxation regimes of cavity QED}
\label{sec:relax_regimes}
In this section we will explore the combined effect of light-matter coupling $g$ and dipole asymmetry $\epsilon$ on the open relaxation dynamics of the system.

For the sake of simplicity, through the whole manuscript we only focus on the relevant dipole-cavity resonant case where $\omega_c = \omega_d$.

\subsection{Zero temperature Liouvillian gap}

\begin{figure}
    \centering
    \includegraphics[width=\columnwidth]{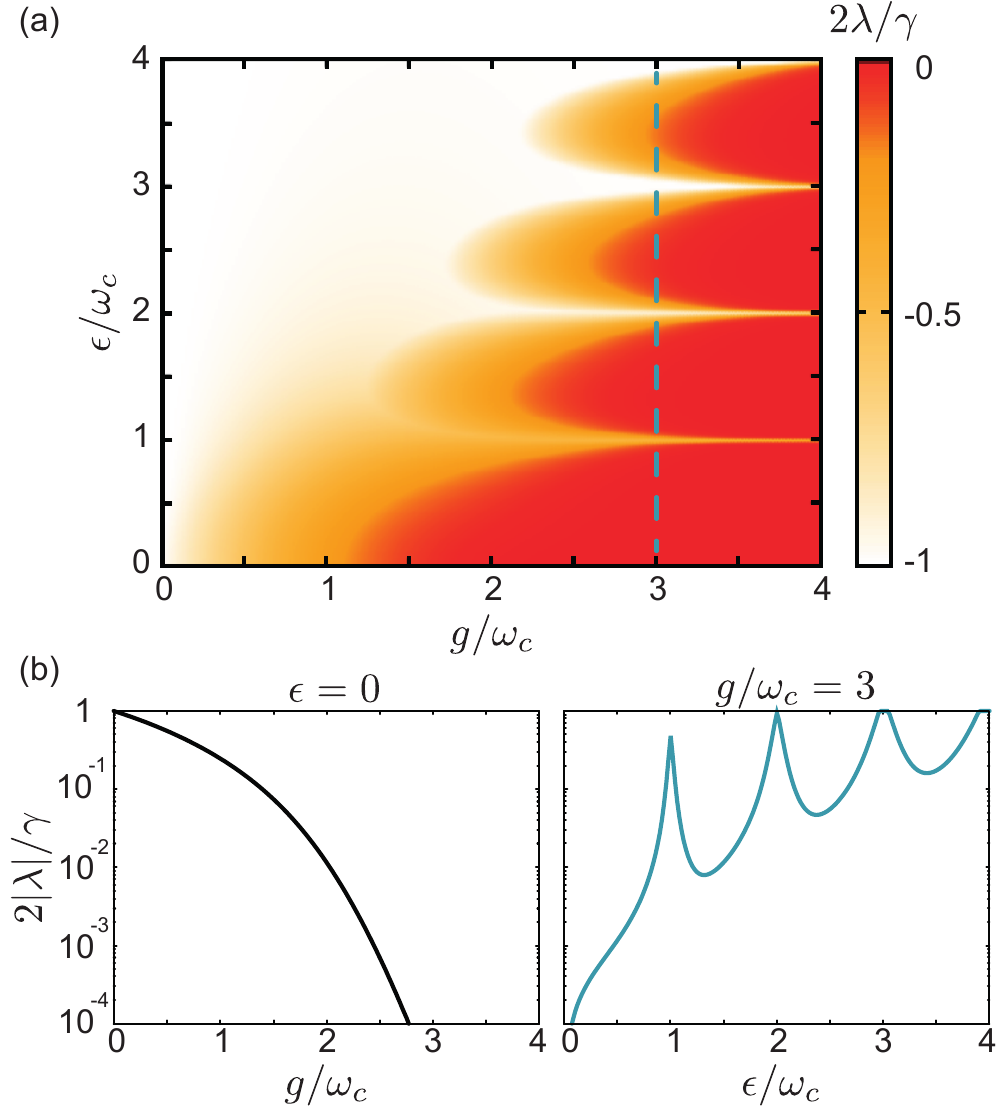}
    \caption{(a) Phase diagram of the Liouvillian gap $\lambda$ as a function of the light-matter coupling $g$ and the dipole asymmetry $\epsilon$. Parameters: $\gamma = \kappa/4 = 0.05 \omega_c$, $\omega_d = \omega_c$.
    (b) A cut of the phase diagram at $\epsilon=0$ as a function of the light-matter coupling $g$, in logscale. (c) A cut of the phase diagram at $g/\omega_c = 3$ as a function of the dipole asymmetry $\epsilon$, in logscale. }
    \label{fig:2}
\end{figure}

To have a first indication about the relaxation properties of the system we consider the Liouvillian gap $\lambda = {\rm Re}[\lambda_1]$ \cite{Cirac_PhysRevA.86.012116,Minganti_PhysRevA.98.042118,Macieszczak_PhysRevLett.116.240404}, obtained from the spectrum $\lbrace{\lambda_n\rbrace}$, $n=0,1,2 \ldots$, of the total Liouvillian operator $\mathcal{L} = \mathcal{L}_H + \mathcal{L}_c + \mathcal{L}_{\rm dip}$ defined from Eq. \eqref{eq:master_eq_total} \cite{qutip_JOHANSSON20131234}.
This quantity provides the slowest relaxation rate of the system, describing the long-time evolution of the system, for which before reaching its thermal steady state the density matrix decays as \cite{Cirac_PhysRevA.86.012116, Macieszczak_PhysRevLett.116.240404}
\begin{equation}
    \lim_{t\rightarrow \infty}\rho(t) \approx \rho_T + \rho_1e^{\lambda t}
\end{equation}
It is worth noticing that using the Liouvillian gap to characterize the relaxation toward equilibrium is not always straightforward and may cause problems in more complex many-body systems \cite{Mori_PhysRevLett.125.230604}.
Anyway we will see that in our case it works without problems or ambiguities,
correctly matching the expected physical predictions and 
giving a correct and clear picture of how relaxation works as a function of our control parameters $(g,\epsilon)$.

We consider only the zero temperature case $T=0$, which is the relevant case for superconducting cavity QED setups \cite{yoshihara_circuitQED_beyond_USC2017}. The same picture holds also for finite temperature, provided that $k_b T \lesssim \hbar \omega_c, \hbar \omega_d$, where $k_b$ is the Boltzmann constant. When the temperature grows larger, and $k_b T > \hbar \omega_c$ USC effects are pushed to much larger light-matter coupling values \cite{Pilar2020thermodynamicsof}.

In Fig. \ref{fig:2}(a) we show the Liouvillan gap as a function of the light-matter coupling and the dipole asymmetry, $\lambda (g, \epsilon )$.
At small light-matter coupling $g \sim 0$, the effect of increasing $\epsilon$ is to progressively rotate the dipole eigenstates from the $s_z$-basis to the $s_x$-basis, decreasing the value of the matrix element in the dipole transition rate in Eq. \eqref{eq:therm_rates}.
However the vanishing matrix element is compensated by the increasing energy difference between the dipole levels, giving larger contribution from the radiative spectral density of the bath, $J_{\rm rad}\sim \omega^3$. This can be seen by explicitly computing the dipole transition rate at $g=0$ using the bare uncoupled dipole states in Eq. \eqref{eq:bare_uncoupled_dipole_states}, for which we have
\begin{equation}
    \begin{split}
        \Gamma_{\rm LR}^{\rm dip} &= \frac{\kappa}{4} \left( 1 + \frac{\epsilon^2}{\omega_d^2}\right)^{3/2}\cos \left( \tan^{-1} \left( \frac{\epsilon}{\omega_d} \right) \right) 
        \\
        &= \frac{\kappa}{4} \sqrt{ 1 + \frac{\epsilon^2}{\omega_d^2}}.
    \end{split}
\end{equation}
In the specific case $\kappa = 4\gamma$ this rate is always larger than the bare photon loss set by $\gamma$ which becomes the slowest relaxation time scale, and so we have $\lambda = - \gamma/2$. 
In such conditions, the Liouvillian gap does not show any structure as long as the light-matter coupling remains small. 
It is worth noticing that replacing the super Ohmic radiative spectral density with a Ohmic spectral density would give a too slow increase of the decay rate as a function of the transition frequency to compensate the effect of the vanishing matrix element of the dipole transition rate. As a result we would have that $(\Gamma_{\rm LR}^{\rm dip})^{\rm Ohm} = \kappa/(4\sqrt{1+\epsilon^2/\omega_d^2})$, and so the overall relaxation rate would decrease as a function of $\epsilon$ at very weak coupling $g / \omega_c\simeq 0$ to then increase again at slightly larger coupling.
In this case the Liouvillian gap $\lambda$ would exhibit a different structure as a function of $(g,\epsilon)$, in the weak coupling limit. Giving rid of this weak-coupling features only considering a radiative bath (and thus a super Ohmic spectral density) for the dipole highlights the effect of the USC, making this choice particularly meaningful.

In the USC regime, $g/\omega_c \gg 1$, for small dipole asymmetry $\epsilon\simeq 0$, the Liouvillian gap goes to zero monothonically with an exponential behaviour $\lambda \sim - \exp[ - g/\omega_c ]$, as is clearly visible from Fig. \ref{fig:2}(b).
Increasing the dipole asymmetry, $\epsilon$, we observe the emergence of lobes where the Liouvillian gap approaches zero $\lambda \sim 0$, separated by a narrow region where relaxation is partially restored and $\lambda \sim - \gamma/2$. This is shown in Fig. \ref{fig:2}(c), where we fixed $g/\omega_c = 3$ and we plot $\lambda$ as a function of $\epsilon$.
Quite surprisingly, these narrow gaps between the lobes appear only when $\epsilon\simeq \omega_c\times k$, where $k=1,2,3\ldots$ is an integer number.
Moreover this lobular structure is present also in the higher Liouvillian eigenstates, suggesting important physical consequences for the system.

\subsection{Relaxation breakdown in the USC regime}

The thermalization exponential slow-down pointed out by the spectral analysis of the Liouvillian $\mathcal{L}$ can be understood as an interplay between the USC spectral properties and transition rates in Eq. \eqref{eq:therm_rates}
(due to the dressing of the jump operators in the USC regime \cite{Solano_PhysRevA.96.013849, Blais_dissipationUSC_first_PhysRevA.84.043832}, see App. \ref{app:master_eq}).
Here we analyze in detail the symmetric case, when $\epsilon = 0$, which will provide the basic tools to understand the whole phase diagram of Fig. \ref{fig:2}(a).

We start by transforming the original Rabi Hamiltonian through the unitary transformation $U_{\rm pol} = \exp \left[ g/\omega_c (a - a^{\dag})s_x \right]$, and obtaining the \emph{Rabi polaron} Hamiltonian ($\hbar = 1$)
\begin{equation}\label{eq:H_rabi_polaron}
    \tilde H_{\rm Rabi} = \omega_c a^{\dag} a + \epsilon s_x + \frac{\omega_d}{2}\left[ \mathcal{D}(g/\omega_c) \tilde s_+ + \mathcal{D}^{\dag}(g/\omega_c) \tilde s_-  \right].
\end{equation}
Here $\tilde s_{\pm} = s_z \pm i s_y$ are the raising/lowering operators along the $s_x$-axis, while $\mathcal{D}(g/\omega_c) = \exp \left[ g/\omega_c (a - a^{\dag} ) \right]$ is the usual displacement operator.

Since both cavity and dipole dissipative operators are unaffected by the polaron transformation $U_{\rm pol}(a- a^{\dag}) U_{\rm pol}^{\dag} = (a- a^{\dag}) ,~ U_{\rm pol} s_x U_{\rm pol}^{\dag} = s_x$, the general master equation defined in Eqs. \eqref{eq:master_eq_total}-\eqref{eq:liouvillian_H}-\eqref{eq:liouvillian_c_dip} is still valid, with the only difference that the eigenstates $|n \rangle, |m \rangle$ appearing in the transition rates in Eq. \eqref{eq:therm_rates} are now replaced with the eigenstates of the Rabi polaron Hamiltonian in Eq. \eqref{eq:H_rabi_polaron}.

\begin{figure}
    \centering
    \includegraphics[width=\columnwidth]{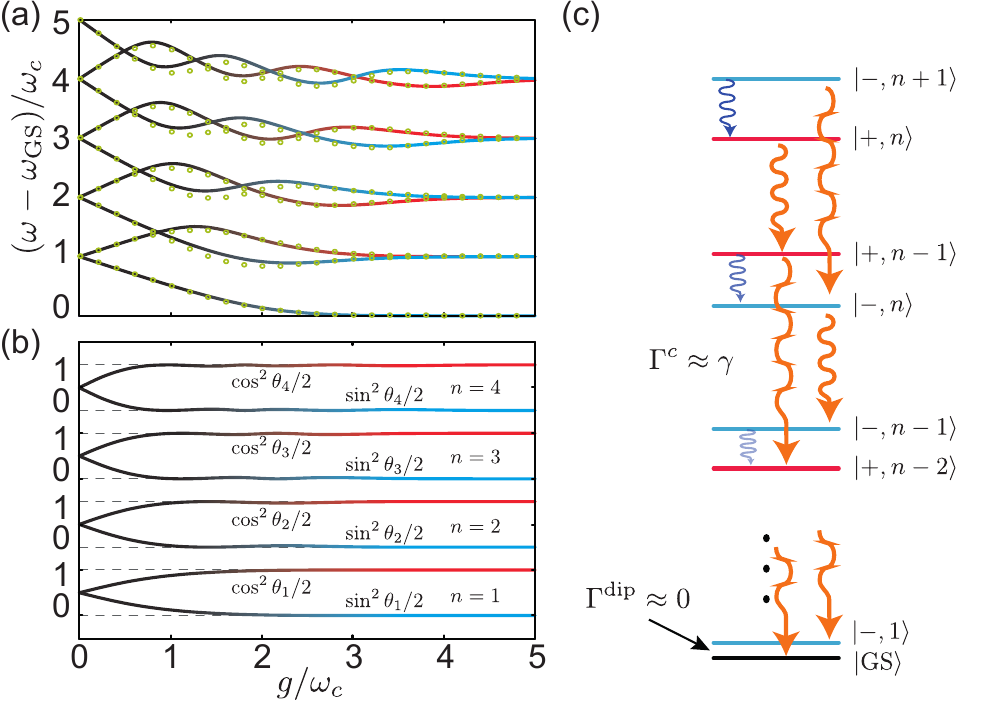}
    \caption{(a) Spectrum of the Rabi model as a function of the light-matter coupling $g$ at fixed $\epsilon=0$. The solid lines are the result of full diagonalization, and the color red/blue are only meant to match the color-code in (c). The yellow dots are given by the analytic Eq. \eqref{eq:rabi_spec_approx_irish} in Appendix \ref{app:gRWA}.
    (b) $\cos \theta_n/2$, $\sin \theta_n/2$ given by Eq. \eqref{eq:sincos_hopfields} for each $n$ block as a function of the light-matter coupling $g$.
    (c) Scheme of the relaxation mechanism. The cavity relaxes jumping mainly between $++$ or $--$ dressed states, while for the dipole is mainly between $+-$ states. The orange curly arrows represent the decay of the photon from an upper state to a lower one, while the blue curly arrows represent the decay of the dipole.
    In the USC limit the dipole does not relax anymore.
    Parameters: $\epsilon = 0$, $\omega_c = \omega_d$.}
    \label{fig:3}
\end{figure}

As reported in \cite{Irish_PhysRevLett.99.173601} and detailed in Appendix \ref{app:gRWA}, the polaron Rabi Hamiltonian supports a generalized rotating-wave approximation (gRWA) and thus follows the structure of the Jaynes-Cummings model, with the approximated conservation of the polaron excitation number $\hat N_{\rm exc}^z = a^{\dag} a + s_z$.
Its eigenstates are then given by the usual dressed states
\begin{equation}
    \begin{split}
        & | +, n \rangle = \cos \frac{\theta_n}{2} | \downarrow, n \rangle + \sin \frac{\theta_n}{2} | \uparrow, n-1 \rangle,
        \\
        & | -, n \rangle = - \sin \frac{\theta_n}{2} | \downarrow, n \rangle + \cos \frac{\theta_n}{2} | \uparrow, n-1 \rangle,
    \end{split}
\end{equation}
where $\theta_n$ is given in Appendix \ref{app:gRWA}. 
The ground-state of the system is simply the uncoupled vacuum state
\begin{equation}
    | {\rm GS} \rangle = | \downarrow, 0 \rangle .
\end{equation}
In order to appreciate the quality of this approximation, in Fig. \ref{fig:3}(a) we compare the spectrum obtained from the exact diagonalization (solid lines) and from the gRWA analytical formula reported in Appendix \ref{app:gRWA} (yellow dots), from which is quite clear that the gRWA gives very good results.

Relaxation can then be understood from the dressed state perspective \cite{Cohen_AtomPhoton_book:91199254} and in Appendix \ref{app:matrix_elm_trans-rate_symm} we explicitly compute the transition rate in Eq. \eqref{eq:therm_rates}.

From the explicit expression for the Hopfield coefficients $\sin,\cos$  (present in the Appendix \ref{app:gRWA}), we find that in the infinite-coupling limit
\begin{equation}
    \begin{split}
        & \lim_{g/\omega_c \rightarrow \infty}\cos \frac{\theta_n}{2} = 1
        \\
        & \lim_{g/\omega_c \rightarrow \infty}\sin \frac{\theta_n}{2} = 0.
    \end{split}
\end{equation}
This is clearly shown in Fig. \ref{fig:3}(b) where we plot the Hopfield coefficient analytically computed through the gRWA  for a few lowest eigenstates.
Using this observation together with the matrix element computed in Appendix \ref{app:matrix_elm_trans-rate_symm} we can build the transition rates in Eq. \eqref{eq:therm_rates}, arriving to the conclusion that the only non-negligible transitions in the USC regime are
\begin{equation}
    \begin{split}
        & \lim_{g/\omega_c \rightarrow \infty}\Gamma_{ (+, n)  (+, n-1)}^c = \gamma\frac{ \omega_{+, n}- \omega_{+, n-1} }{\omega_c} \approx \gamma
        \\
        & \lim_{g/\omega_c \rightarrow \infty}\Gamma_{ (-, n)  (-, n-1)}^c = \gamma\frac{ \omega_{-, n}- \omega_{-, n-1} }{\omega_c} \approx \gamma
        \\
        & \lim_{g/\omega_c \rightarrow \infty}\Gamma_{(-, n )( + , n-1)}^{\rm \, dip} = \kappa \left(\frac{ \omega_{- , n}- \omega_{+ , n-1} }{\omega_c}\right)^3 \approx 0.
    \end{split}
\end{equation}
The fact that the dipole transition rate goes to zero $\Gamma_{(-, n ) (+ , n-1)}^{\rm \, dip} \approx 0$ follows from the approximate degeneracy of the states $| -, n \rangle, | +, n-1 \rangle$ in the USC limit, for which $\omega_{- , n}- \omega_{+ , n-1} \approx 0$, while $\omega_{+ , n}- \omega_{+ , n-1} \approx \omega_{- , n}- \omega_{- , n-1} \approx \omega_c$.

\begin{figure}
    \centering
    \includegraphics[width=\columnwidth]{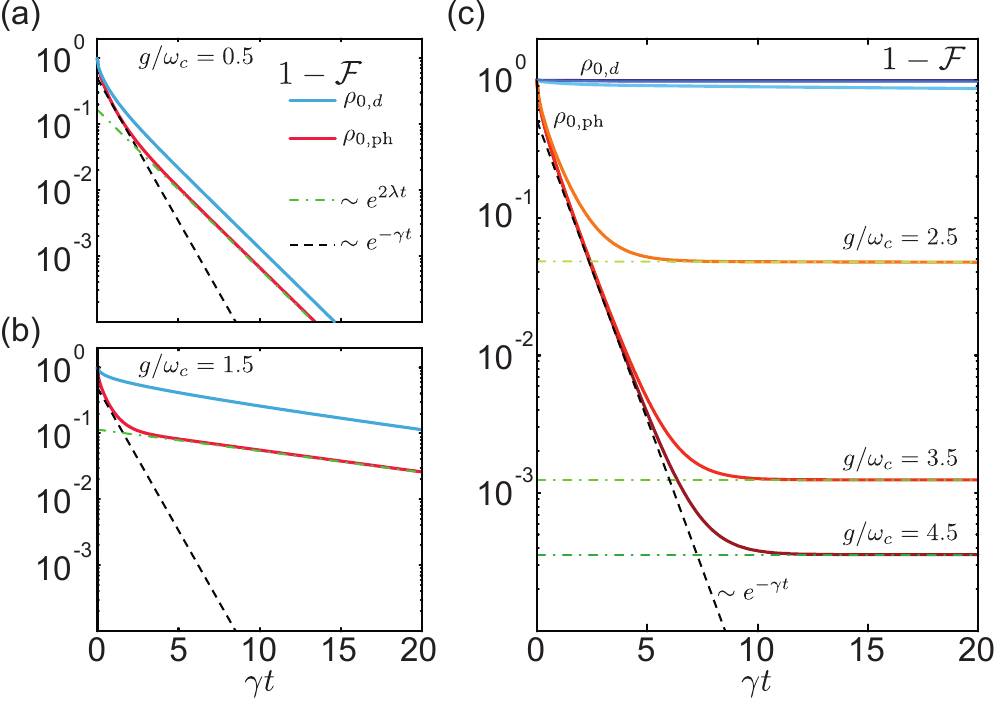}
    \caption{Infidelity time-evolution.
    (a-b) Weak and intermediate coupling regime, the infidelity is calculated starting with the initial density matrix $\rho_{0,d}$ (blue solid line) and $\rho_{0,\rm ph}$ (red solid line). The black dashed line highlights the bare decay scaling $s_0\exp[-\gamma t]$ while the green dot-dashed line marks the scaling given by the Liouvillian gap $s'_0\exp[2\lambda t]$ (here $s_0, s'_0$ are arbitrary offsets). The light-matter coupling is given in the panels.
    (c) USC regime, the infidelity is calculated starting with the initial density matrix $\rho_{0,d}$ (weak-blue, mid-blue, deep-blue solid lines) and $\rho_{0,\rm ph}$ (orange, red, dark-red solid lines). For both initial state the couplings are $g/\omega_c=2.5, 3.5, 4.5$ going from the lighter to the darker color. The mid-blue and deep-blue lines representing the infedelity starting from $\rho_{0,d}$ are almost overlapping and not well distinguishable.
    Parameters: $\omega_c=\omega_d$, $\epsilon=0$, $\gamma = \kappa/4=0.1\omega_c$, $T=0$. }
    \label{fig:4}
\end{figure}

Here we realize that the USC Liouvillian gap suppression observed in Fig. \ref{fig:2}(a-b) is only due to a suppression of the dipole transition rates only, while the cavity transition rates return to their bare uncoupled values when the USC regime is reached.
In the infinite coupling limit the exponential slowdown become a proper cavity-induced breakdown of the relaxation of the dipole, resulting in a localization transition similar to what happens in the so-called spin-boson model \cite{Leggett_RevModPhys.59.1}.
The schematic representation of the remaining relaxation channels is shown in Fig. \ref{fig:3}(c).
It is important to stress that what described above holds only in the infinite coupling limit, and for finite values of the light-matter coupling $g$, the long-time dynamics is always given by the finite Liouvillian gap, both for the dipole and the cavity. 
However, if we consider the relaxation of a single photon (in the polaron frame) in the USC regime, initializing the system in the state $\rho_{0,\rm ph} = |1_{\rm ph}, \downarrow\rangle \langle 1_{\rm ph}, \downarrow |$ we observe a transient dynamics where the system relaxes as a bare cavity photon as $\exp [-\gamma t ]$, and arriving progressively closer to the equilibrium state before entering in the long-time dynamics settled by the suppressed Liouvillian gap. On contrary initializing the state in a pure dipole excitation (in the polaron frame) $\rho_{0,d} = |0_{\rm ph}, \uparrow\rangle \langle0_{\rm ph}, \uparrow|$, the system enters almost immediately in the long-time dynamics, remaining frozen there.
This is well visible from the time evolution of the infidelity with respect to the thermal state (or, at $T=0$, the groundstate) $1-\mathcal{F}=1-{\rm Tr}[\sqrt{\sqrt{\rho(t)}\rho_{T}\sqrt{\rho(t)}}]$ \cite{qutip_JOHANSSON20131234}, that is shown in Fig. \ref{fig:4}.
In particular in Fig. \ref{fig:4}(a-b) we show, for comparison, the time evolution at weak and intermediate coupling. It is well visible the slower relaxation at higher coupling, but still the two different states decay to the groundstate in a similar way. In Fig. \ref{fig:4}(c) on contrary there is a strong asymmetry in the dipole and photon state decay and it is clear that asymptotically a single polaron photon decays with its bare decay rate, while a polaron dipole excitation is completely frozen and does not decay.
This behaviour is not specific for the infedelity only, but it is common for most of the observables and states of this system.

This result can be physically interpreted from a polaronic perspective:
the USC cavity vacuum heavily dresses the dipole with  virtual photons, which inhibit its ability to tunnel from one side to the other of its double well potential.
Because of the radiative nature (but in the Ohmic case as well) of its dissipation mechanism, the dipole can loose energy only moving between the two wells (i.e. tunneling), and the faster it moves the stronger it dissipates.
In this regime of heavy dressing by virtual-photon tunneling becomes extremely slow and so the dipole's rate to release energy in the bath.

\section{USC multi-photon resonant tunneling}
\label{sec:USC_res_tunnel}
In this section we are going to explore more in detail the nature of the gaps between the relaxation-slowdown lobes in Fig. \ref{fig:2}(a).
In these narrow regions the system can relax as is almost unaffected by the USC suppression of tunneling described in the previous section. 
However, if we artificially remove the cavity dissipation, $\gamma = 0$, we see that these narrow gaps disappear. This suggests that the suppression of tunneling described above is still present for $\epsilon\neq 0$, but a new resonant mechanism appears, allowing the dipole to tunnel again by exchanging photons with the cavity.
This effect is the cavity analogous of resonant tunneling in electronic setups interacting with vibrational degrees of freedom, and thus establishing a connection between USC cavity QED and Franck-Condon physics in molecular-electronic setups \cite{Andreev_PhysRevB.74.205438, Joachim_phonon_electron_tranport_PhysRevLett.114.016804, Eaves_phonon_resonant_tunnel_2016_PhysRevLett.116.186603}.

After analyzing the relaxation properties from the spectral features of the system, as in the previous section, we show that signature of this physics are also present in quantities that are not strictly related to relaxation and real-time dynamics, such as transmission spectra.

\subsection{Diagonalization of the asymmetric Rabi Hamiltonian}
\label{sec:diag_asymm_rabi}
\begin{figure}
    \centering
    \includegraphics[width=\columnwidth]{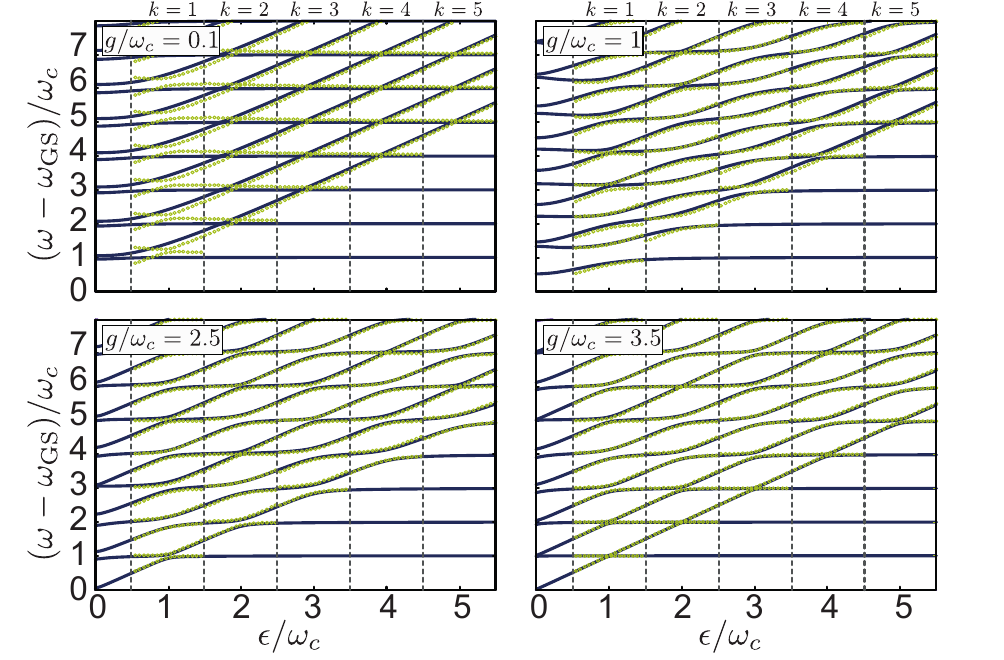}
    \caption{Spectrum of the Rabi model as a function of the dipole asymmetry $\epsilon$ for various $g/\omega_c = 0.1, 1, 2.5, 3.5$ light-matter couplings. The solid lines are the result of exact diagonalization while the yellow dot are given by the analitical formula in Eq. \eqref{eq:Rabi_spectrum_anal_eps}. For $k>1$ the yellow dots do not cover the lower lines. This is because our approximation treats these eigenstates as bare photon state for which the energy is trivially $n\omega_c$. In order to highlight the part of the spectrum where cavity and dipole are effectively coupled we do not put the yellow dots on these trivial eigenvalues. }
    Parameters $\omega_d = \omega_c$.
    \label{fig:5}
\end{figure}

As in the symmetric case $\epsilon= 0$, also the asymmetric Rabi model, $\epsilon \neq 0$ is approximately block-diagonal, as a consequence of the general form of the displacement operators. 
However here the situation is more complicated and we cannot find a unique formula that fits the whole spectrum for every $(\epsilon, g)$, but we can only have analytic expressions valid near to each resonance.

We start by noticing that Eq. \eqref{eq:H_rabi_polaron} is written in a form that calls for the gRWA, provided that the system has an \emph{asymmetric resonance} $\epsilon \simeq \omega_c \times  k$, with $k=1,2,\ldots$.
Differently from the usual Jaynes-Cummings model, and the gRWA of the symmetric Rabi model, here we need to take the dipole basis as an eigenstate of $s_x$.
Moreover, considering higher resonances at $\epsilon= \omega_c, 2\omega_c, 3\omega_c \ldots$ is well motivated by the fact that the displacement operator contains all power of creation/annihilation operators, giving access to multi-photon processes with higher frequencies. 
This is indeed well visible considering the normal-order expansion \cite{Glauber_PhysRev.177.1857}
\begin{equation}\label{eq:displacement_normal_order}
    \mathcal{D}(x) = e^{-x^2/2} \sum_{n,m=0}\frac{(x a^{\dag})^n}{n!}\frac{(-xa)^m}{m!}.
\end{equation}
From this expression is also clear that the non-linear interaction term in the polaron Hamiltonian in Eq. \eqref{eq:H_rabi_polaron} is exponentially suppressed by the factor $\sim \omega_d e^{-g^2/(2\omega_c^2)}$. As a consequence, when 
\begin{equation}
    \epsilon, \omega_c > \omega_d e^{-g^2/(2\omega_c^2)}
\end{equation}
the polaron light-matter interaction becomes perturbative, and we can adopt the gRWA.
Notice that this correspond to keep only the terms $n<m$ with $\omega_{c}(m-n) \simeq \epsilon$ in Eq. \eqref{eq:displacement_normal_order}, so, even if the interaction is perturbative, is still multi-photon and thus highly non-linear.

The asymmetric polaron Rabi Hamiltonian can then be approximately diagonalized around each $k$-resonance by projecting it on the states $\lbrace{ | \leftarrow, n \rangle , | \rightarrow, n-k \rangle \rbrace}$ and the ground-state is simply given by $|{\rm GS} \rangle \approx  | \leftarrow, 0 \rangle$.
As for the symmetric case explained in App. \ref{app:gRWA} this treatment is equivalent to a quasi-degenerate pertubation theory on polaron interaction Hamiltonian.

The Hamiltonian can be then expressed succinctly in a matrix form, as the sum of $2\times 2$ blocks
\begin{equation}\label{eq:block_Ham_k-res}
\begin{split}
    &\tilde H_{\rm Rabi}^k \approx \sum_{n=1}^{\infty} \frac{\omega_ck - \epsilon}{2}\sigma_x^{(n,k)} + \frac{\omega_d}{2}\mathcal{D}_{n\,n-k}\, \sigma_z^{(n,k)}
    \\
    &+ \frac{2\omega_c n - \omega_ck}{2}\mathds{1}_{(n,k)},
\end{split}
\end{equation}
where $\sigma_{x,y,z}^{(n,k)}$ are the Pauli matrices for each $n=1,2,\ldots$ block for the $k$-resonance, while
\begin{equation}
    \mathcal{D}_{n\, n-k} = \frac{g^{k}}{\omega_c^{k}} e^{-\frac{g^2}{2\omega_c^2}}L_{n-k}^{(k)}\left( g^2/\omega_c^2 \right) \sqrt{\frac{(n-k)!}{n!}}
\end{equation}
is the $n,n-k$ matrix element of the displacement operator \cite{Glauber_PhysRev.177.1857}. Here $L_{m}^{(l)}(x)$ is the special Laguerre polynomials.
The excited eigenfrequencies are then given by
\begin{equation}\label{eq:Rabi_spectrum_anal_eps}
\begin{split}
    &\omega_{k,n,\pm}^{\rm R} = \omega_c \left( n - \frac{k}{2} \right) \pm \frac{1}{2}\sqrt{ \left( \omega_c k - \epsilon \right)^2 + \omega_d^2 \mathcal{D}_{n\,n-k}^2 }.
\end{split}
\end{equation}
Since the displacement operator has diagonal matrix element different from zero $\mathcal{D}_{nn}\neq 0$, one should consider the dipole basis states composed by dipole states oriented along $\sim \cos \phi s_x + \sin \phi s_z$, with a certain angle $\phi$ given by $\mathcal{D}_{nn}$.
Including these corrections makes the analytical formula in general quite complicated, having a simple expression only for the ground-state, which is
\begin{equation}
    \omega^{\rm R}_{0,0} = -\frac{\sqrt{\epsilon^2 + \omega_d^2 e^{-g^2/\omega_c^2}}}{2}.
\end{equation}
However, in the USC regime $\phi \sim 0$ is a small angle and we can thus neglect it, proceeding with the simple $s_x$ picture developed above. 

In Fig. \ref{fig:5} we compare the real spectrum to the one obtained from the gRWA at each resonant point. In the USC limit, when $g/\omega_c \gg 1$ the agreement is very good.

The eigenstates are now given in terms of a multi-photon version of the $s_x$-polarized Jaynes-Cummings dressed states, fully characterized by the Hopfield coefficients $\cos \theta_{(k,n)}/2, \sin \theta_{(k,n)}/2$, generalizing the symmetric case in Appendix \ref{app:gRWA}.
When the resonance condition $\epsilon = \omega_c \times k$ is satiesfied, the system eigenstates become 
\begin{equation}
    \begin{split}
        & | +_{(k,n)} \rangle = \frac{1}{\sqrt{2}}\left( | \leftarrow, n \rangle +  | \rightarrow, n-k \rangle \right)
        \\
        & | -_{(k,n)} \rangle = \frac{1}{\sqrt{2}}\left( | \leftarrow, n \rangle - | \rightarrow, n-k \rangle \right),
    \end{split}
\end{equation}
and the ground state is $| {\rm GS} \rangle = | \leftarrow, 0 \rangle$.

Repeating the analysis on the matrix elements in Appendix \ref{app:matrix_elm_trans-rate_symm}, we realize that the $s_x$ operator can only connect dressed states of the same $(k,n)$ block, for which the only non-diagonal non-zero matrix element is 
\begin{equation}\label{eq:sx_matrix_elm_asymm}
    \braket{+_{(k,n)}| s_x | -_{(k,n)} } = \cos \frac{\theta_{(k,n)}}{2}\sin \frac{\theta_{(k,n)}}{2}.
\end{equation}
Each block is disconnected by the others and the ground-state is disconnected from all other states. Therefore relaxation toward equilibrium is still suppressed from the USC also when $\epsilon\neq 0$.

On contrary, the cavity is still able to efficiently dissipate. 
So, when hitting a $k$-resonance, also the dipole can lose energy by exchanging $k$-photons with the cavity, which are consequently flushed out.
This resonant tunneling effect provides a relaxation channel for the dipole, as depicted in Fig. \ref{fig:6}(a) and give the proper explanation for the gaps between the lobes observed in Fig. \ref{fig:2}(a).


We conclude this subsection by highlighting that: in the polaron frame, the USC open dynamics is mainly given by a polaronic version of the standard text-book dressed state master equation dynamics \cite{Cohen_AtomPhoton_book:91199254}.

\subsection{Multi-photon oscillations and cavity-mediated relaxation}

\begin{figure}
    \centering
    \includegraphics[width=\columnwidth]{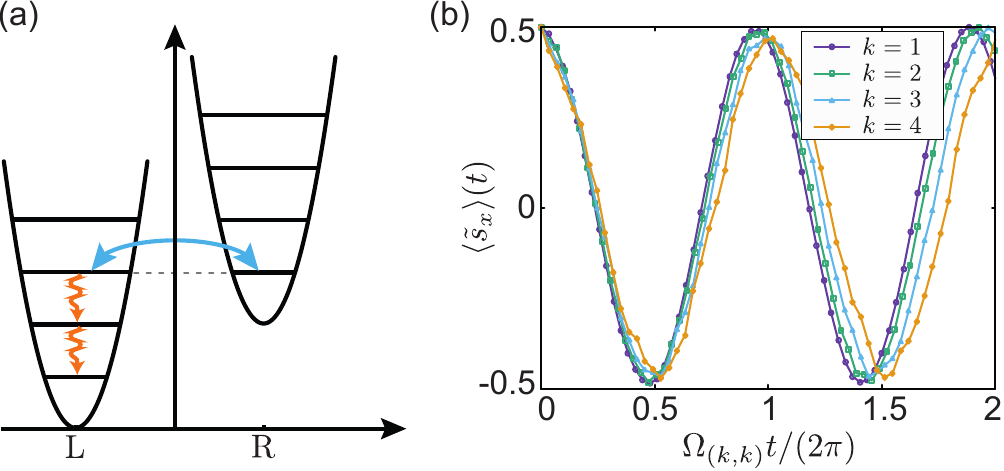}
    \caption{(a) Schematic view of the resonant tunnel mechanism. In the USC regime, the dipole can switch well by exchanging $k$-photons with the cavity. (b) Rabi oscillations data collapse. Each curve is labelled by its resonant index $k$ and represents $\braket{\tilde s_x} = e^{k\gamma/2 t}(\braket{s_x}+1/2) - 1/2$. For each $k$-curve the time is normalized on its respective $k$-Rabi frequency, $\Omega_{(k,k)}/(2\pi)$. In this way it is clear how our analytical description fits very well the full numerics.
    Parameters: $\omega_d = \omega_c$, $g/\omega_c = 3$, $\gamma=\kappa/4=0.002\omega_c$.}
    \label{fig:6}
\end{figure}

Here we illustrate how the polaronic dressed state picture emerges clearly in the full time-dependent dynamics. 
As a striking example we show that the system undergoes to damped Rabi oscillations, as in traditional cavity QED systems described by the Jaynes-Cummings model.
However here, depending from the resonance condition, the Rabi oscillations involve multiple photons \cite{ashhab_nori_PhysRevA.81.042311, Savasta_2015_multiPhotonRabi_PhysRevA.92.063830, Ken_2015_PhysRevA.92.023842} and must be interpreted as tunneling oscillations for the dipole.

From the block-Hamiltonian in Eq. \eqref{eq:block_Ham_k-res} we can derive the Rabi frequency of the $k$-resonance multi-photon Rabi oscillations reading
\begin{equation}
    \Omega_{(k,n)} = \omega_d \frac{g^{k}}{\omega_c^{k}} e^{-\frac{g^2}{2\omega_c^2}}L_{n-k}^{(k)}\left( g^2/\omega_c^2 \right) \sqrt{\frac{(n-k)!}{n!}},
\end{equation}
where $n\geq k$ is the total number of photons involved.

Differently from usual Rabi oscillations in cavity QED, here the dipole oscillates between the right and left states of its asymmetric double well potential, for which we can call them \emph{tunneling oscillations}.
The relevant quantity to follow is then $\braket{s_x}(t)$ (on contrary to traditional Rabi oscillations, visible looking at $\braket{s_z}(t)$, in standard notation).

We then numerically simulate $\braket{s_x}(t)$ starting from the initial state $|\psi_0 \rangle = |\rightarrow, 0\rangle$ (in the polaron frame). When $\gamma < \Omega_{(k,k)}$, we observe a very good fit on the curve
\begin{equation}
    \braket{s_x}(t) \approx e^{-\frac{k\gamma}{2} t} \frac{\cos \left[  \Omega_{(k,k)}t \right] + 1}{2} - \frac{1}{2}.
\end{equation}
Notice that the overall decay rate is given by $\sim k\times \gamma/2$, with a factor $k$. This takes into account that the photon decay increase linearly with the number of photons involved, which, in this case is properly $k$.
In Fig. \ref{fig:6}(b) we show the Rabi oscillations data collapse for various resonant values $\epsilon = \omega_c, 2\omega_c, 3\omega_c, 4\omega_c$. For each $k$ the curve is plotted against its normalized time $\tilde t = \Omega_{(k,k)}t/(2\pi)$ and is normalized to remove the exponential decay accordingly to $\braket{\tilde s_x} = e^{k\gamma/2 t}(\braket{s_x}+1/2) - 1/2$.


\subsection{Response functions and higher-order processes}

\begin{figure}
    \centering
    \includegraphics[width=\columnwidth]{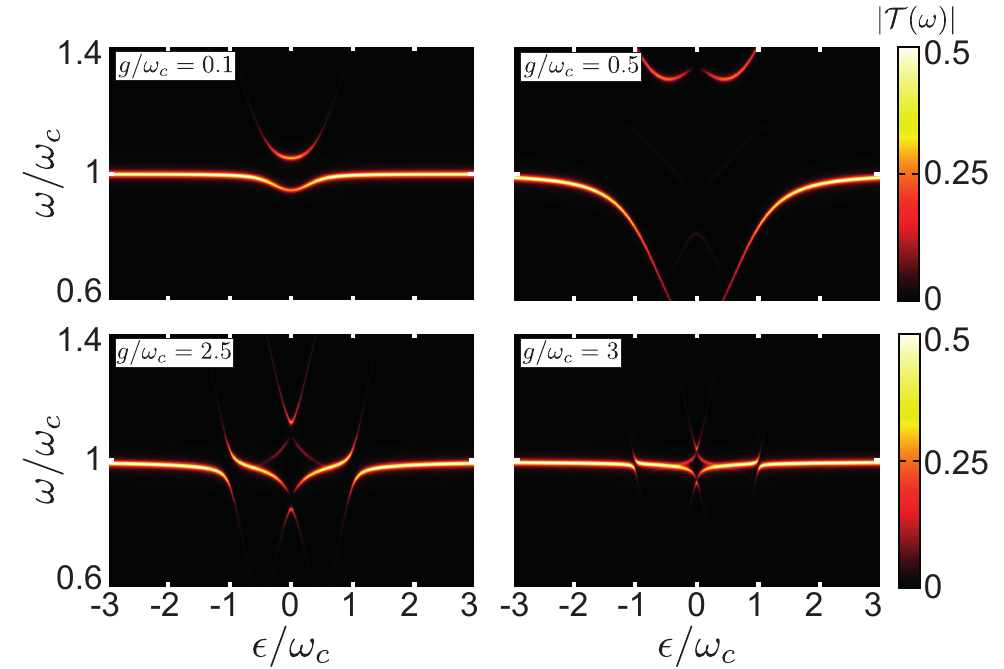}
    \caption{Current transmission $|\mathcal{T}(\omega )|$ as a function of the dipole asymmetry $\epsilon$ and the probe frequency $\omega$, for various $g/\omega_c = 0.1, 0.5, 2,2.5$ light-matter couplings. 
    Parameters: $\omega_d = \omega_c$, $k_b T = 0.2 \hbar \omega_c$, $Q=\omega_c/\gamma=10^2$.}
    \label{fig:7}
\end{figure}

Here we take a quick detour from the investigation of the relaxation properties of the system and we focus our attention more specifically on how the spectral features analyzed so far manifest themselves through standard transmission measurements.

This is particularly important because current experiments, for instance in circuit QED, cannot easily probe the time-dependent dynamics and thus have no direct access to measuring how the light-matter coupling affects relaxation.
Nevertheless, since we have seen that relaxation is in the end determined by the eigenstates of the system, measuring some specific spectral features can be an indirect indication that the system follows the physics described in the previous sections.

We start considering a weak probe current entering in the LC-circuit and we look for the transmitted current.
With the help of linear response theory (see Appendix \ref{app:linear_resp}), the current response is mainly given by the cavity structure factor
\begin{equation}\label{eq:cavity_struct}
    \mathcal{S}_c(\omega ) = \frac{\hbar Z_{LC}}{2} \sum_{n,m}\frac{e^{-\hbar \omega_{n}/(k_b T)}}{\mathcal{Z}} \left| \braket{n| a - a^{\dag} |m} \right|^2\delta(\omega - \omega_{mn}).
\end{equation}
Here $\mathcal{Z}=\sum_n e^{-\hbar \omega_{n}/(k_b T)}$ is the thermal equilibrium partition function of the system, and $Z_{LC}$ is the characteristic cavity impedance parameter defined in App. \ref{sec:cQED_ham}.
The system circuit impedance is then defined as
\begin{equation}
    Z_{\rm sys}(\omega ) = -\frac{i \omega \mathcal{S}_c(\omega )}{\hbar},
\end{equation}
and consequently the current transmission function
\begin{equation}
    \frac{I_{\rm out}}{I_{\rm in}} = \mathcal{T}(\omega)= \frac{Q^{-1}}{Q^{-1} + Z_{\rm LC}/Z_{\rm sys}(\omega )}.
\end{equation}
Here $Q=\omega_c/\gamma$ is the LC cavity quality factor.

In Fig. \ref{fig:7} we show the current transmission $|\mathcal{T}(\omega)|$ as a function of the dipole asymmetry $\epsilon$ and the probe frequency $\omega$. To mimic experimental conditions, we consider a fixed temperature $k_b T\simeq 0.2 \hbar \omega_c$. 

For coupling strength up to $g/\omega_c \lesssim 0.5$ the transmission spectrum exhibits the usual Jaynes-Cummings polaritonic (or dressed state) behaviour. At $\epsilon\neq 0$ the response mainly follows the bare LC circuit response, while the maximum hybridization is at $\epsilon=0$, with maximum Rabi splitting between the upper and lower dressed state branches.
This is well visible in the first panel of Fig. \ref{fig:7}.

For increasing coupling strength, as in the second panel of Fig. \ref{fig:7}, where $g/\omega_c = 1$, the upper branch of the  transmission spectrum starts to vanish exactly at $\epsilon=0$, signalling that we are entering the USC regime.

At even larger couplings the transmission is drastically changed. This is well visible from the two lower panels in Fig.  \ref{fig:7}, where $g/\omega_c = 2.5, 3$.
In the region around $\epsilon=0$ the transmission becomes much smaller, and the two branches related to the Jaynes-Cummings dressed states are gone. Instead the $k=1$ avoided crossing due to the resonant tunneling is well visible around $|\epsilon |/\omega_c = 1$.
The $k>1$ higher resonances, on contrary, are not well visible, since they are covered by the bare photon resonance between the lower levels $n<k$.
It is worth noticing that the ultrastrong-coupling spectral features shown here, and in particular the $k=1$ resonance, are already visible in recent experiments with superconducting circuits \cite{Gross_first_circuitUSC_2010, yoshihara_circuitQED_beyond_USC2017, Yoshihara_PhysRevA.95.053824, you_natcom_2023Asymm_Rabi_exp}.

Another interesting quantity to probe the spectrum of the system is provided by the dipole structure factor
\begin{equation}\label{eq:dipole_struct_factor}
    \mathcal{S}_{\rm dip}(\omega ) =  2\hbar Z_{\rm dip} \sum_{n,m} \frac{e^{-\hbar \omega_{n}/(k_b T)}}{\mathcal{Z}} \left| \braket{n| s_x |m} \right|^2\delta(\omega - \omega_{mn}).
\end{equation}
Here the characteristic dipole impedance $Z_{\rm dip}$ is defined in App. \ref{sec:TLA}.
From the linear response theory perspective, $\mathcal{S}_{\rm dip}(\omega )$ quantifies the dipole radiation response to a direct drive of the dipole.
Because of the consideration done in Sec. \ref{sec:diag_asymm_rabi}, is clear that this quantity is strongly suppressed in the USC regime, at low temperature.
If we only stick to the dressed state picture we should observe vanishing transitions for $T\rightarrow 0$, due to the fact that the dipole matrix element between the ground-state and the first block is zero
\begin{equation}
    \braket{\pm_{(k,1)}|s_x|\rm GS} = 0.
\end{equation}
Thus in this framework only transitions beyond the dressed state picture are visible. It is important to stress that this transitions are also present in the cavity transmission, but, since they are much weaker they are much better visualized without the presence of the dressed state transitions which are dominant in the cavity transmission. 

\begin{figure}
    \centering
    \includegraphics[width=\columnwidth]{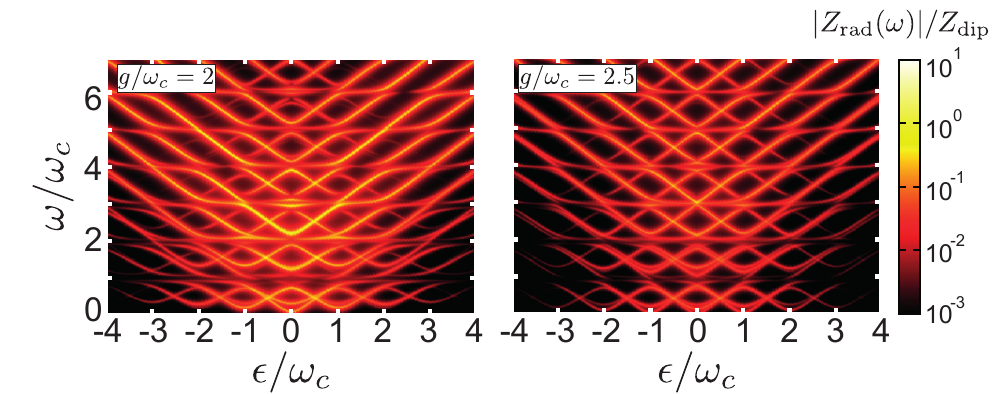}
    \caption{Dipole radiation impedance $|Z_{\rm rad}(\omega)|$ as a function of the dipole asymmetry $\epsilon$ and the probe frequency $\omega$ for various $g/\omega_c=2,2.5$ light-matter couplings. Parameters: $\omega_d=\omega_c$, $k_bT=0.5\hbar \omega_c$. In this plot we assumed a linewidth $\gamma_{\mathcal{S}_{\rm dip}} = 0.05 \omega_c$.}
    \label{fig:8}
\end{figure}

Similarly to the system circuit impedance we can define a dipole radiation impedance as
\begin{equation}
    Z_{\rm rad}(\omega) = -\frac{i\omega \mathcal{S}_{\rm dip}(\omega)}{\hbar}.
\end{equation}
In Fig. \ref{fig:8} we show the radiation impedance $Z_{\rm rad}(\omega )$ in logscale, as a function of the dipole asymmetry $\epsilon$ and the probe frequency $\omega$ (with an artificial linewidth $\gamma_{\mathcal{S}_{\rm dip}}$ to smear out the delta function in Eq. \eqref{eq:dipole_struct_factor}).
On contrary to the previous case of the cavity response, at frequencies $\omega\sim \omega_c$ the dipole response is strongly suppressed in favour of higher frequencies transitions that emerge with a diamond-like pattern.

The dipole matrix elements giving the amplitude for these transitions are much weaker than the cavity-current matrix elements for the $k$-resonant transitions between dressed states, since are given by beyond gRWA corrections.
They are the USC cavity equivalent of the vibronic transitions responsible of the Coulomb diamond structure in the Franck-Condon blockade voltage/current characteristic \cite{Andreev_PhysRevB.74.205438}.

\section{Cascaded relaxation in multi-well dipole}
\label{sec:discussion_cascaded}
Finally we comment on the possibility to extend our results to the case of a multi-well dipole.
The relaxation dynamics of this system is particularly interesting because it can be interpreted as a prototype of a transport problem through an extended system: intuitively, in a tilted multi-well potential a particle would relax from a higher well to the lowest one, but this means that this particle is also transported from side to side of the system.

Despite that a full coverage of cavity-modified relaxation or transport in an extended system is well beyond the scope of this paper, we can still use the concepts developed above to give an initial intuition which sets the basis for future investigations.

\subsection{The extended Dicke model}

We model the multiple-well dipole generalizing the two-level approximation to $(N+1)$-level, where each level represent a potential well.
In this way, the dipole is simply described by spin-$N/2$ operators, $S_{x,y,z}$ that generalizes the spin description of Sec. \ref{sec:TLA}.
In particular the eigenvalues of $S_x$, $|m_x\rangle$, are interpreted as localized states in the $m_x$-th well, while the $S_z$ operator creates some tunnelling between them.
This model is very similar to the famous Wannier-Stark ladder model, where the only difference is in the non-homogeneous hopping rates, settled by the $S_z$ matrix element between Dicke states along the $S_x$-direction.

While the dissipations, and the master equation, are derived in the same way as before, just replacing $s_{x,y,z} \mapsto S_{x,y,z}$ everywhere,
the light-matter Hamiltonian is no more given by the Rabi model of Eq. \eqref{eq:ham_Rabi}. 
Indeed, when taking the two-level approximation of Eq. \eqref{eq:ham_cavityQED} we had discarded the $x^2$-term, which is only a constant within the two-level subspace, $x^2\approx 4x_{10}^2s_x^2=4x_{10}^2\mathds{1}$. 
For a multi-level dipole, described by a spin-$N/2$ system if $N>1$ we have that $S_x^2\neq \mathds{1}/4$, and thus the correct cavity QED Hamiltonian within the $(N+1)$-level subspace is given by the so-called \emph{extended Dicke model} (EDM) \cite{Tuomas_PhysRevA.94.033850, DeBernardis_PhysRevA.97.043820}
\begin{equation}
    H_{\rm EDM} = \omega_c a^{\dag} a + \omega_d S_z + g \left(a+a^{\dag}\right)S_x + \frac{g^2}{\omega_c}S_x^2 + \epsilon S_x.
\end{equation}

Performing the polaron transformation in the same way as for the Rabi model, we arrive to the polaron (EDM) \cite{DeBernardis_PhysRevA.97.043820}:
\begin{equation}
    \tilde{H}_{\rm EDM} = \omega_c a^{\dag} a + \epsilon S_x + \frac{\omega_d}{2}\left[ \mathcal{D}(g/\omega_c) \tilde S_+ + \mathcal{D}^{\dag}(g/\omega_c) \tilde S_-  \right],
\end{equation}
where, again, $\tilde{S}_-= S_z - i S_y$.
In the USC regime $g\gg \omega_c$, and for non-negligible asymmetry $\epsilon\neq 0$, in the limit of large spin, $N \gg 1$, we can use the Holstein-Primakoff approximation \cite{HP_approx_PhysRev.58.1098} on the $S_x$-direction, for which $S_x \approx -N/2 + b^{\dag}b$, and $\tilde{S}_-\approx \sqrt{N}b$.
The polaron EDM can then be approximated by
\begin{equation}\label{eq:ham_EDM_HP}
\begin{split}
    \tilde{H}_{\rm EDM} \approx \tilde{H}_{\rm EDM}^{\rm HP} &=  \omega_c a^{\dag} a + \epsilon b^{\dag}b + 
    \\
    & + \frac{\omega_d\sqrt{N}}{2}\left[ \mathcal{D}(g/\omega_c) b^{\dag} + \mathcal{D}^{\dag}(g/\omega_c) b  \right].
\end{split}
\end{equation}

\subsection{Relaxation dynamics of the EDM}

The considerations done for the Rabi model in Sec. \ref{sec:diag_asymm_rabi} are still valid, in particular regarding the possibility to discard the counter-rotating terms in the displacement operators and the suppression of tunneling. 
It is then clear that one can use the same dressed state approach to diagonalize the polaron EDM as well.
In particular, considering the relaxation from the initial state $|0,m\rangle = (b^{\dag})^m/\sqrt{m!}|0,0\rangle $ the resonant tunnelling effect gives rise to a cavity-mediated cascaded dynamics to the ground-state, where, depending from the resonance condition $\epsilon = k \times  \omega_c$, $n_{\rm ph} \approx k\times  m$ photons are released.

To have a more quantitative understanding we consider the limit of strong cavity dissipations with respect to the k-resonance splitting, $\gamma \gg \Omega_{(k,k)}$.
In this regime we can adiabaticaly eliminate the cavity in favor of an effective master equation for the dipole only \cite{zoller_quantum_world_2_doi:10.1142/p983}.
Following the previous analysis on cavity and dipole transition rates, we completely neglect the dipole dissipations, while we take as a jump operator of the cavity its bare annihilation operator $c=a$. Again, this is well motivated by the analysis performed above.
Using the approximated form of the EDM in Eq. \eqref{eq:ham_EDM_HP} and assuming that at each time the total density matrix of the system is $\rho(t) \approx \rho_d(t)\otimes\rho^{\rm th}_c$ (here $\rho^{\rm th}_c$ is the thermal density matrix for the bare cavity at temperature $T$) we have that
\begin{equation}\label{eq:dipole_eff_master_equation}
    \begin{split}
        &\partial_t \rho_d = -i\left[\epsilon b^{\dag}b, \rho_d\right] + \frac{\Gamma_T(\epsilon)}{2}\left( 2b\rho_d b^{\dag} - \left[b^{\dag}b, \rho_d\right]_+ \right) 
        \\
        & + \frac{\Gamma_T(-\epsilon)}{2}\left( 2b^{\dag}\rho_d b - \left[bb^{\dag}, \rho_d\right]_+ \right).
    \end{split}
\end{equation}
Similarly to non-linear optomechanics setups \cite{peter_PhysRevLett.107.063601,yuri_minoguchi2019environmentinduced}, the cooling and heating rates are given by
\begin{equation}
    \begin{split}
        &\Gamma_T(\omega) = \frac{\omega_d^2N}{2}\times
        \\
        &\times {\rm Re}\left[\int_0^{\infty} dt \left(\braket{\mathcal{D}\left(t,x\right) \mathcal{D}^{\dag}\left(x\right)} - \braket{\mathcal{D}(x)}^2\right) e^{i\omega t}\right],
    \end{split}
\end{equation}
where $H_c = \omega_c a^{\dag}a$, $x=g/\omega_c$ and $\mathcal{D}(t,x)=e^{i H_c t}\mathcal{D}\left(x\right)e^{-i H_c t}$. 
Since that the average $\braket{\cdot}$ is intended over the cavity thermal state $\rho_c^{\rm th}$, we can explicitly compute this quantity \cite{peter_PhysRevLett.107.063601, yuri_minoguchi2019environmentinduced, Pilar2020thermodynamicsof}
\begin{equation}
    \begin{split}
        &\Gamma_T(\omega) = \frac{\omega_d^2N}{\gamma} e^{-x^2\left(1+2N_T(\omega_c )\right)} \times
        \\
        &\times \sum_{q,r \neq 0}\frac{x^{2r} N_T^r(\omega_c )}{r!}\frac{x^{2q}(1+N_T(\omega_c ))^q}{q!}\frac{\gamma^2/4}{\left(\omega - \omega_c(q-r)\right)^2+\frac{\gamma^2}{4}}.
    \end{split}
\end{equation}
Here $N_{T}(\omega_c )=1/(e^{\hbar \omega_c/(k_bT)}-1)$ is the cavity thermal population.
From this expression it is particularly evident the multi-photon character of this cavity assisted relaxation mechanism, where the dipole can relax by emitting $q$-photons into the cavity, and, at the same time, can be re-excited by absorbing $r$-photons from the cavity (if the temperature is non-zero $T>0$).

\begin{figure}
    \centering
    \includegraphics[width=\columnwidth]{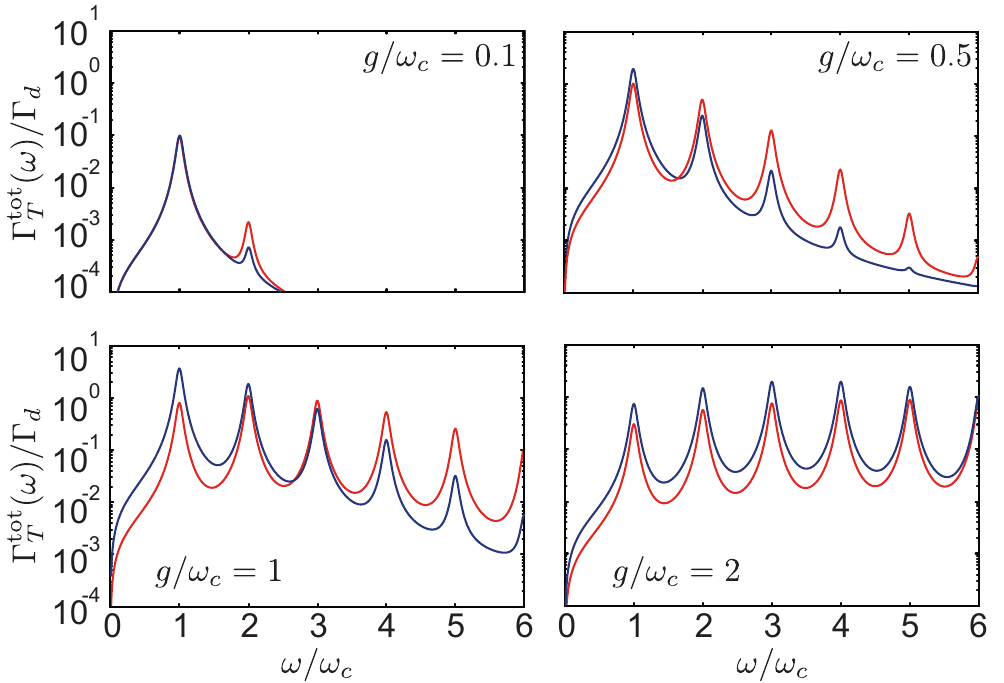}
    \caption{Total relaxation rate $\Gamma^{\rm tot}_T(\omega)$ defined in Eq. \eqref{eq:EDM_tot_relax_rate} as a function of frequency $\omega$ for various $g/\omega_c = 0.1, 0.5, 1,2$ coupling strengths (normalized over $\Gamma_d = \omega_d^2 N/\gamma$). The blue line corresponds to $T = 0$, while the red line is for $k_b T = 2\, \hbar\omega_c$. Parameters: $\gamma/\omega_c = 0.1$. }
    \label{fig:9}
\end{figure}

As in the standard theory of laser cooling the total relaxation rate is given by 
\begin{equation}\label{eq:EDM_tot_relax_rate}
    \Gamma^{\rm tot}_T = \Gamma_T(\epsilon)-\Gamma_T(-\epsilon).
\end{equation}
At $T=0$ and close to resonance $\epsilon\simeq \omega_c\times k$, the total relaxation rate is approximately given by $\Gamma_{T=0}^{\rm tot} \approx \Omega_{(k,k)}^2N/\gamma$, which is the USC version of the Purcell effect.
In Fig. \ref{fig:9} we show some examples of the total relaxation rate for $T=0$ and for $T>0$ at different coupling strengths.
Interestingly larger temperature may help in activating the higher $k$-resonances even in the non-USC regime, $g/\omega_c \ll 1$, resembling the behaviour of optomechanical laser cooling setups \cite{Peter_PhysRevB.82.165320}. We highlight the fact that at weak coupling this description does not hold, since it is based on the assumption that dipole tunneling is suppressed by the USC regime. However we found interesting to show the total relaxation rate $\Gamma^{\rm tot}_T$ also in this regime for completness.

As anticipated in the beginning of this section, the relaxation dynamics of this multi-well setup can be seen as a way to study how the incoherent transport is modified by the cavity.
Staying at the single particle level we can interpreted an excitation produced by the dipole operator $b^{\dag}$ as the particle moving one well up in energy, and the dipole ground-state as the state where only the lowest energy well is occupied.
Following this line of thoughts, we can say that the system has good transport properties if it rapidly thermalizes sufficiently close to its ground-state.
Since that the saturation number of the steady state of Eq. \eqref{eq:dipole_eff_master_equation}, $\braket{b^{\dag} b}_{\rm ss}= \mathcal{N}_0$, around each $k$-resonance $\epsilon = \omega_c \times k$ is the dipole thermal occupation
\begin{equation}
    \mathcal{N}_0= \frac{\Gamma_T(-\epsilon)}{\Gamma_T(\epsilon)-\Gamma_T(-\epsilon)} \approx N_{T}(\omega_c \times k),
\end{equation}
we also need that the temperature $T$ is small enough so thermal photons cannot push the particle (the dipole excitation) on an upper energy level.

In summary, the cavity USC suppresses tunnelling also in a multi-well dipole scenario, inhibiting the dipole's relaxation and its ability to transport excitations from one well to the other. Fast relaxation (and transport) is possible only when the tilted multi-well dipole is resonant with the cavity having access to the multi-photon resonant tunnelling process. 


\section{Conclusion}
\label{sec:conclusion}

In conclusion, we studied the relaxation properties of a simple (but paradigmatic) cavity QED setup in the ultrastrong coupling regime described by the asymmetric quantum Rabi model.
Here the Bosonic cavity mode is provided by an LC resonant circuit while the  two-level atom is given by an asymmetric dipole inside the capacitor of the LC circuit.
We introduce dissipation by considering the cavity coupled to a Ohmic transmission line, while the dipole dissipates into radiating modes with super-Ohmic spectral density. 
The system's dynamics is thus described by a thermalizing master equation, valid at arbitrary light-matter coupling strengths and arbitrary dipole asymmetry.
From the Liouvillian gap we obtained the longest relaxation rate of the system, that we can also consider its asymptotic thermalization rate. 
From this quantity emerges clearly that the effect of the USC is to slow-down the system's thermalization by an exponential suppression of the Liouvillian gap. However, for special values of the dipole asymmetry the standard relaxation is restored and the system can thermalize accordingly to its bare relaxation rates.

To understand this behaviour of the Liouvillian gap and to link it to the physical observables of the system,
we employed a generalized rotating-wave approximation (gRWA) \cite{Irish_PhysRevLett.99.173601}, valid in the so-called polaron frame. Within this approximation we showed that is possible to analytically diagonalize the asymmetric Rabi model, even in the USC regime, where the eigenstates are given by a polaronic multi-photon version of the usual Jaynes-Cummings dressed states. 
In this way, we were able to compute the relaxation rates in the USC regime analytically, explicitly showing the exponential slow-down of thermalization due to an effective suppression of the dipole tunnelling dynamics, while the cavity, remaining effectively uncoupled from the dipole, can still efficiently relax.

When the dipole asymmetry is resonant with the cavity, the dipole dynamics is revitalized, by a cavity assisted tunnelling, where the dipole resonantly tunnels from one well to the other by releasing multiple photons. 
Since photons can then relax out of the cavity, this process gives an effective relaxation channel also for the dipole.

After showing that these phenomenon can be observed indirectly from the cavity transmission or the dipole impedance, we highlight a link to the Franck-Condon physics of electronic transport through a molecular dot\cite{Andreev_PhysRevB.74.205438, Valmorra2021_natCom_Vacuum_field_induced_transport}.

At the end, we commented on the possibility of extending this non-linear resonant processes to a multi-well dipole. 
A simple toy model to describe this situation is provided by the extended Dicke model, introduced originally to study a multiple qubit ultrastrongly coupled to a single LC cavity \cite{Tuomas_PhysRevA.94.033850,DeBernardis_PhysRevA.97.043820}.
From this setup is clear that the USC resonant tunnel dynamics can affect also a multi-well system, giving rise to a resonant cascaded multi-photon process.
We argue that this cascaded effect could be observed in cavity modified transport experiments with multiple electronic quantum dots, or in superconducting circuit devices with only minor modifications of the already existing platforms \cite{yoshihara_circuitQED_beyond_USC2017, Valmorra2021_natCom_Vacuum_field_induced_transport}. 
This suggests that these findings could thus provide an interesting playground to study an implementation for cascaded-laser electronic devices operating in the USC regime in GHz or THz range. 

\acknowledgements
We thanks Gianluca Rastelli, Iacopo Carusotto, Peter Rabl, Alberto Biella, Fabrizio Minganti, Alberto Nardin, Gian Marcello Andolina Luca Giacomelli for very helpful and insightful discussions.
We acknowledge financial support from the Provincia Autonoma di Trento from the Q@TN initiative.

\appendix

\section{A dipole in a asymmetric double well potential}
\label{sec:TLA}
The dipole dynamics is described as a single particle with mass $m$ in a potential
\begin{equation}
    H_{\rm dipole} = \frac{p^2}{2m} + V(x).
\end{equation}
The dipole has charge $+q$ on an extreme and $-q$ on the other extreme, and $qx$ is its dipole moment.
So, the dipole displacement $x$ is its main dynamical variable and $p$ its canonical momentum.

As depicted in Fig. \ref{fig:1}(b), we consider only the paradigmatic case of a dipole described by a double-well potential, very similarly to Refs. \cite{DeBernardis_PhysRevA.98.053819,Savasta_NatPhys_2019}.
Considering only its low energy dynamics we are basically studying the electromagnetism of quantum tunnelling.

Differently from Ref. \cite{DeBernardis_PhysRevA.97.043820} here we introduce a linear tilt that makes the height of the two wells asymmetric. The total dipole's potential is then given by
\begin{equation}
    V(x) = - \frac{\mu_2^2}{2}x^2 + \frac{\mu_4^4}{4}x^4 + q\mathcal{E}x.
\end{equation}
The linear tilt $\sim q\mathcal{E}x$ is physically implemented by a bias static external electric field of amplitude $\mathcal{E}$ and has no influence on the LC resonator dynamics.

Whenever $\mu_2/\mu_4 \gg 1$ and $q\mathcal{E}/\mu_2 \ll 1$, the two lowest levels are below the central barrier and well separated from the other energy levels.
We can then truncate the dipole's Hilbert space keeping only these two lowest energy levels \cite{DeBernardis_PhysRevA.97.043820}. 
We perform the two level approximation (TLA) on the dipole Hamiltonian projecting on the eigenstates without tilt $\mathcal{E} = 0$, and we obtain
\begin{equation}\label{eq:H_dipole_TLA}
    H_{\rm dipole}^{\rm TLA} = \hbar \omega_d s_z + \hbar \epsilon s_x.
\end{equation}
Here we have introduced the pseudospin operators $s_a = \sigma_a/2$ ($\sigma_a$ are the usual Pauli matrices). 
The dipole frequency $\omega_d$ is the energy difference between the two lowest states without the tilt $\mathcal{E}=0$, and $\epsilon = 2q\mathcal{E}x_{10}/\hbar$, where $x_{10} = \braket{1 | x | 0}$ is the dipole matrix element between the two lowest dipole states. The dipole operator now is given by $x \approx x_{10} \sigma_x$.

When the tilt is on, the energy splitting between the two eigenstates of Eq. \eqref{eq:H_dipole_TLA} is given by $\omega_{\epsilon} = \sqrt{\omega_d^2 + \epsilon^2}$, while the eigenfunctions are partially localized on the left or right well, with a small, but non-negligible, overlap with the opposite well, Fig. \ref{fig:1}(b).
In the two-level language these states are given by
\begin{equation}\label{eq:bare_uncoupled_dipole_states}
    \begin{split}
        &| {\rm L}_{\epsilon} \rangle = \cos \frac{\theta_{\epsilon}}{2} | \downarrow \rangle + \sin \frac{\theta_{\epsilon}}{2} | \uparrow \rangle
        \\
        &| {\rm R}_{\epsilon} \rangle = -\sin \frac{\theta_{\epsilon}}{2} | \downarrow \rangle + \cos \frac{\theta_{\epsilon}}{2} | \uparrow \rangle,
    \end{split}
\end{equation}
where $\tan \left( \theta_{\epsilon} \right) = \epsilon/\omega_d$.

We can also associate a characteristic impedance to the dipole by considering $Z_{\rm dip} = \hbar/q^2 f_{10}$, and $f_{10} = 2m\omega_{0}|x_{10}|^2/\hbar$ is the oscillator strength of the two-level dipole transition.
Introducing this parameter is particularly convenient when discussing the linear response theory, and creates a nice parallelism with the circuit description of the cavity given in what follows.


\section{General cavity QED Hamiltonian}
\label{sec:cQED_ham}
The Hamiltonian of the full cavity QED system can be written summing up the dipole energy and the total energy stored in the electromagnetic field
\begin{equation}
    H_{\rm cQED} = H_{\rm em} + H_{\rm dipole}.
\end{equation}

For an LC-resonant system, the electromagnetic energy is described by
\begin{equation}
    H_{\rm em} = \frac{CU^2}{2} + \frac{\Phi^2}{2L},
\end{equation}
where $U$ is the total voltage drop across the capacitor $C$, $\Phi$ is the magnetic flux through the inductance $L$.

When the dipole is inside the capacitor the total voltage $U$ is no more the right canonical variable conjugate to $\Phi$. In order to have the correct canonical description we need to introduce the total capacitor charge variable $Q$, such that $\left[\Phi , Q \right] = i\hbar$. 
Without the presence of the dipole the total charge and the voltage drop are directly proportional through the usual relation $U = CQ$. 
When the dipole is inside the capacitor the charge responsible for the voltage drop is modified by the presence of the charge induced by the dipole on the metallic plates. 
This induced charge does not contribute to any voltage drop, and must be removed \cite{Tuomas_PhysRevA.94.033850, DeBernardis_PhysRevA.97.043820}. The voltage drop becomes
\begin{equation}
    U = C \left( Q - Q_{\rm ind} \right).
\end{equation}
For an ideal capacitor we have $Q_{\rm ind} \simeq q x/d$ \cite{DeBernardis_PhysRevA.97.043820}, where $d$ is the distance between the capacitor plates.

We introduce now the cavity creation/annihilation operators through the relations
\begin{equation}
    \begin{split}
        &\Phi = i\frac{\Phi_0}{\sqrt{2}}\left( a - a^{\dag}  \right)
        \\
        & Q = \frac{Q_0}{\sqrt{2}}\left( a + a^{\dag}\right).
    \end{split}
\end{equation}
Here $\Phi_0 = \sqrt{\hbar Z_{LC}}$, $Q_0 = \sqrt{\hbar / Z_{LC}}$ and $Z_{LC} = \sqrt{L/C}$.
The cavity QED reads
\begin{equation}\label{eq:ham_cavityQED}
    \begin{split}
        H_{\rm cQED} = \hbar \omega_c a^{\dag} a + H_{\rm dipole} + F_0 x \left( a + a^{\dag} \right) + \frac{F_0^2}{\hbar \omega_c} x^2, 
    \end{split}
\end{equation}
where $\omega_c = 1/\sqrt{LC}$ and we introduced the zero-point electric force $F_0 = \sqrt{\hbar \omega_c/(2 C d ^2)q^2}$.

As detailed in Refs. \cite{DeBernardis_PhysRevA.98.053819, Savasta_NatPhys_2019}, implementing the TLA described in Section \ref{sec:TLA}, we can now approximate the cavity QED Hamiltonian with the so-called \emph{quantum Rabi model} ($\hbar = 1$)
\begin{equation}\label{eq:app_ham_Rabi}
\begin{split}
    H_{\rm cQED} \approx H_{\rm Rabi} = \omega_c a^{\dag} a +  \omega_d s_z + \epsilon s_x  + g \left( a + a^{\dag}\right) s_x,
\end{split}
\end{equation}
where the light-matter coupling is given by $g = 2F_0 x_{10}$.

We will see in particular that the transverse term $\sim \epsilon s_x$, that breaks the $\mathbb{Z}_2$ symmetry of the usual Rabi model, is fundamental in our development, becoming a switch between slow and fast dissipation of the dipole.
In circuit QED this term emerges quite naturally through a bias in the external magnetic flux and is typically used in the observation of the spectral features of the  USC regime \cite{yoshihara_circuitQED_beyond_USC2017, Yoshihara_PhysRevA.95.053824}.

\section{Linear damping}
\label{app:linear_damp}

This section reviews the standard derivation of the Langevin equation for open quantum systems as it is developed in standard textbook \cite{petruccione2002theory,zoller_quantum_world_2_doi:10.1142/p983}, that sets the basis for the formalism that we use through the  whole paper.

We consider a generic system, described by the hamiltonian $H_{\rm sys}$, coupled to a bath of harmonic oscillators (which may represent the electromagnetic field outside of a cavity, or a resistance in a circuit):
\begin{equation}\label{eq:ham_generic_sys_bath_cailderaLegget}
H = H_{\rm sys} + \sum_k \left[ \frac{P_k^2}{2m_k} + \frac{1}{2}m_k \omega_k^2 \left( Y_k - \frac{c_k}{m_k \omega_k^2} X \right)^2 \right] .
\end{equation}
The equations of motion for a generic system operator $A$ are given by
\begin{equation}\label{eq:app_generic_sysopA_eqmotion}
\begin{split}
    & \partial_t A = -i \left[ A, H_{\rm sys} \right] + i \sum_k \frac{c_k}{2} \left( Y_k \left[ A,X \right] + \left[ A, X \right]Y_k \right) +
    \\
    &- i\sum_k \frac{c_k^2}{2m_k \omega_k^2}\left[A, X^2 \right]
\end{split}
\end{equation}
\begin{equation}
\partial_t Y_k = \frac{P_k}{m_k}, ~~~~ \partial_t P_k = - m_k \omega_k^2 Y_k + c_k X .
\end{equation}
The formal solution of the bath's equations is given by
\begin{equation}\label{app:eq:linbath_Xop}
Y_k(t) = Y_k^{homg.}(t) + \frac{c_k}{m_k \omega_k} \int_{t_0}^t dt' \sin(\omega_k (t-t'))X (t'),
\end{equation}
where
\begin{equation}
Y_k^{homg.}(t) = Y_k(t_0) \cos(\omega_k (t - t_0)) + \frac{P_k(t_0)}{m_k\omega_k}\sin(\omega_k (t-t_0)).
\end{equation}
Plugging back this solution is \eqref{eq:app_generic_sysopA_eqmotion}, and integrating by part, we get the quantum Langevin equation, describing the whole open-dissipative dynamics of our quantum system
\begin{equation}\label{eq:app_inout_LngEq_linearBath}
\begin{split}
\partial_t A(t) & = -i \left[ A(t), H_{\rm sys} \right] 
\\
&+ \frac{i}{2} \left( \xi(t) \left[ A(t), X(t) \right] + \left[ A(t), X(t) \right]\xi(t) \right)
\\
& - \frac{i}{2} \left[ \int_{t_0}^t K(t-t') \partial_{t'} X (t') dt'\, ,\,  \left[A(t), X(t) \right] \right]_+ .
\end{split}
\end{equation}
Here $[\cdot , \cdot]_+$ is the anti-commutator, and
\begin{equation}
\begin{split}
\xi(t) & = \sum_k c_k \left( Y_k^{homg.}(t) - \frac{c_k}{m_k \omega_k^2}X (t_0) \cos(\omega_k (t -t_0)) \right)
\\
K(t) & = \sum_k \frac{c_k^2}{m_k \omega_k^2}\cos(\omega_k t),
\end{split}
\end{equation}
are, respectively, the \emph{quantum noise term} and the \emph{dissipative kernel}.
We notice that the last term in \eqref{eq:app_generic_sysopA_eqmotion} is exactly cancelled by the term proportional to $ K (t)$ coming out by the integration by part.
From the fluctuations-dissipation theorem we obtain the specific value of the quantum noise correlator \cite{petruccione2002theory}. In the high-temperature limit it reads
\begin{equation}
\frac{1}{2}\langle{ \left[ \xi(t), \xi(t')\right]_+ \rangle} \simeq 2 k_b T K(t-t').
\end{equation}

The in/out relation are given by \cite{petruccione2002theory}
\begin{equation}
Y^{out}(t) = Y^{in}(t) - \int_{-\infty}^{+\infty} K(t - t') \dot X (t') d t',
\end{equation}
where $Y^{in}(t) = \sum_k c_kY_k^{homg.}(t)$.

A useful way to treat the dissipation without having all the details of the bath is to introduce the bath spectral density 
\begin{equation}
J(\omega) = \frac{\pi}{2} \sum_k \frac{c_k^2}{m_k \omega_k} \delta(\omega - \omega_k),
\end{equation}
and recast the dissipator in the form
\begin{equation}
K(t) = \int_0^{\infty} \frac{d\omega}{\pi} \frac{J(\omega)}{\omega}\cos(\omega t).
\end{equation}
Now all bath properties are encoded in the spectral density $J(\omega)$.

\section{Thermalizing master equation}
\label{app:master_eq}

We consider here the master equation suitable to study relaxation and thermalization processes in cavity QED under the ultra-strong coupling regime. 
For this purpose we consider the treatment used in \cite{Blais_dissipationUSC_first_PhysRevA.84.043832}.
We do not repeat the derivation here, but we stress that the physical assumptions are almost the same as used in deriving the Langevin equation in Appendix \ref{app:linear_damp}, with the further assumption that the coupling between the system and the bath is very small. 
In particular this latter one ensures that we can implement the rotating wave approximation between the system and the bath, proceeding with the standard textbook derivation.

The crucial step here is to isolate the components of the system coupling operator $X$ that rotates with positive (negative) frequencies.
This can be done as follows:
given an Hamiltonian $H_{\rm sys}$ and a (or multiple) system operator(s) $X$, we express them on the system eigenbasis $X = \sum_{n,m} \braket{n | X | m} | n \rangle \langle m |$. The jump operators are then given by the set $\lbrace{ c_{nm} = \braket{n | X | m} | n \rangle \langle m |\, , ~ {\rm such~that} ~ n<m \rbrace}$. In the Heinseberg picture these jump operators evolve with positive frequencies. This allows to implement the rotating wave approximation in the standard system-bath linear Hamiltonian in Eq. \eqref{eq:ham_generic_sys_bath_cailderaLegget}.

The master equation is then given by
\begin{equation}
    \partial_t \rho = \mathcal{L}_H(\rho ) + \mathcal{L}_{D}(\rho ),
\end{equation}
where the conservative time evolution is generated by
\begin{equation}
    \mathcal{L}_H(\rho ) = -i \left[ H_{\rm sys} , \rho \right],
\end{equation}
while dissipations are given by
\begin{equation}\label{eq:dissipative_liouvillian}
    \begin{split}
        & \mathcal{L}_{D}(\rho ) = \sum_{n < m} \left[ 1 + N_T (\omega_{mn}) \right] \Gamma_{nm} D \left( | n \rangle \langle m |, \rho  \right) +
        \\
        & + \sum_{n < m}  N_T (\omega_{mn}) \Gamma_{nm}  D \left( | m \rangle \langle n |, \rho   \right).
    \end{split}
\end{equation}
Here 
\begin{equation}
    D\left( c, \rho  \right) = c\, \rho\, c^{\dag} - \frac{1}{2}\left[c^{\dag} c \, , \, \rho\right]_+
\end{equation}
is the usual dissipator super-operator \cite{zoller_quantum_world_2_doi:10.1142/p983}, and
\begin{equation}
    N_T (\omega ) = \frac{1}{e^{\omega/(k_B T)} - 1}
\end{equation}
is the bosonic thermal population, where $k_B$ is the Boltzmann constant.
The thermalization rates are given by
\begin{equation}
    \Gamma_{nm} = J( |\omega_{mn}| )\left| \braket{n | X | m} \right|^2 .
\end{equation}
It is important to keep in mind here that there is another implicit assumption to correctly use this master equation: the energy levels are well resolved with respect to the bath induced linewidth, meaning that $\Gamma_{nm}\ll \omega_{nm}$. Considering only weak losses and not too weak nor not too strong coupling, we can consider it always satisfied in our development.

Considering the thermal density matrix $\rho_{\rm ss} = e^{- H_{\rm sys}/(k_B T)}/\mathcal{Z}$, where $\mathcal{Z} = {\rm Tr}[\rho_{\rm ss}]$, one can easily prove that is the steady state of the system.

\section{Physical dissipators}
\label{app:physical_dissipators}
\subsection{Cavity dissipation}

A standard way to introduce dissipation in an LC circuit is to couple it to a transmission line \cite{zoller_quantum_world_2_doi:10.1142/p983}. When is traced out from the dynamics, the transmission line plays the role of a resistive element, effectively realizing the scheme described in Fig. \ref{fig:1}(c). 
An input voltage can inject current in the system, and the resistance damps the excited oscillations of the LC-circuit. 
This scheme provides a basic input/output theory describing the system's read out from the energy dissipated into the resistance.

The formal description of this setup assumes a linear coupling to a multi-mode bath of harmonic modes, as described in Eq. \eqref{eq:ham_generic_sys_bath_cailderaLegget} of Appendix \ref{app:linear_damp}.
To realize the resistive circuit, since $\Phi = L I$ is linked to the current passing through the inductor (and so through the whole circuit)
the system operator coupled to the bath is \cite{zoller_quantum_world_2_doi:10.1142/p983, Pilar2020thermodynamicsof, garcia_ripoll_book_10261_285945}
\begin{equation}
    \hat X = \Phi = i\frac{\Phi_0}{\sqrt{2}}\left( a - a^{\dag}  \right).
\end{equation}

To reproduce a standard Ohmic resistance we assume the standard linear spectral density (neglecting for now the correct dimensional units)
\begin{equation}
    J_{\rm Ohm}(\omega ) \sim \gamma \omega.
\end{equation}

Eliminating the bath's dynamics, we obtain the equations of motion of the circuit in terms of the Langevin equation in Eq. \eqref{eq:app_inout_LngEq_linearBath}, and, considering $A=\dot \Phi$ in Eq. \eqref{eq:app_inout_LngEq_linearBath}, we have that
\begin{equation}
\ddot{\Phi} \sim - \gamma \dot{\Phi},
\end{equation}
correctly matching the standard Kirchhoff equations of a resistive circuit.

To implement the thermalizing master equation used in the main text, we introduce the jump operators corresponding to this decay channel as
\begin{equation}\label{eq:jump_op_dress_cav}
        c_{nm}^c = \braket{n | a - a^{\dag} | m} | n \rangle \langle m |,
\end{equation}
where $m>n$ and $| n \rangle$ are the eigenstates of the whole system.
In such a way that they correctly describe the relaxation process from higher to lower energy states, even when the system is ultrastrongly coupled, see App. \ref{app:master_eq}.

Since in our theory the coefficient $\gamma$ is a free parameter, we absorb the dimensional quantity $\Phi_0/\sqrt{2}$ in the definition of the spectral density, such that
\begin{equation}
    \frac{\omega_c|\Phi_0|^2}{2} J_{\rm Ohm}(\omega)\longmapsto J_{\rm Ohm}(\omega).
\end{equation}

\subsection{Dipole dissipation}

In our simplified picture, the main source of dissipation of an oscillating electric dipole is given by radiative emission into free space electromagnetic modes.
Indeed, even if the dipole is strongly coupled to the sub-wavelength mode of the LC cavity, still interacts with the all the other transverse electromagnetic modes. These have typically a small effect on the coherent dynamics \cite{saezblazquez2022observe}, but they provide a decay channel for the dipole.

Because of the harmonic dynamics of the electromagnetic field, and its general linear coupling with the dipole, we can again model the dipole dissipation as a linear damping, as described in Appendix \ref{app:linear_damp}.
In this case the system operator coupled to the dissipative bath is given by the dipole moment 
\begin{equation}
    \hat X = x \approx 2x_{10}s_x
\end{equation}

The bath spectral function is given from the spectral function of the transverse electromagnetic modes. In free-space this would be given approximately by (neglecting the dimensionality, the correct dimensional units will be reintroduced in the next section)
\begin{equation}
    J_{\rm rad} (\omega ) \sim \kappa \omega^3.
\end{equation}
Considering the Langevin equation in Eq. \eqref{eq:app_inout_LngEq_linearBath} for the dipole moment velocity $A=\dot x$, using this spectral density we find
\begin{equation}
    \ddot x \sim  \kappa \dddot x,
\end{equation}
recovering the Abraham-Lorentz formula for a radiating dipole \cite{dalibard:jpa-00209544}.

More generally for our developments, one can choose any spectral density for the dipole dissipative bath of the shape $J_{\rm rad}\sim \omega^{\nu}$, with $\nu \geq 1$. 
Having $\nu > 1$ gives particularly simple results.

As for the cavity, the jump operators corresponding to this decay channel are
\begin{equation}\label{eq:jump_op_dress_cav}
        c_{nm}^{\rm dip} = \braket{n | s_x | m} | n \rangle \langle m |.
\end{equation}
where $m>n$ and $| n \rangle$ are the eigenstates of the whole system.
In such a way that they correctly describe the relaxation process from higher to lower energy states, even when the system is ultrastrongly coupled, see App. \ref{app:master_eq}.

Since also here $\kappa$ is a free parameter, we absorb the dimensional quantity $2 x_{10}$ in the definition of the spectral density, such that
\begin{equation}
    4 \omega_d |x_{10}|^2 J_{\rm rad}(\omega ) \longmapsto J_{\rm rad}(\omega ).
\end{equation}

\section{Diagonalization of the resonant-symmetric Rabi Hamiltonian}
\label{app:gRWA}
In this section we perform the approximated diagonalization of the Rabi model in the regime where
\begin{equation}
    \begin{split}
        & \omega_d \simeq \omega_c
        \\
        & \epsilon = 0.
    \end{split}
\end{equation}
We first transform the original Rabi Hamiltonian through the unitary transformation $U_{\rm pol} = \exp \left[ g/\omega_c (a - a^{\dag})s_x \right]$, obtaining the Rabi polaron Hamiltonian in the form
\begin{equation}\label{eq:H_rabi_polaron_app}
\begin{split}
     \tilde H_{\rm Rabi} &= \omega_c a^{\dag} a + \omega_d\Big[ \cosh \left( g/\omega_c (a - a^{\dag}) \right) s_z 
    \\
    &+ i\sinh \left( g/\omega_c (a - a^{\dag}) \right) s_y  \Big].
\end{split}
\end{equation}
Here $\cosh$, and $\sinh$ operators can be expressed in terms of the displacement operator $\mathcal{D}(x) = \exp \left[ x (a - a^{\dag} ) \right]$.

In this frame one can then perform a generalized rotating wave approximation  \cite{Irish_PhysRevLett.99.173601}, following from the fact that the Hamiltonian in this basis is approximately block diagonal. Each block is spanned by the states $\lbrace{ | \uparrow, n-1 \rangle , | \downarrow, n \rangle \rbrace}$, where $n=1,2\ldots$, and the ground-state is given by the polaron vacuum state $| \downarrow, 0 \rangle$.

This block-diagonal structure is ultimately linked to the matrix elements of the displacements operators in Eq. \eqref{eq:H_rabi_polaron_app}, which are known to be exponentially suppressed as $\braket{n|\mathcal{D}(g/\omega_c)|m} \sim e^{-g^2/(2\omega_c^2)}$ \cite{Glauber_PhysRev.177.1857}.
Moreover because of parity selection rule of the $\cosh (g/\omega_c(a-a^{\dag}))s_z$, $\sinh (g/\omega_c(a-a^{\dag}))s_y$ operators only second nearest neighbour blocks are coupled. Combining this two observations we have that the most relevant transitions are within each block, for which we need only two matrix element of the displacement operator per block:
\begin{equation}
    \begin{split}
        & \mathcal{D}_{n\, n-1} = \frac{g}{\omega_c} \sqrt{\frac{(n-1)!}{n!}} e^{-\frac{g^2}{2\omega_c^2}}L_{n-1}^{(1)}\left( g^2/\omega_c^2 \right), 
        \\
        & \mathcal{D}_{n\, n} = e^{-\frac{g^2}{2\omega_c^2}}L_{n}^{(0)}\left( g^2/\omega_c^2 \right),
    \end{split}
\end{equation}
where $L_{n}^{(\alpha)}(x)$ are the special Laguerre polynomials.
Notice that, since $L_{n}^{(\alpha)}(0) = (n+\alpha )!/(n!)$, we have that $L_{n-1}^{(1)}(0) = n$, recovering the usual Jaynes-Cummings picture at weak-coupling.

We can then rewrite the polaron Rabi Hamiltonian as a block-diagonal matrix,
$\tilde H_{\rm Rabi} \approx \sum_n \tilde H_{\rm Rabi}^n$, where each block reads
\begin{equation}
\tilde H_{\rm Rabi}^n = 
    \begin{pmatrix}
    A_n & C_n \\
    C_n & B_n\\
    \end{pmatrix}
\end{equation}
where 
\begin{equation}
    \begin{split}
        & A_n = \omega_c n - \frac{\omega_de^{-g^2/(2\omega_c^2)}}{2}L_{n}^{(0)}\left(g^2/\omega_c^2 \right),
        \\
        & B_n = \omega_c (n-1) + \frac{\omega_de^{-g^2/(2\omega_c^2)}}{2}L_{n-1}^{(0)}\left(g^2/\omega_c^2 \right),
        \\
        & C_n = \frac{g}{\omega_c}\frac{\omega_de^{-g^2/(2\omega_c^2)}}{2} \sqrt{\frac{(n-1)!}{n!}} L_{n-1}^{(1)}\left(g^2/\omega_c^2 \right).
    \end{split}
\end{equation}

The spectrum is
\begin{equation}\label{eq:rabi_spec_approx_irish}
    \omega_{\pm, n} = \frac{A_n + B_n}{2} \pm \sqrt{\frac{\left( A_n + B_n \right)^2}{4} + C_n^2 - A_nB_n }
\end{equation}
and the eigenstates are
\begin{equation}
    \begin{split}
        & | +, n \rangle = \cos \frac{\theta_n}{2} | \downarrow, n \rangle + \sin \frac{\theta_n}{2} | \uparrow, n-1 \rangle,
        \\
        & | -, n \rangle = -\sin \frac{\theta_n}{2} | \downarrow, n \rangle + \cos \frac{\theta_n}{2} | \uparrow, n-1 \rangle,
    \end{split}
\end{equation}
where
\begin{equation}\label{eq:sincos_hopfields}
    \begin{split}
        & \cos \frac{\theta_n}{2} = \pm \sqrt{\frac{1}{2}\left( 1 + \frac{A_n - B_n}{\sqrt{\left( A_n - B_n \right)^2 + 4 C_n^2}} \right)}
        \\
        & \sin \frac{\theta_n}{2} = \pm \sqrt{\frac{1}{2}\left( 1 - \frac{A_n - B_n}{\sqrt{\left( A_n - B_n \right)^2 + 4 C_n^2}} \right)}
    \end{split}
\end{equation}

This approximate solution of the symmetric Rabi model is valid in all coupling regimes for each value of $g$. However its validity is restricted to the cases when $\omega_d \lesssim \omega_c$ \cite{Irish_PhysRevLett.99.173601}.

\section{Matrix element and transition rates of the symmetric Rabi model}
\label{app:matrix_elm_trans-rate_symm}

As for standard dressed states, the allowed transitions are only between states of neighbouring blocks, with $(n, n\pm 1)$-excitations, and the only relevant matrix elements contributing to the transition rates of the cavity are
\begin{equation}
    \begin{split}
        \langle +, n | \left( a^{\dag} - a \right) | -, n-1 \rangle & =  \sqrt{n-1} \cos \frac{\theta_{n-1}}{2} \sin \frac{\theta_n}{2}  + 
        \\
        & - \sqrt{n}  \cos \frac{\theta_n}{2} \sin \frac{\theta_{n-1}}{2} ,
    \end{split}
\end{equation}
\begin{equation}
    \begin{split}
        \langle -, n | \left( a^{\dag} - a \right) | +, n-1 \rangle & =  \sqrt{n-1} \sin \frac{\theta_{n-1}}{2} \cos \frac{\theta_{n}}{2} + 
        \\
        & - \sqrt{n} \sin \frac{\theta_{n}}{2} \cos \frac{\theta_{n-1}}{2} ,
    \end{split}
\end{equation}
\begin{equation}
    \begin{split}
        \langle +, n | \left( a^{\dag} - a \right) | +, n-1 \rangle & =  \sqrt{n-1} \sin \frac{\theta_{n-1}}{2} \sin \frac{\theta_{n}}{2} + 
        \\
        & + \sqrt{n} \cos \frac{\theta_{n}}{2} \cos \frac{\theta_{n-1}}{2} ,
    \end{split}
\end{equation}
\begin{equation}
    \begin{split}
        \langle -, n | \left( a^{\dag} - a \right) | -, n-1 \rangle & =  \sqrt{n-1} \cos \frac{\theta_{n-1}}{2} \cos \frac{\theta_{n}}{2} + 
        \\
        & + \sqrt{n} \sin \frac{\theta_{n}}{2} \sin \frac{\theta_{n-1}}{2} ,
    \end{split}
\end{equation}
and for the ground-state
\begin{equation}
    \begin{split}
        & \langle \downarrow, 0 | \left( a^{\dag} - a \right) | +, 1 \rangle = -\cos \frac{\theta_1}{2},
        \\
        & \langle \downarrow, 0 | \left( a^{\dag} - a \right) | -, 1 \rangle = \sin \frac{\theta_1}{2}.
    \end{split}
\end{equation}
For the dipole we have a complementary situation
\begin{equation}
    \begin{split}
        & \langle +, n | s_x | -, n-1 \rangle = -\frac{1}{2} \sin \frac{\theta_{n-1}}{2}\sin \frac{\theta_{n}}{2},
    \end{split}
\end{equation}

\begin{equation}
    \begin{split}
        & \langle -, n | s_x | +, n-1 \rangle = \frac{1}{2} \cos \frac{\theta_{n-1}}{2}\cos \frac{\theta_{n}}{2},
    \end{split}
\end{equation}

\begin{equation}
    \begin{split}
        & \langle +, n | s_x | +, n-1 \rangle = \frac{1}{2} \cos \frac{\theta_{n-1}}{2}\sin \frac{\theta_{n}}{2},
    \end{split}
\end{equation}

\begin{equation}
    \begin{split}
        & \langle -, n | s_x | -, n-1 \rangle = -\frac{1}{2} \sin \frac{\theta_{n-1}}{2}\cos \frac{\theta_{n}}{2},
    \end{split}
\end{equation}
and the ground-state
\begin{equation}
    \begin{split}
        & \langle \downarrow, 0 | s_x | +, 1 \rangle = \sin \frac{\theta_1}{2},
        \\
        & \langle \downarrow, 0 | s_x | -, 1 \rangle = \cos \frac{\theta_1}{2}.
    \end{split}
\end{equation}

\section{Linear response and absorption spectra}
\label{app:linear_resp}
Exciting the cavity corresponds to insert a current in the circuit, which can be interpreted as a parallel LC filter. 
We can then define a system circuit impedance $Z_{\rm sys}(\omega)$ for the LC circuit, which takes into account the presence of the dipole in the capacitor.
By using the standard composition of circuit impedance, we can derive the response input/output relation from the current flowing in the resistance (the transmission line)
\begin{equation}
    \frac{I_{out}}{I_{in}} = \frac{Z_{\rm sys}(\omega)}{R + Z_{\rm sys}(\omega)},
\end{equation}
where $R=Z_{\rm LC}Q$ is the Ohmic resistance of the transmission line coupled to the LC cavity, and $Q=\omega_c/\gamma$ is the LC cavity quality factor.
In Fig. \ref{fig:1}(c) it is shown the general scheme of our circuit approach.

The system impedance is defined by considering the relation between voltage and current flowing through the circuit, $V = Z I$, which gives 
\begin{equation}
    Z_{\rm sys}(\omega) = \frac{\braket{\dot \Phi }(\omega)}{ I_{\rm in}(\omega )}.
\end{equation} 
When the input current is very small we can invoke linear response theory \cite{Clerck_RevModPhys.82.1155}, for which we have
\begin{equation}
\chi_{V I} = \lim_{I_{\rm in} \rightarrow 0} \frac{\braket{\dot \Phi}(\omega )}{I_{\rm in}(\omega )},
\end{equation}
where $\chi_{V I}$ is the voltage-current linear response function \cite{Clerck_RevModPhys.82.1155}.
From here it follows an operative definition of the system impedance as
\begin{equation}
Z_{\rm sys}(\omega) = -i\omega \chi_{I I}(\omega),
\end{equation}
where
\begin{equation}
\chi_{I I} = \lim_{I_{\rm in} \rightarrow 0} \frac{ \braket{\Phi}(\omega )}{I_{\rm in}(\omega )}
\end{equation}
is the current-current linear response function.

The current-current linear response function can be calculated in many ways, but the simplest one is to use the cavity structure factor
\begin{equation}
    \mathcal{S}_c(\omega ) = \sum_{n,m}\frac{e^{-\hbar \omega_{n}/(k_b T)}}{\mathcal{Z}} \left| \braket{n| \Phi |m} \right|^2\delta(\omega - \omega_{mn}). 
\end{equation}
The system impedance is then given by $Z_{\rm sys}(\omega) = -i\omega\mathcal{S}_c(\omega ).$

\bibliographystyle{mybibstyle}
 
\bibliography{references}

\begin{thebibliography}{74}%
\makeatletter
\providecommand \@ifxundefined [1]{%
 \@ifx{#1\undefined}
}%
\providecommand \@ifnum [1]{%
 \ifnum #1\expandafter \@firstoftwo
 \else \expandafter \@secondoftwo
 \fi
}%
\providecommand \@ifx [1]{%
 \ifx #1\expandafter \@firstoftwo
 \else \expandafter \@secondoftwo
 \fi
}%
\providecommand \natexlab [1]{#1}%
\providecommand \emph  [1]{``#1''}%
\providecommand \bibnamefont  [1]{#1}%
\providecommand \bibfnamefont [1]{#1}%
\providecommand \citenamefont [1]{#1}%
\providecommand \href@noop [0]{\@secondoftwo}%
\providecommand \href [0]{\begingroup \@sanitize@url \@href}%
\providecommand \@href[1]{\@@startlink{#1}\@@href}%
\providecommand \@@href[1]{\endgroup#1\@@endlink}%
\providecommand \@sanitize@url [0]{\catcode `\\12\catcode `\$12\catcode
  `\&12\catcode `\#12\catcode `\^12\catcode `\_12\catcode `\%12\relax}%
\providecommand \@@startlink[1]{}%
\providecommand \@@endlink[0]{}%
\providecommand \url  [0]{\begingroup\@sanitize@url \@url }%
\providecommand \@url [1]{\endgroup\@href {#1}{\urlprefix }}%
\providecommand \urlprefix  [0]{URL }%
\providecommand \Eprint [0]{\href }%
\providecommand \doibase [0]{http://dx.doi.org/}%
\providecommand \selectlanguage [0]{\@gobble}%
\providecommand \bibinfo  [0]{\@secondoftwo}%
\providecommand \bibfield  [0]{\@secondoftwo}%
\providecommand \translation [1]{[#1]}%
\providecommand \BibitemOpen [0]{}%
\providecommand \bibitemStop [0]{}%
\providecommand \bibitemNoStop [0]{.\EOS\space}%
\providecommand \EOS [0]{\spacefactor3000\relax}%
\providecommand \BibitemShut  [1]{\csname bibitem#1\endcsname}%
\let\auto@bib@innerbib\@empty
\bibitem [{\citenamefont {Landau}\ and\ \citenamefont
  {Lifshitz}(2013)}]{landau2013statistical}%
  \BibitemOpen
  \bibfield  {author} {\bibinfo {author} {\bibfnamefont {L.}~\bibnamefont
  {Landau}}\ and\ \bibinfo {author} {\bibfnamefont {E.}~\bibnamefont
  {Lifshitz}},\ }\href {https://books.google.ch/books?id=VzgJN-XPTRsC} {\emph
  {\bibinfo {title} {Statistical Physics: Volume 5}}},\ \bibinfo {number} {Bd.
  5}\ (\bibinfo  {publisher} {Elsevier Science},\ \bibinfo {year}
  {2013})\BibitemShut {NoStop}%
\bibitem [{\citenamefont {Piskulich}\ \emph {et~al.}(2019)\citenamefont
  {Piskulich}, \citenamefont {Mesele},\ and\ \citenamefont
  {Thompson}}]{Piskulich_2019_activation_energy_ACS}%
  \BibitemOpen
  \bibfield  {author} {\bibinfo {author} {\bibfnamefont {Z.~A.}\ \bibnamefont
  {Piskulich}}, \bibinfo {author} {\bibfnamefont {O.~O.}\ \bibnamefont
  {Mesele}}, \ and\ \bibinfo {author} {\bibfnamefont {W.~H.}\ \bibnamefont
  {Thompson}},\ }\bibfield  {title} {\emph {\bibinfo {title} {Activation
  Energies and Beyond},}\ }\bibfield  {booktitle} {\emph {\bibinfo {booktitle}
  {The Journal of Physical Chemistry A}},\ }\href {\doibase
  10.1021/acs.jpca.9b03967} {\bibfield  {journal} {\bibinfo  {journal} {The
  Journal of Physical Chemistry A}\ }\textbf {\bibinfo {volume} {123}},\
  \bibinfo {pages} {7185} (\bibinfo {year} {2019})}\ \BibitemShut {NoStop}%
\bibitem [{\citenamefont
  {Merzbacher}(2002)}]{Merzbacher2002_History_Quantum_Tunneling_doi:10.1063/1.1510281}%
  \BibitemOpen
  \bibfield  {author} {\bibinfo {author} {\bibfnamefont {E.}~\bibnamefont
  {Merzbacher}},\ }\bibfield  {title} {\emph {\bibinfo {title} {The Early
  History of Quantum Tunneling},}\ }\href {\doibase 10.1063/1.1510281}
  {\bibfield  {journal} {\bibinfo  {journal} {Physics Today}\ }\textbf
  {\bibinfo {volume} {55}},\ \bibinfo {pages} {44} (\bibinfo {year} {2002})}\
  \BibitemShut {NoStop}%
\bibitem [{\citenamefont {Caldeira}\ and\ \citenamefont
  {Leggett}(1981)}]{Leggett_PhysRevLett.46.211}%
  \BibitemOpen
  \bibfield  {author} {\bibinfo {author} {\bibfnamefont {A.~O.}\ \bibnamefont
  {Caldeira}}\ and\ \bibinfo {author} {\bibfnamefont {A.~J.}\ \bibnamefont
  {Leggett}},\ }\bibfield  {title} {\emph {\bibinfo {title} {Influence of
  Dissipation on Quantum Tunneling in Macroscopic Systems},}\ }\href {\doibase
  10.1103/PhysRevLett.46.211} {\bibfield  {journal} {\bibinfo  {journal} {Phys.
  Rev. Lett.}\ }\textbf {\bibinfo {volume} {46}},\ \bibinfo {pages} {211}
  (\bibinfo {year} {1981})}\ \BibitemShut {NoStop}%
\bibitem [{\citenamefont {Leggett}\ \emph {et~al.}(1987)\citenamefont
  {Leggett}, \citenamefont {Chakravarty}, \citenamefont {Dorsey}, \citenamefont
  {Fisher}, \citenamefont {Garg},\ and\ \citenamefont
  {Zwerger}}]{Leggett_RevModPhys.59.1}%
  \BibitemOpen
  \bibfield  {author} {\bibinfo {author} {\bibfnamefont {A.~J.}\ \bibnamefont
  {Leggett}}, \bibinfo {author} {\bibfnamefont {S.}~\bibnamefont
  {Chakravarty}}, \bibinfo {author} {\bibfnamefont {A.~T.}\ \bibnamefont
  {Dorsey}}, \bibinfo {author} {\bibfnamefont {M.~P.~A.}\ \bibnamefont
  {Fisher}}, \bibinfo {author} {\bibfnamefont {A.}~\bibnamefont {Garg}}, \ and\
  \bibinfo {author} {\bibfnamefont {W.}~\bibnamefont {Zwerger}},\ }\bibfield
  {title} {\emph {\bibinfo {title} {Dynamics of the dissipative two-state
  system},}\ }\href {\doibase 10.1103/RevModPhys.59.1} {\bibfield  {journal}
  {\bibinfo  {journal} {Rev. Mod. Phys.}\ }\textbf {\bibinfo {volume} {59}},\
  \bibinfo {pages} {1} (\bibinfo {year} {1987})}\ \BibitemShut {NoStop}%
\bibitem [{\citenamefont {Milonni}(1994)}]{milonni1994quantum}%
  \BibitemOpen
  \bibfield  {author} {\bibinfo {author} {\bibfnamefont {P.}~\bibnamefont
  {Milonni}},\ }\href {https://books.google.it/books?id=P83vAAAAMAAJ} {\emph
  {\bibinfo {title} {The Quantum Vacuum: An Introduction to Quantum
  Electrodynamics}}}\ (\bibinfo  {publisher} {Elsevier Science},\ \bibinfo
  {year} {1994})\BibitemShut {NoStop}%
\bibitem [{\citenamefont {Hutchison}\ \emph {et~al.}(2012)\citenamefont
  {Hutchison}, \citenamefont {Schwartz}, \citenamefont {Genet}, \citenamefont
  {Devaux},\ and\ \citenamefont {Ebbesen}}]{Ebbesen_2012_reactionRate-cQED}%
  \BibitemOpen
  \bibfield  {author} {\bibinfo {author} {\bibfnamefont {J.~A.}\ \bibnamefont
  {Hutchison}}, \bibinfo {author} {\bibfnamefont {T.}~\bibnamefont {Schwartz}},
  \bibinfo {author} {\bibfnamefont {C.}~\bibnamefont {Genet}}, \bibinfo
  {author} {\bibfnamefont {E.}~\bibnamefont {Devaux}}, \ and\ \bibinfo {author}
  {\bibfnamefont {T.~W.}\ \bibnamefont {Ebbesen}},\ }\bibfield  {title} {\emph
  {\bibinfo {title} {Modifying Chemical Landscapes by Coupling to Vacuum
  Fields},}\ }\href {\doibase https://doi.org/10.1002/anie.201107033}
  {\bibfield  {journal} {\bibinfo  {journal} {Angewandte Chemie International
  Edition}\ }\textbf {\bibinfo {volume} {51}},\ \bibinfo {pages} {1592}
  (\bibinfo {year} {2012})}\ \BibitemShut {NoStop}%
\bibitem [{\citenamefont {Paravicini-Bagliani}\ \emph
  {et~al.}(2019{\natexlab{a}})\citenamefont {Paravicini-Bagliani},
  \citenamefont {Appugliese}, \citenamefont {Richter}, \citenamefont
  {Valmorra}, \citenamefont {Keller}, \citenamefont {Beck}, \citenamefont
  {Bartolo}, \citenamefont {R{\"o}ssler}, \citenamefont {Ihn}, \citenamefont
  {Ensslin}, \citenamefont {Ciuti}, \citenamefont {Scalari},\ and\
  \citenamefont {Faist}}]{Faist_magneto_transport_2019NatPhys}%
  \BibitemOpen
  \bibfield  {author} {\bibinfo {author} {\bibfnamefont {G.~L.}\ \bibnamefont
  {Paravicini-Bagliani}}, \bibinfo {author} {\bibfnamefont {F.}~\bibnamefont
  {Appugliese}}, \bibinfo {author} {\bibfnamefont {E.}~\bibnamefont {Richter}},
  \bibinfo {author} {\bibfnamefont {F.}~\bibnamefont {Valmorra}}, \bibinfo
  {author} {\bibfnamefont {J.}~\bibnamefont {Keller}}, \bibinfo {author}
  {\bibfnamefont {M.}~\bibnamefont {Beck}}, \bibinfo {author} {\bibfnamefont
  {N.}~\bibnamefont {Bartolo}}, \bibinfo {author} {\bibfnamefont
  {C.}~\bibnamefont {R{\"o}ssler}}, \bibinfo {author} {\bibfnamefont
  {T.}~\bibnamefont {Ihn}}, \bibinfo {author} {\bibfnamefont {K.}~\bibnamefont
  {Ensslin}}, \bibinfo {author} {\bibfnamefont {C.}~\bibnamefont {Ciuti}},
  \bibinfo {author} {\bibfnamefont {G.}~\bibnamefont {Scalari}}, \ and\
  \bibinfo {author} {\bibfnamefont {J.}~\bibnamefont {Faist}},\ }\bibfield
  {title} {\emph {\bibinfo {title} {Magneto-transport controlled by Landau
  polariton states},}\ }\href {\doibase 10.1038/s41567-018-0346-y} {\bibfield
  {journal} {\bibinfo  {journal} {Nature Physics}\ }\textbf {\bibinfo {volume}
  {15}},\ \bibinfo {pages} {186} (\bibinfo {year} {2019}{\natexlab{a}})}\
  \BibitemShut {NoStop}%
\bibitem [{\citenamefont {Valmorra}\ \emph {et~al.}(2021)\citenamefont
  {Valmorra}, \citenamefont {Yoshida}, \citenamefont {Contamin}, \citenamefont
  {Messelot}, \citenamefont {Massabeau}, \citenamefont {Delbecq}, \citenamefont
  {Dartiailh}, \citenamefont {Desjardins}, \citenamefont {Cubaynes},
  \citenamefont {Leghtas}, \citenamefont {Hirakawa}, \citenamefont {Tignon},
  \citenamefont {Dhillon}, \citenamefont {Balibar}, \citenamefont {Mangeney},
  \citenamefont {Cottet},\ and\ \citenamefont
  {Kontos}}]{Valmorra2021_natCom_Vacuum_field_induced_transport}%
  \BibitemOpen
  \bibfield  {author} {\bibinfo {author} {\bibfnamefont {F.}~\bibnamefont
  {Valmorra}}, \bibinfo {author} {\bibfnamefont {K.}~\bibnamefont {Yoshida}},
  \bibinfo {author} {\bibfnamefont {L.~C.}\ \bibnamefont {Contamin}}, \bibinfo
  {author} {\bibfnamefont {S.}~\bibnamefont {Messelot}}, \bibinfo {author}
  {\bibfnamefont {S.}~\bibnamefont {Massabeau}}, \bibinfo {author}
  {\bibfnamefont {M.~R.}\ \bibnamefont {Delbecq}}, \bibinfo {author}
  {\bibfnamefont {M.~C.}\ \bibnamefont {Dartiailh}}, \bibinfo {author}
  {\bibfnamefont {M.~M.}\ \bibnamefont {Desjardins}}, \bibinfo {author}
  {\bibfnamefont {T.}~\bibnamefont {Cubaynes}}, \bibinfo {author}
  {\bibfnamefont {Z.}~\bibnamefont {Leghtas}}, \bibinfo {author} {\bibfnamefont
  {K.}~\bibnamefont {Hirakawa}}, \bibinfo {author} {\bibfnamefont
  {J.}~\bibnamefont {Tignon}}, \bibinfo {author} {\bibfnamefont
  {S.}~\bibnamefont {Dhillon}}, \bibinfo {author} {\bibfnamefont
  {S.}~\bibnamefont {Balibar}}, \bibinfo {author} {\bibfnamefont
  {J.}~\bibnamefont {Mangeney}}, \bibinfo {author} {\bibfnamefont
  {A.}~\bibnamefont {Cottet}}, \ and\ \bibinfo {author} {\bibfnamefont
  {T.}~\bibnamefont {Kontos}},\ }\bibfield  {title} {\emph {\bibinfo {title}
  {Vacuum-field-induced THz transport gap in a carbon nanotube quantum dot},}\
  }\href {\doibase 10.1038/s41467-021-25733-x} {\bibfield  {journal} {\bibinfo
  {journal} {Nature Communications}\ }\textbf {\bibinfo {volume} {12}},\
  \bibinfo {pages} {5490} (\bibinfo {year} {2021})}\ \BibitemShut {NoStop}%
\bibitem [{\citenamefont {Appugliese}\ \emph {et~al.}(2022)\citenamefont
  {Appugliese}, \citenamefont {Enkner}, \citenamefont {Paravicini-Bagliani},
  \citenamefont {Beck}, \citenamefont {Reichl}, \citenamefont {Wegscheider},
  \citenamefont {Scalari}, \citenamefont {Ciuti},\ and\ \citenamefont
  {Faist}}]{faist_Science_2022}%
  \BibitemOpen
  \bibfield  {author} {\bibinfo {author} {\bibfnamefont {F.}~\bibnamefont
  {Appugliese}}, \bibinfo {author} {\bibfnamefont {J.}~\bibnamefont {Enkner}},
  \bibinfo {author} {\bibfnamefont {G.~L.}\ \bibnamefont
  {Paravicini-Bagliani}}, \bibinfo {author} {\bibfnamefont {M.}~\bibnamefont
  {Beck}}, \bibinfo {author} {\bibfnamefont {C.}~\bibnamefont {Reichl}},
  \bibinfo {author} {\bibfnamefont {W.}~\bibnamefont {Wegscheider}}, \bibinfo
  {author} {\bibfnamefont {G.}~\bibnamefont {Scalari}}, \bibinfo {author}
  {\bibfnamefont {C.}~\bibnamefont {Ciuti}}, \ and\ \bibinfo {author}
  {\bibfnamefont {J.}~\bibnamefont {Faist}},\ }\bibfield  {title} {\emph
  {\bibinfo {title} {Breakdown of topological protection by cavity vacuum
  fields in the integer quantum Hall effect},}\ }\href {\doibase
  10.1126/science.abl5818} {\bibfield  {journal} {\bibinfo  {journal}
  {Science}\ }\textbf {\bibinfo {volume} {375}},\ \bibinfo {pages} {1030}
  (\bibinfo {year} {2022})}\ \BibitemShut {NoStop}%
\bibitem [{\citenamefont {Flick}\ \emph {et~al.}(2017)\citenamefont {Flick},
  \citenamefont {Ruggenthaler}, \citenamefont {Appel},\ and\ \citenamefont
  {Rubio}}]{Rubio_2017_pnas.1615509114}%
  \BibitemOpen
  \bibfield  {author} {\bibinfo {author} {\bibfnamefont {J.}~\bibnamefont
  {Flick}}, \bibinfo {author} {\bibfnamefont {M.}~\bibnamefont {Ruggenthaler}},
  \bibinfo {author} {\bibfnamefont {H.}~\bibnamefont {Appel}}, \ and\ \bibinfo
  {author} {\bibfnamefont {A.}~\bibnamefont {Rubio}},\ }\bibfield  {title}
  {\emph {\bibinfo {title} {Atoms and molecules in cavities, from weak to
  strong coupling in quantum-electrodynamics (QED) chemistry},}\ }\href
  {\doibase 10.1073/pnas.1615509114} {\bibfield  {journal} {\bibinfo  {journal}
  {Proceedings of the National Academy of Sciences}\ }\textbf {\bibinfo
  {volume} {114}},\ \bibinfo {pages} {3026} (\bibinfo {year} {2017})}\
  \BibitemShut {NoStop}%
\bibitem [{\citenamefont {Fregoni}\ \emph {et~al.}(2022)\citenamefont
  {Fregoni}, \citenamefont {Garcia-Vidal},\ and\ \citenamefont
  {Feist}}]{Feist_2022_ACS_review_polaritonic_chemistry}%
  \BibitemOpen
  \bibfield  {author} {\bibinfo {author} {\bibfnamefont {J.}~\bibnamefont
  {Fregoni}}, \bibinfo {author} {\bibfnamefont {F.~J.}\ \bibnamefont
  {Garcia-Vidal}}, \ and\ \bibinfo {author} {\bibfnamefont {J.}~\bibnamefont
  {Feist}},\ }\bibfield  {title} {\emph {\bibinfo {title} {Theoretical
  Challenges in Polaritonic Chemistry},}\ }\bibfield  {booktitle} {\emph
  {\bibinfo {booktitle} {ACS Photonics}},\ }\href {\doibase
  10.1021/acsphotonics.1c01749} {\bibfield  {journal} {\bibinfo  {journal} {ACS
  Photonics}\ }\textbf {\bibinfo {volume} {9}},\ \bibinfo {pages} {1096}
  (\bibinfo {year} {2022})}\ \BibitemShut {NoStop}%
\bibitem [{\citenamefont {Sch{\"a}fer}\ \emph {et~al.}(2022)\citenamefont
  {Sch{\"a}fer}, \citenamefont {Flick}, \citenamefont {Ronca}, \citenamefont
  {Narang},\ and\ \citenamefont
  {Rubio}}]{Rubio_NatCom_2022_ShiningLightCavityChemestry}%
  \BibitemOpen
  \bibfield  {author} {\bibinfo {author} {\bibfnamefont {C.}~\bibnamefont
  {Sch{\"a}fer}}, \bibinfo {author} {\bibfnamefont {J.}~\bibnamefont {Flick}},
  \bibinfo {author} {\bibfnamefont {E.}~\bibnamefont {Ronca}}, \bibinfo
  {author} {\bibfnamefont {P.}~\bibnamefont {Narang}}, \ and\ \bibinfo {author}
  {\bibfnamefont {A.}~\bibnamefont {Rubio}},\ }\bibfield  {title} {\emph
  {\bibinfo {title} {Shining light on the microscopic resonant mechanism
  responsible for cavity-mediated chemical reactivity},}\ }\href {\doibase
  10.1038/s41467-022-35363-6} {\bibfield  {journal} {\bibinfo  {journal}
  {Nature Communications}\ }\textbf {\bibinfo {volume} {13}},\ \bibinfo {pages}
  {7817} (\bibinfo {year} {2022})}\ \BibitemShut {NoStop}%
\bibitem [{\citenamefont {Paravicini-Bagliani}\ \emph
  {et~al.}(2019{\natexlab{b}})\citenamefont {Paravicini-Bagliani},
  \citenamefont {Appugliese}, \citenamefont {Richter}, \citenamefont
  {Valmorra}, \citenamefont {Keller}, \citenamefont {Beck}, \citenamefont
  {Bartolo}, \citenamefont {R{\"o}ssler}, \citenamefont {Ihn}, \citenamefont
  {Ensslin}, \citenamefont {Ciuti}, \citenamefont {Scalari},\ and\
  \citenamefont {Faist}}]{ciuti_magnetotransport_nature}%
  \BibitemOpen
  \bibfield  {author} {\bibinfo {author} {\bibfnamefont {G.~L.}\ \bibnamefont
  {Paravicini-Bagliani}}, \bibinfo {author} {\bibfnamefont {F.}~\bibnamefont
  {Appugliese}}, \bibinfo {author} {\bibfnamefont {E.}~\bibnamefont {Richter}},
  \bibinfo {author} {\bibfnamefont {F.}~\bibnamefont {Valmorra}}, \bibinfo
  {author} {\bibfnamefont {J.}~\bibnamefont {Keller}}, \bibinfo {author}
  {\bibfnamefont {M.}~\bibnamefont {Beck}}, \bibinfo {author} {\bibfnamefont
  {N.}~\bibnamefont {Bartolo}}, \bibinfo {author} {\bibfnamefont
  {C.}~\bibnamefont {R{\"o}ssler}}, \bibinfo {author} {\bibfnamefont
  {T.}~\bibnamefont {Ihn}}, \bibinfo {author} {\bibfnamefont {K.}~\bibnamefont
  {Ensslin}}, \bibinfo {author} {\bibfnamefont {C.}~\bibnamefont {Ciuti}},
  \bibinfo {author} {\bibfnamefont {G.}~\bibnamefont {Scalari}}, \ and\
  \bibinfo {author} {\bibfnamefont {J.}~\bibnamefont {Faist}},\ }\bibfield
  {title} {\emph {\bibinfo {title} {Magneto-transport controlled by Landau
  polariton states},}\ }\href {\doibase 10.1038/s41567-018-0346-y} {\bibfield
  {journal} {\bibinfo  {journal} {Nature Physics}\ }\textbf {\bibinfo {volume}
  {15}},\ \bibinfo {pages} {186} (\bibinfo {year} {2019}{\natexlab{b}})}\
  \BibitemShut {NoStop}%
\bibitem [{\citenamefont {Arwas}\ and\ \citenamefont
  {Ciuti}(2023)}]{Ciuti_transport_2023_PhysRevB.107.045425}%
  \BibitemOpen
  \bibfield  {author} {\bibinfo {author} {\bibfnamefont {G.}~\bibnamefont
  {Arwas}}\ and\ \bibinfo {author} {\bibfnamefont {C.}~\bibnamefont {Ciuti}},\
  }\bibfield  {title} {\emph {\bibinfo {title} {Quantum electron transport
  controlled by cavity vacuum fields},}\ }\href {\doibase
  10.1103/PhysRevB.107.045425} {\bibfield  {journal} {\bibinfo  {journal}
  {Phys. Rev. B}\ }\textbf {\bibinfo {volume} {107}},\ \bibinfo {pages}
  {045425} (\bibinfo {year} {2023})}\ \BibitemShut {NoStop}%
\bibitem [{\citenamefont {De~Bernardis}\ \emph
  {et~al.}(2018{\natexlab{a}})\citenamefont {De~Bernardis}, \citenamefont
  {Jaako},\ and\ \citenamefont {Rabl}}]{DeBernardis_PhysRevA.97.043820}%
  \BibitemOpen
  \bibfield  {author} {\bibinfo {author} {\bibfnamefont {D.}~\bibnamefont
  {De~Bernardis}}, \bibinfo {author} {\bibfnamefont {T.}~\bibnamefont {Jaako}},
  \ and\ \bibinfo {author} {\bibfnamefont {P.}~\bibnamefont {Rabl}},\
  }\bibfield  {title} {\emph {\bibinfo {title} {Cavity quantum electrodynamics
  in the nonperturbative regime},}\ }\href {\doibase
  10.1103/PhysRevA.97.043820} {\bibfield  {journal} {\bibinfo  {journal} {Phys.
  Rev. A}\ }\textbf {\bibinfo {volume} {97}},\ \bibinfo {pages} {043820}
  (\bibinfo {year} {2018}{\natexlab{a}})}\ \BibitemShut {NoStop}%
\bibitem [{\citenamefont {Schuler}\ \emph {et~al.}(2020)\citenamefont
  {Schuler}, \citenamefont {Bernardis}, \citenamefont {Läuchli},\ and\
  \citenamefont {Rabl}}]{Schuler_SciPostPhys.9.5.066}%
  \BibitemOpen
  \bibfield  {author} {\bibinfo {author} {\bibfnamefont {M.}~\bibnamefont
  {Schuler}}, \bibinfo {author} {\bibfnamefont {D.~D.}\ \bibnamefont
  {Bernardis}}, \bibinfo {author} {\bibfnamefont {A.~M.}\ \bibnamefont
  {Läuchli}}, \ and\ \bibinfo {author} {\bibfnamefont {P.}~\bibnamefont
  {Rabl}},\ }\bibfield  {title} {\emph {\bibinfo {title} {{The vacua of dipolar
  cavity quantum electrodynamics}},}\ }\href {\doibase
  10.21468/SciPostPhys.9.5.066} {\bibfield  {journal} {\bibinfo  {journal}
  {SciPost Phys.}\ }\textbf {\bibinfo {volume} {9}},\ \bibinfo {pages} {066}
  (\bibinfo {year} {2020})}\ \BibitemShut {NoStop}%
\bibitem [{\citenamefont {Ashida}\ \emph {et~al.}(2020)\citenamefont {Ashida},
  \citenamefont {\ifmmode \dot{I}\else \.{I}\fi{}mamo\ifmmode~\breve{g}\else
  \u{g}\fi{}lu}, \citenamefont {Faist}, \citenamefont {Jaksch}, \citenamefont
  {Cavalleri},\ and\ \citenamefont {Demler}}]{Demler_PhysRevX.10.041027}%
  \BibitemOpen
  \bibfield  {author} {\bibinfo {author} {\bibfnamefont {Y.}~\bibnamefont
  {Ashida}}, \bibinfo {author} {\bibfnamefont {A.~m.~c.}\ \bibnamefont
  {\ifmmode \dot{I}\else \.{I}\fi{}mamo\ifmmode~\breve{g}\else \u{g}\fi{}lu}},
  \bibinfo {author} {\bibfnamefont {J.}~\bibnamefont {Faist}}, \bibinfo
  {author} {\bibfnamefont {D.}~\bibnamefont {Jaksch}}, \bibinfo {author}
  {\bibfnamefont {A.}~\bibnamefont {Cavalleri}}, \ and\ \bibinfo {author}
  {\bibfnamefont {E.}~\bibnamefont {Demler}},\ }\bibfield  {title} {\emph
  {\bibinfo {title} {Quantum Electrodynamic Control of Matter: Cavity-Enhanced
  Ferroelectric Phase Transition},}\ }\href {\doibase
  10.1103/PhysRevX.10.041027} {\bibfield  {journal} {\bibinfo  {journal} {Phys.
  Rev. X}\ }\textbf {\bibinfo {volume} {10}},\ \bibinfo {pages} {041027}
  (\bibinfo {year} {2020})}\ \BibitemShut {NoStop}%
\bibitem [{\citenamefont {Andolina}\ \emph {et~al.}(2020)\citenamefont
  {Andolina}, \citenamefont {Pellegrino}, \citenamefont {Giovannetti},
  \citenamefont {MacDonald},\ and\ \citenamefont
  {Polini}}]{GMAndolina_PhysRevB.102.125137}%
  \BibitemOpen
  \bibfield  {author} {\bibinfo {author} {\bibfnamefont {G.~M.}\ \bibnamefont
  {Andolina}}, \bibinfo {author} {\bibfnamefont {F.~M.~D.}\ \bibnamefont
  {Pellegrino}}, \bibinfo {author} {\bibfnamefont {V.}~\bibnamefont
  {Giovannetti}}, \bibinfo {author} {\bibfnamefont {A.~H.}\ \bibnamefont
  {MacDonald}}, \ and\ \bibinfo {author} {\bibfnamefont {M.}~\bibnamefont
  {Polini}},\ }\bibfield  {title} {\emph {\bibinfo {title} {Theory of photon
  condensation in a spatially varying electromagnetic field},}\ }\href
  {\doibase 10.1103/PhysRevB.102.125137} {\bibfield  {journal} {\bibinfo
  {journal} {Phys. Rev. B}\ }\textbf {\bibinfo {volume} {102}},\ \bibinfo
  {pages} {125137} (\bibinfo {year} {2020})}\ \BibitemShut {NoStop}%
\bibitem [{\citenamefont {Rom\'an-Roche}\ \emph {et~al.}(2021)\citenamefont
  {Rom\'an-Roche}, \citenamefont {Luis},\ and\ \citenamefont
  {Zueco}}]{Zueco_PhysRevLett.127.167201}%
  \BibitemOpen
  \bibfield  {author} {\bibinfo {author} {\bibfnamefont {J.}~\bibnamefont
  {Rom\'an-Roche}}, \bibinfo {author} {\bibfnamefont {F.}~\bibnamefont {Luis}},
  \ and\ \bibinfo {author} {\bibfnamefont {D.}~\bibnamefont {Zueco}},\
  }\bibfield  {title} {\emph {\bibinfo {title} {Photon Condensation and
  Enhanced Magnetism in Cavity QED},}\ }\href {\doibase
  10.1103/PhysRevLett.127.167201} {\bibfield  {journal} {\bibinfo  {journal}
  {Phys. Rev. Lett.}\ }\textbf {\bibinfo {volume} {127}},\ \bibinfo {pages}
  {167201} (\bibinfo {year} {2021})}\ \BibitemShut {NoStop}%
\bibitem [{\citenamefont {Schlawin}\ \emph {et~al.}(2019)\citenamefont
  {Schlawin}, \citenamefont {Cavalleri},\ and\ \citenamefont
  {Jaksch}}]{Cavalleri_PhysRevLett.122.133602}%
  \BibitemOpen
  \bibfield  {author} {\bibinfo {author} {\bibfnamefont {F.}~\bibnamefont
  {Schlawin}}, \bibinfo {author} {\bibfnamefont {A.}~\bibnamefont {Cavalleri}},
  \ and\ \bibinfo {author} {\bibfnamefont {D.}~\bibnamefont {Jaksch}},\
  }\bibfield  {title} {\emph {\bibinfo {title} {Cavity-Mediated Electron-Photon
  Superconductivity},}\ }\href {\doibase 10.1103/PhysRevLett.122.133602}
  {\bibfield  {journal} {\bibinfo  {journal} {Phys. Rev. Lett.}\ }\textbf
  {\bibinfo {volume} {122}},\ \bibinfo {pages} {133602} (\bibinfo {year}
  {2019})}\ \BibitemShut {NoStop}%
\bibitem [{\citenamefont {Hausinger}\ and\ \citenamefont
  {Grifoni}(2008)}]{Hausinger_2008}%
  \BibitemOpen
  \bibfield  {author} {\bibinfo {author} {\bibfnamefont {J.}~\bibnamefont
  {Hausinger}}\ and\ \bibinfo {author} {\bibfnamefont {M.}~\bibnamefont
  {Grifoni}},\ }\bibfield  {title} {\emph {\bibinfo {title} {Dissipative
  dynamics of a biased qubit coupled to a harmonic oscillator: analytical
  results beyond the rotating wave approximation},}\ }\href {\doibase
  10.1088/1367-2630/10/11/115015} {\bibfield  {journal} {\bibinfo  {journal}
  {New Journal of Physics}\ }\textbf {\bibinfo {volume} {10}},\ \bibinfo
  {pages} {115015} (\bibinfo {year} {2008})}\ \BibitemShut {NoStop}%
\bibitem [{\citenamefont {Mercurio}\ \emph {et~al.}(2022)\citenamefont
  {Mercurio}, \citenamefont {Macr\`{\i}}, \citenamefont {Gustin}, \citenamefont
  {Hughes}, \citenamefont {Savasta},\ and\ \citenamefont
  {Nori}}]{Savasta_RabiIncoherentPump_PhysRevResearch.4.023048}%
  \BibitemOpen
  \bibfield  {author} {\bibinfo {author} {\bibfnamefont {A.}~\bibnamefont
  {Mercurio}}, \bibinfo {author} {\bibfnamefont {V.}~\bibnamefont
  {Macr\`{\i}}}, \bibinfo {author} {\bibfnamefont {C.}~\bibnamefont {Gustin}},
  \bibinfo {author} {\bibfnamefont {S.}~\bibnamefont {Hughes}}, \bibinfo
  {author} {\bibfnamefont {S.}~\bibnamefont {Savasta}}, \ and\ \bibinfo
  {author} {\bibfnamefont {F.}~\bibnamefont {Nori}},\ }\bibfield  {title}
  {\emph {\bibinfo {title} {Regimes of cavity QED under incoherent excitation:
  From weak to deep strong coupling},}\ }\href {\doibase
  10.1103/PhysRevResearch.4.023048} {\bibfield  {journal} {\bibinfo  {journal}
  {Phys. Rev. Res.}\ }\textbf {\bibinfo {volume} {4}},\ \bibinfo {pages}
  {023048} (\bibinfo {year} {2022})}\ \BibitemShut {NoStop}%
\bibitem [{\citenamefont {Settineri}\ \emph {et~al.}(2018)\citenamefont
  {Settineri}, \citenamefont {Macr\'{\i}}, \citenamefont {Ridolfo},
  \citenamefont {Di~Stefano}, \citenamefont {Kockum}, \citenamefont {Nori},\
  and\ \citenamefont
  {Savasta}}]{Savasta_relaxation_generalized_MasterEq_PhysRevA.98.053834}%
  \BibitemOpen
  \bibfield  {author} {\bibinfo {author} {\bibfnamefont {A.}~\bibnamefont
  {Settineri}}, \bibinfo {author} {\bibfnamefont {V.}~\bibnamefont
  {Macr\'{\i}}}, \bibinfo {author} {\bibfnamefont {A.}~\bibnamefont {Ridolfo}},
  \bibinfo {author} {\bibfnamefont {O.}~\bibnamefont {Di~Stefano}}, \bibinfo
  {author} {\bibfnamefont {A.~F.}\ \bibnamefont {Kockum}}, \bibinfo {author}
  {\bibfnamefont {F.}~\bibnamefont {Nori}}, \ and\ \bibinfo {author}
  {\bibfnamefont {S.}~\bibnamefont {Savasta}},\ }\bibfield  {title} {\emph
  {\bibinfo {title} {Dissipation and thermal noise in hybrid quantum systems in
  the ultrastrong-coupling regime},}\ }\href {\doibase
  10.1103/PhysRevA.98.053834} {\bibfield  {journal} {\bibinfo  {journal} {Phys.
  Rev. A}\ }\textbf {\bibinfo {volume} {98}},\ \bibinfo {pages} {053834}
  (\bibinfo {year} {2018})}\ \BibitemShut {NoStop}%
\bibitem [{\citenamefont {Schaeverbeke}\ \emph {et~al.}(2019)\citenamefont
  {Schaeverbeke}, \citenamefont {Avriller}, \citenamefont {Frederiksen},\ and\
  \citenamefont {Pistolesi}}]{Pistolesi_2019_PhysRevLett.123.246601}%
  \BibitemOpen
  \bibfield  {author} {\bibinfo {author} {\bibfnamefont {Q.}~\bibnamefont
  {Schaeverbeke}}, \bibinfo {author} {\bibfnamefont {R.}~\bibnamefont
  {Avriller}}, \bibinfo {author} {\bibfnamefont {T.}~\bibnamefont
  {Frederiksen}}, \ and\ \bibinfo {author} {\bibfnamefont {F.}~\bibnamefont
  {Pistolesi}},\ }\bibfield  {title} {\emph {\bibinfo {title} {Single-Photon
  Emission Mediated by Single-Electron Tunneling in Plasmonic Nanojunctions},}\
  }\href {\doibase 10.1103/PhysRevLett.123.246601} {\bibfield  {journal}
  {\bibinfo  {journal} {Phys. Rev. Lett.}\ }\textbf {\bibinfo {volume} {123}},\
  \bibinfo {pages} {246601} (\bibinfo {year} {2019})}\ \BibitemShut {NoStop}%
\bibitem [{\citenamefont {Kelly}\ \emph {et~al.}(2021)\citenamefont {Kelly},
  \citenamefont {Rey},\ and\ \citenamefont
  {Marino}}]{Shane_PhysRevLett.126.133603}%
  \BibitemOpen
  \bibfield  {author} {\bibinfo {author} {\bibfnamefont {S.~P.}\ \bibnamefont
  {Kelly}}, \bibinfo {author} {\bibfnamefont {A.~M.}\ \bibnamefont {Rey}}, \
  and\ \bibinfo {author} {\bibfnamefont {J.}~\bibnamefont {Marino}},\
  }\bibfield  {title} {\emph {\bibinfo {title} {Effect of Active Photons on
  Dynamical Frustration in Cavity QED},}\ }\href {\doibase
  10.1103/PhysRevLett.126.133603} {\bibfield  {journal} {\bibinfo  {journal}
  {Phys. Rev. Lett.}\ }\textbf {\bibinfo {volume} {126}},\ \bibinfo {pages}
  {133603} (\bibinfo {year} {2021})}\ \BibitemShut {NoStop}%
\bibitem [{\citenamefont {Magazz\`u}\ \emph {et~al.}(2021)\citenamefont
  {Magazz\`u}, \citenamefont {Forn-D\'{\i}az},\ and\ \citenamefont
  {Grifoni}}]{Grifoni_2021_PhysRevA.104.053711}%
  \BibitemOpen
  \bibfield  {author} {\bibinfo {author} {\bibfnamefont {L.}~\bibnamefont
  {Magazz\`u}}, \bibinfo {author} {\bibfnamefont {P.}~\bibnamefont
  {Forn-D\'{\i}az}}, \ and\ \bibinfo {author} {\bibfnamefont {M.}~\bibnamefont
  {Grifoni}},\ }\bibfield  {title} {\emph {\bibinfo {title} {Transmission
  spectra of the driven, dissipative Rabi model in the ultrastrong-coupling
  regime},}\ }\href {\doibase 10.1103/PhysRevA.104.053711} {\bibfield
  {journal} {\bibinfo  {journal} {Phys. Rev. A}\ }\textbf {\bibinfo {volume}
  {104}},\ \bibinfo {pages} {053711} (\bibinfo {year} {2021})}\ \BibitemShut
  {NoStop}%
\bibitem [{\citenamefont {Kelly}\ \emph {et~al.}(2022)\citenamefont {Kelly},
  \citenamefont {Thompson}, \citenamefont {Rey},\ and\ \citenamefont
  {Marino}}]{Shane_PhysRevResearch.4.L042032}%
  \BibitemOpen
  \bibfield  {author} {\bibinfo {author} {\bibfnamefont {S.~P.}\ \bibnamefont
  {Kelly}}, \bibinfo {author} {\bibfnamefont {J.~K.}\ \bibnamefont {Thompson}},
  \bibinfo {author} {\bibfnamefont {A.~M.}\ \bibnamefont {Rey}}, \ and\
  \bibinfo {author} {\bibfnamefont {J.}~\bibnamefont {Marino}},\ }\bibfield
  {title} {\emph {\bibinfo {title} {Resonant light enhances phase coherence in
  a cavity QED simulator of fermionic superfluidity},}\ }\href {\doibase
  10.1103/PhysRevResearch.4.L042032} {\bibfield  {journal} {\bibinfo  {journal}
  {Phys. Rev. Res.}\ }\textbf {\bibinfo {volume} {4}},\ \bibinfo {pages}
  {L042032} (\bibinfo {year} {2022})}\ \BibitemShut {NoStop}%
\bibitem [{\citenamefont {Chen}\ \emph {et~al.}(2022)\citenamefont {Chen},
  \citenamefont {Che}, \citenamefont {Chen}, \citenamefont {Wang},\ and\
  \citenamefont {Ren}}]{Ren_2022_PhysRevResearch.4.013152}%
  \BibitemOpen
  \bibfield  {author} {\bibinfo {author} {\bibfnamefont {Z.-H.}\ \bibnamefont
  {Chen}}, \bibinfo {author} {\bibfnamefont {H.-X.}\ \bibnamefont {Che}},
  \bibinfo {author} {\bibfnamefont {Z.-K.}\ \bibnamefont {Chen}}, \bibinfo
  {author} {\bibfnamefont {C.}~\bibnamefont {Wang}}, \ and\ \bibinfo {author}
  {\bibfnamefont {J.}~\bibnamefont {Ren}},\ }\bibfield  {title} {\emph
  {\bibinfo {title} {Tuning nonequilibrium heat current and two-photon
  statistics via composite qubit-resonator interaction},}\ }\href {\doibase
  10.1103/PhysRevResearch.4.013152} {\bibfield  {journal} {\bibinfo  {journal}
  {Phys. Rev. Res.}\ }\textbf {\bibinfo {volume} {4}},\ \bibinfo {pages}
  {013152} (\bibinfo {year} {2022})}\ \BibitemShut {NoStop}%
\bibitem [{\citenamefont {Ciuti}\ \emph {et~al.}(2005)\citenamefont {Ciuti},
  \citenamefont {Bastard},\ and\ \citenamefont
  {Carusotto}}]{Ciuti_USC_original_PhysRevB.72.115303}%
  \BibitemOpen
  \bibfield  {author} {\bibinfo {author} {\bibfnamefont {C.}~\bibnamefont
  {Ciuti}}, \bibinfo {author} {\bibfnamefont {G.}~\bibnamefont {Bastard}}, \
  and\ \bibinfo {author} {\bibfnamefont {I.}~\bibnamefont {Carusotto}},\
  }\bibfield  {title} {\emph {\bibinfo {title} {Quantum vacuum properties of
  the intersubband cavity polariton field},}\ }\href {\doibase
  10.1103/PhysRevB.72.115303} {\bibfield  {journal} {\bibinfo  {journal} {Phys.
  Rev. B}\ }\textbf {\bibinfo {volume} {72}},\ \bibinfo {pages} {115303}
  (\bibinfo {year} {2005})}\ \BibitemShut {NoStop}%
\bibitem [{\citenamefont {Schlawin}\ \emph {et~al.}(2022)\citenamefont
  {Schlawin}, \citenamefont {Kennes},\ and\ \citenamefont
  {Sentef}}]{Sentef_cavity_qmaterial_2022_5.0083825}%
  \BibitemOpen
  \bibfield  {author} {\bibinfo {author} {\bibfnamefont {F.}~\bibnamefont
  {Schlawin}}, \bibinfo {author} {\bibfnamefont {D.~M.}\ \bibnamefont
  {Kennes}}, \ and\ \bibinfo {author} {\bibfnamefont {M.~A.}\ \bibnamefont
  {Sentef}},\ }\bibfield  {title} {\emph {\bibinfo {title} {Cavity quantum
  materials},}\ }\href {\doibase 10.1063/5.0083825} {\bibfield  {journal}
  {\bibinfo  {journal} {Applied Physics Reviews}\ }\textbf {\bibinfo {volume}
  {9}},\ \bibinfo {pages} {011312} (\bibinfo {year} {2022})}\ \BibitemShut
  {NoStop}%
\bibitem [{\citenamefont {Bloch}\ \emph {et~al.}(2022)\citenamefont {Bloch},
  \citenamefont {Cavalleri}, \citenamefont {Galitski}, \citenamefont {Hafezi},\
  and\ \citenamefont {Rubio}}]{Hafezi_review_electronPhotons}%
  \BibitemOpen
  \bibfield  {author} {\bibinfo {author} {\bibfnamefont {J.}~\bibnamefont
  {Bloch}}, \bibinfo {author} {\bibfnamefont {A.}~\bibnamefont {Cavalleri}},
  \bibinfo {author} {\bibfnamefont {V.}~\bibnamefont {Galitski}}, \bibinfo
  {author} {\bibfnamefont {M.}~\bibnamefont {Hafezi}}, \ and\ \bibinfo {author}
  {\bibfnamefont {A.}~\bibnamefont {Rubio}},\ }\bibfield  {title} {\emph
  {\bibinfo {title} {Strongly correlated electron--photon systems},}\ }\href
  {\doibase 10.1038/s41586-022-04726-w} {\bibfield  {journal} {\bibinfo
  {journal} {Nature}\ }\textbf {\bibinfo {volume} {606}},\ \bibinfo {pages}
  {41} (\bibinfo {year} {2022})}\ \BibitemShut {NoStop}%
\bibitem [{\citenamefont {Garcia-Vidal}\ \emph {et~al.}(2021)\citenamefont
  {Garcia-Vidal}, \citenamefont {Ciuti},\ and\ \citenamefont
  {Ebbesen}}]{Ebbesen_review_science.abd0336}%
  \BibitemOpen
  \bibfield  {author} {\bibinfo {author} {\bibfnamefont {F.~J.}\ \bibnamefont
  {Garcia-Vidal}}, \bibinfo {author} {\bibfnamefont {C.}~\bibnamefont {Ciuti}},
  \ and\ \bibinfo {author} {\bibfnamefont {T.~W.}\ \bibnamefont {Ebbesen}},\
  }\bibfield  {title} {\emph {\bibinfo {title} {Manipulating matter by strong
  coupling to vacuum fields},}\ }\href {\doibase 10.1126/science.abd0336}
  {\bibfield  {journal} {\bibinfo  {journal} {Science}\ }\textbf {\bibinfo
  {volume} {373}},\ \bibinfo {pages} {eabd0336} (\bibinfo {year} {2021})}\
  \BibitemShut {NoStop}%
\bibitem [{\citenamefont {Andolina}\ \emph {et~al.}(2019)\citenamefont
  {Andolina}, \citenamefont {Pellegrino}, \citenamefont {Giovannetti},
  \citenamefont {MacDonald},\ and\ \citenamefont
  {Polini}}]{Andolina_nogo_PhysRevB.100.121109}%
  \BibitemOpen
  \bibfield  {author} {\bibinfo {author} {\bibfnamefont {G.~M.}\ \bibnamefont
  {Andolina}}, \bibinfo {author} {\bibfnamefont {F.~M.~D.}\ \bibnamefont
  {Pellegrino}}, \bibinfo {author} {\bibfnamefont {V.}~\bibnamefont
  {Giovannetti}}, \bibinfo {author} {\bibfnamefont {A.~H.}\ \bibnamefont
  {MacDonald}}, \ and\ \bibinfo {author} {\bibfnamefont {M.}~\bibnamefont
  {Polini}},\ }\bibfield  {title} {\emph {\bibinfo {title} {Cavity quantum
  electrodynamics of strongly correlated electron systems: A no-go theorem for
  photon condensation},}\ }\href {\doibase 10.1103/PhysRevB.100.121109}
  {\bibfield  {journal} {\bibinfo  {journal} {Phys. Rev. B}\ }\textbf {\bibinfo
  {volume} {100}},\ \bibinfo {pages} {121109} (\bibinfo {year} {2019})}\
  \BibitemShut {NoStop}%
\bibitem [{\citenamefont {Galego}\ \emph {et~al.}(2019)\citenamefont {Galego},
  \citenamefont {Climent}, \citenamefont {Garcia-Vidal},\ and\ \citenamefont
  {Feist}}]{Feist_CasimirPolderMolecule_PhysRevX.9.021057}%
  \BibitemOpen
  \bibfield  {author} {\bibinfo {author} {\bibfnamefont {J.}~\bibnamefont
  {Galego}}, \bibinfo {author} {\bibfnamefont {C.}~\bibnamefont {Climent}},
  \bibinfo {author} {\bibfnamefont {F.~J.}\ \bibnamefont {Garcia-Vidal}}, \
  and\ \bibinfo {author} {\bibfnamefont {J.}~\bibnamefont {Feist}},\ }\bibfield
   {title} {\emph {\bibinfo {title} {Cavity Casimir-Polder Forces and Their
  Effects in Ground-State Chemical Reactivity},}\ }\href {\doibase
  10.1103/PhysRevX.9.021057} {\bibfield  {journal} {\bibinfo  {journal} {Phys.
  Rev. X}\ }\textbf {\bibinfo {volume} {9}},\ \bibinfo {pages} {021057}
  (\bibinfo {year} {2019})}\ \BibitemShut {NoStop}%
\bibitem [{\citenamefont {Andolina}\ \emph {et~al.}(2022)\citenamefont
  {Andolina}, \citenamefont {Pasquale}, \citenamefont {Pellegrino},
  \citenamefont {Torre}, \citenamefont {Koppens},\ and\ \citenamefont
  {Polini}}]{andolina2022deep}%
  \BibitemOpen
  \bibfield  {author} {\bibinfo {author} {\bibfnamefont {G.~M.}\ \bibnamefont
  {Andolina}}, \bibinfo {author} {\bibfnamefont {A.~D.}\ \bibnamefont
  {Pasquale}}, \bibinfo {author} {\bibfnamefont {F.~M.~D.}\ \bibnamefont
  {Pellegrino}}, \bibinfo {author} {\bibfnamefont {I.}~\bibnamefont {Torre}},
  \bibinfo {author} {\bibfnamefont {F.~H.~L.}\ \bibnamefont {Koppens}}, \ and\
  \bibinfo {author} {\bibfnamefont {M.}~\bibnamefont {Polini}},\ }\emph
  {\bibinfo {title} {Can deep sub-wavelength cavities induce Amperean
  superconductivity in a 2D material?}}\ \href@noop {} {\Eprint
  {http://arxiv.org/abs/2210.10371} {arXiv:2210.10371 [cond-mat.supr-con]}
  (\bibinfo {year} {2022})}\BibitemShut {NoStop}%
\bibitem [{\citenamefont {S\'aez-Bl\'azquez}\ \emph {et~al.}(2023)\citenamefont
  {S\'aez-Bl\'azquez}, \citenamefont {de~Bernardis}, \citenamefont {Feist},\
  and\ \citenamefont {Rabl}}]{saezblazquez2022observe}%
  \BibitemOpen
  \bibfield  {author} {\bibinfo {author} {\bibfnamefont {R.}~\bibnamefont
  {S\'aez-Bl\'azquez}}, \bibinfo {author} {\bibfnamefont {D.}~\bibnamefont
  {de~Bernardis}}, \bibinfo {author} {\bibfnamefont {J.}~\bibnamefont {Feist}},
  \ and\ \bibinfo {author} {\bibfnamefont {P.}~\bibnamefont {Rabl}},\
  }\bibfield  {title} {\emph {\bibinfo {title} {Can We Observe Nonperturbative
  Vacuum Shifts in Cavity QED?}}\ }\href {\doibase
  10.1103/PhysRevLett.131.013602} {\bibfield  {journal} {\bibinfo  {journal}
  {Phys. Rev. Lett.}\ }\textbf {\bibinfo {volume} {131}},\ \bibinfo {pages}
  {013602} (\bibinfo {year} {2023})}\ \BibitemShut {NoStop}%
\bibitem [{\citenamefont {Koch}\ \emph {et~al.}(2006)\citenamefont {Koch},
  \citenamefont {von Oppen},\ and\ \citenamefont
  {Andreev}}]{Andreev_PhysRevB.74.205438}%
  \BibitemOpen
  \bibfield  {author} {\bibinfo {author} {\bibfnamefont {J.}~\bibnamefont
  {Koch}}, \bibinfo {author} {\bibfnamefont {F.}~\bibnamefont {von Oppen}}, \
  and\ \bibinfo {author} {\bibfnamefont {A.~V.}\ \bibnamefont {Andreev}},\
  }\bibfield  {title} {\emph {\bibinfo {title} {Theory of the Franck-Condon
  blockade regime},}\ }\href {\doibase 10.1103/PhysRevB.74.205438} {\bibfield
  {journal} {\bibinfo  {journal} {Phys. Rev. B}\ }\textbf {\bibinfo {volume}
  {74}},\ \bibinfo {pages} {205438} (\bibinfo {year} {2006})}\ \BibitemShut
  {NoStop}%
\bibitem [{\citenamefont {Leturcq}\ \emph {et~al.}(2009)\citenamefont
  {Leturcq}, \citenamefont {Stampfer}, \citenamefont {Inderbitzin},
  \citenamefont {Durrer}, \citenamefont {Hierold}, \citenamefont {Mariani},
  \citenamefont {Schultz}, \citenamefont {von Oppen},\ and\ \citenamefont
  {Ensslin}}]{Oppen_NatPhys_FrankCondonExp_2009}%
  \BibitemOpen
  \bibfield  {author} {\bibinfo {author} {\bibfnamefont {R.}~\bibnamefont
  {Leturcq}}, \bibinfo {author} {\bibfnamefont {C.}~\bibnamefont {Stampfer}},
  \bibinfo {author} {\bibfnamefont {K.}~\bibnamefont {Inderbitzin}}, \bibinfo
  {author} {\bibfnamefont {L.}~\bibnamefont {Durrer}}, \bibinfo {author}
  {\bibfnamefont {C.}~\bibnamefont {Hierold}}, \bibinfo {author} {\bibfnamefont
  {E.}~\bibnamefont {Mariani}}, \bibinfo {author} {\bibfnamefont {M.~G.}\
  \bibnamefont {Schultz}}, \bibinfo {author} {\bibfnamefont {F.}~\bibnamefont
  {von Oppen}}, \ and\ \bibinfo {author} {\bibfnamefont {K.}~\bibnamefont
  {Ensslin}},\ }\bibfield  {title} {\emph {\bibinfo {title} {Franck--Condon
  blockade in suspended carbon nanotube quantum dots},}\ }\href {\doibase
  10.1038/nphys1234} {\bibfield  {journal} {\bibinfo  {journal} {Nature
  Physics}\ }\textbf {\bibinfo {volume} {5}},\ \bibinfo {pages} {327} (\bibinfo
  {year} {2009})}\ \BibitemShut {NoStop}%
\bibitem [{\citenamefont {Cui}\ \emph {et~al.}(2015)\citenamefont {Cui},
  \citenamefont {Tosoni}, \citenamefont {Schneider}, \citenamefont {Pacchioni},
  \citenamefont {Nilius},\ and\ \citenamefont
  {Freund}}]{Joachim_phonon_electron_tranport_PhysRevLett.114.016804}%
  \BibitemOpen
  \bibfield  {author} {\bibinfo {author} {\bibfnamefont {Y.}~\bibnamefont
  {Cui}}, \bibinfo {author} {\bibfnamefont {S.}~\bibnamefont {Tosoni}},
  \bibinfo {author} {\bibfnamefont {W.-D.}\ \bibnamefont {Schneider}}, \bibinfo
  {author} {\bibfnamefont {G.}~\bibnamefont {Pacchioni}}, \bibinfo {author}
  {\bibfnamefont {N.}~\bibnamefont {Nilius}}, \ and\ \bibinfo {author}
  {\bibfnamefont {H.-J.}\ \bibnamefont {Freund}},\ }\bibfield  {title} {\emph
  {\bibinfo {title} {Phonon-Mediated Electron Transport through CaO Thin
  Films},}\ }\href {\doibase 10.1103/PhysRevLett.114.016804} {\bibfield
  {journal} {\bibinfo  {journal} {Phys. Rev. Lett.}\ }\textbf {\bibinfo
  {volume} {114}},\ \bibinfo {pages} {016804} (\bibinfo {year} {2015})}\
  \BibitemShut {NoStop}%
\bibitem [{\citenamefont {Vdovin}\ \emph {et~al.}(2016)\citenamefont {Vdovin},
  \citenamefont {Mishchenko}, \citenamefont {Greenaway}, \citenamefont {Zhu},
  \citenamefont {Ghazaryan}, \citenamefont {Misra}, \citenamefont {Cao},
  \citenamefont {Morozov}, \citenamefont {Makarovsky}, \citenamefont
  {Fromhold}, \citenamefont {Patan\`e}, \citenamefont {Slotman}, \citenamefont
  {Katsnelson}, \citenamefont {Geim}, \citenamefont {Novoselov},\ and\
  \citenamefont
  {Eaves}}]{Eaves_phonon_resonant_tunnel_2016_PhysRevLett.116.186603}%
  \BibitemOpen
  \bibfield  {author} {\bibinfo {author} {\bibfnamefont {E.~E.}\ \bibnamefont
  {Vdovin}}, \bibinfo {author} {\bibfnamefont {A.}~\bibnamefont {Mishchenko}},
  \bibinfo {author} {\bibfnamefont {M.~T.}\ \bibnamefont {Greenaway}}, \bibinfo
  {author} {\bibfnamefont {M.~J.}\ \bibnamefont {Zhu}}, \bibinfo {author}
  {\bibfnamefont {D.}~\bibnamefont {Ghazaryan}}, \bibinfo {author}
  {\bibfnamefont {A.}~\bibnamefont {Misra}}, \bibinfo {author} {\bibfnamefont
  {Y.}~\bibnamefont {Cao}}, \bibinfo {author} {\bibfnamefont {S.~V.}\
  \bibnamefont {Morozov}}, \bibinfo {author} {\bibfnamefont {O.}~\bibnamefont
  {Makarovsky}}, \bibinfo {author} {\bibfnamefont {T.~M.}\ \bibnamefont
  {Fromhold}}, \bibinfo {author} {\bibfnamefont {A.}~\bibnamefont {Patan\`e}},
  \bibinfo {author} {\bibfnamefont {G.~J.}\ \bibnamefont {Slotman}}, \bibinfo
  {author} {\bibfnamefont {M.~I.}\ \bibnamefont {Katsnelson}}, \bibinfo
  {author} {\bibfnamefont {A.~K.}\ \bibnamefont {Geim}}, \bibinfo {author}
  {\bibfnamefont {K.~S.}\ \bibnamefont {Novoselov}}, \ and\ \bibinfo {author}
  {\bibfnamefont {L.}~\bibnamefont {Eaves}},\ }\bibfield  {title} {\emph
  {\bibinfo {title} {Phonon-Assisted Resonant Tunneling of Electrons in
  Graphene--Boron Nitride Transistors},}\ }\href {\doibase
  10.1103/PhysRevLett.116.186603} {\bibfield  {journal} {\bibinfo  {journal}
  {Phys. Rev. Lett.}\ }\textbf {\bibinfo {volume} {116}},\ \bibinfo {pages}
  {186603} (\bibinfo {year} {2016})}\ \BibitemShut {NoStop}%
\bibitem [{\citenamefont {Jaako}\ \emph {et~al.}(2016)\citenamefont {Jaako},
  \citenamefont {Xiang}, \citenamefont {Garcia-Ripoll},\ and\ \citenamefont
  {Rabl}}]{Tuomas_PhysRevA.94.033850}%
  \BibitemOpen
  \bibfield  {author} {\bibinfo {author} {\bibfnamefont {T.}~\bibnamefont
  {Jaako}}, \bibinfo {author} {\bibfnamefont {Z.-L.}\ \bibnamefont {Xiang}},
  \bibinfo {author} {\bibfnamefont {J.~J.}\ \bibnamefont {Garcia-Ripoll}}, \
  and\ \bibinfo {author} {\bibfnamefont {P.}~\bibnamefont {Rabl}},\ }\bibfield
  {title} {\emph {\bibinfo {title} {Ultrastrong-coupling phenomena beyond the
  Dicke model},}\ }\href {\doibase 10.1103/PhysRevA.94.033850} {\bibfield
  {journal} {\bibinfo  {journal} {Phys. Rev. A}\ }\textbf {\bibinfo {volume}
  {94}},\ \bibinfo {pages} {033850} (\bibinfo {year} {2016})}\ \BibitemShut
  {NoStop}%
\bibitem [{\citenamefont {Forn-D\'{\i}az}\ \emph {et~al.}(2019)\citenamefont
  {Forn-D\'{\i}az}, \citenamefont {Lamata}, \citenamefont {Rico}, \citenamefont
  {Kono},\ and\ \citenamefont {Solano}}]{Solano_review_RevModPhys.91.025005}%
  \BibitemOpen
  \bibfield  {author} {\bibinfo {author} {\bibfnamefont {P.}~\bibnamefont
  {Forn-D\'{\i}az}}, \bibinfo {author} {\bibfnamefont {L.}~\bibnamefont
  {Lamata}}, \bibinfo {author} {\bibfnamefont {E.}~\bibnamefont {Rico}},
  \bibinfo {author} {\bibfnamefont {J.}~\bibnamefont {Kono}}, \ and\ \bibinfo
  {author} {\bibfnamefont {E.}~\bibnamefont {Solano}},\ }\bibfield  {title}
  {\emph {\bibinfo {title} {Ultrastrong coupling regimes of light-matter
  interaction},}\ }\href {\doibase 10.1103/RevModPhys.91.025005} {\bibfield
  {journal} {\bibinfo  {journal} {Rev. Mod. Phys.}\ }\textbf {\bibinfo {volume}
  {91}},\ \bibinfo {pages} {025005} (\bibinfo {year} {2019})}\ \BibitemShut
  {NoStop}%
\bibitem [{\citenamefont {Frisk~Kockum}\ \emph {et~al.}(2019)\citenamefont
  {Frisk~Kockum}, \citenamefont {Miranowicz}, \citenamefont {De~Liberato},
  \citenamefont {Savasta},\ and\ \citenamefont
  {Nori}}]{USC_Review_Nature_2019}%
  \BibitemOpen
  \bibfield  {author} {\bibinfo {author} {\bibfnamefont {A.}~\bibnamefont
  {Frisk~Kockum}}, \bibinfo {author} {\bibfnamefont {A.}~\bibnamefont
  {Miranowicz}}, \bibinfo {author} {\bibfnamefont {S.}~\bibnamefont
  {De~Liberato}}, \bibinfo {author} {\bibfnamefont {S.}~\bibnamefont
  {Savasta}}, \ and\ \bibinfo {author} {\bibfnamefont {F.}~\bibnamefont
  {Nori}},\ }\bibfield  {title} {\emph {\bibinfo {title} {Ultrastrong coupling
  between light and matter},}\ }\href {\doibase 10.1038/s42254-018-0006-2}
  {\bibfield  {journal} {\bibinfo  {journal} {Nature Reviews Physics}\ }\textbf
  {\bibinfo {volume} {1}},\ \bibinfo {pages} {19} (\bibinfo {year} {2019})}\
  \BibitemShut {NoStop}%
\bibitem [{\citenamefont {Yoshihara}\ \emph {et~al.}(2022)\citenamefont
  {Yoshihara}, \citenamefont {Ashhab}, \citenamefont {Fuse}, \citenamefont
  {Bamba},\ and\ \citenamefont {Semba}}]{Yoshihara_natcom_2022}%
  \BibitemOpen
  \bibfield  {author} {\bibinfo {author} {\bibfnamefont {F.}~\bibnamefont
  {Yoshihara}}, \bibinfo {author} {\bibfnamefont {S.}~\bibnamefont {Ashhab}},
  \bibinfo {author} {\bibfnamefont {T.}~\bibnamefont {Fuse}}, \bibinfo {author}
  {\bibfnamefont {M.}~\bibnamefont {Bamba}}, \ and\ \bibinfo {author}
  {\bibfnamefont {K.}~\bibnamefont {Semba}},\ }\bibfield  {title} {\emph
  {\bibinfo {title} {Hamiltonian of a flux qubit-LC oscillator circuit in the
  deep--strong-coupling regime},}\ }\href {\doibase 10.1038/s41598-022-10203-1}
  {\bibfield  {journal} {\bibinfo  {journal} {Scientific Reports}\ }\textbf
  {\bibinfo {volume} {12}},\ \bibinfo {pages} {6764} (\bibinfo {year} {2022})}\
  \BibitemShut {NoStop}%
\bibitem [{\citenamefont {Gardiner}\ and\ \citenamefont
  {Zoller}(2015)}]{zoller_quantum_world_2_doi:10.1142/p983}%
  \BibitemOpen
  \bibfield  {author} {\bibinfo {author} {\bibfnamefont {C.}~\bibnamefont
  {Gardiner}}\ and\ \bibinfo {author} {\bibfnamefont {P.}~\bibnamefont
  {Zoller}},\ }\href {\doibase 10.1142/p983} {\emph {\bibinfo {title} {The
  Quantum World of Ultra-Cold Atoms and Light Book II: The Physics of
  Quantum-Optical Devices}}}\ (\bibinfo  {publisher} {IMPERIAL COLLEGE PRESS},\
  \bibinfo {year} {2015})\ \Eprint
  {http://arxiv.org/abs/https://www.worldscientific.com/doi/pdf/10.1142/p983}
  {https://www.worldscientific.com/doi/pdf/10.1142/p983}\BibitemShut {NoStop}%
\bibitem [{\citenamefont {Kessler}\ \emph {et~al.}(2012)\citenamefont
  {Kessler}, \citenamefont {Giedke}, \citenamefont {Imamoglu}, \citenamefont
  {Yelin}, \citenamefont {Lukin},\ and\ \citenamefont
  {Cirac}}]{Cirac_PhysRevA.86.012116}%
  \BibitemOpen
  \bibfield  {author} {\bibinfo {author} {\bibfnamefont {E.~M.}\ \bibnamefont
  {Kessler}}, \bibinfo {author} {\bibfnamefont {G.}~\bibnamefont {Giedke}},
  \bibinfo {author} {\bibfnamefont {A.}~\bibnamefont {Imamoglu}}, \bibinfo
  {author} {\bibfnamefont {S.~F.}\ \bibnamefont {Yelin}}, \bibinfo {author}
  {\bibfnamefont {M.~D.}\ \bibnamefont {Lukin}}, \ and\ \bibinfo {author}
  {\bibfnamefont {J.~I.}\ \bibnamefont {Cirac}},\ }\bibfield  {title} {\emph
  {\bibinfo {title} {Dissipative phase transition in a central spin system},}\
  }\href {\doibase 10.1103/PhysRevA.86.012116} {\bibfield  {journal} {\bibinfo
  {journal} {Phys. Rev. A}\ }\textbf {\bibinfo {volume} {86}},\ \bibinfo
  {pages} {012116} (\bibinfo {year} {2012})}\ \BibitemShut {NoStop}%
\bibitem [{\citenamefont {Minganti}\ \emph {et~al.}(2018)\citenamefont
  {Minganti}, \citenamefont {Biella}, \citenamefont {Bartolo},\ and\
  \citenamefont {Ciuti}}]{Minganti_PhysRevA.98.042118}%
  \BibitemOpen
  \bibfield  {author} {\bibinfo {author} {\bibfnamefont {F.}~\bibnamefont
  {Minganti}}, \bibinfo {author} {\bibfnamefont {A.}~\bibnamefont {Biella}},
  \bibinfo {author} {\bibfnamefont {N.}~\bibnamefont {Bartolo}}, \ and\
  \bibinfo {author} {\bibfnamefont {C.}~\bibnamefont {Ciuti}},\ }\bibfield
  {title} {\emph {\bibinfo {title} {Spectral theory of Liouvillians for
  dissipative phase transitions},}\ }\href {\doibase
  10.1103/PhysRevA.98.042118} {\bibfield  {journal} {\bibinfo  {journal} {Phys.
  Rev. A}\ }\textbf {\bibinfo {volume} {98}},\ \bibinfo {pages} {042118}
  (\bibinfo {year} {2018})}\ \BibitemShut {NoStop}%
\bibitem [{\citenamefont {Macieszczak}\ \emph {et~al.}(2016)\citenamefont
  {Macieszczak}, \citenamefont {Gu\ifmmode \mbox{\c{t}}\else
  \c{t}\fi{}\ifmmode~\u{a}\else \u{a}\fi{}}, \citenamefont {Lesanovsky},\ and\
  \citenamefont {Garrahan}}]{Macieszczak_PhysRevLett.116.240404}%
  \BibitemOpen
  \bibfield  {author} {\bibinfo {author} {\bibfnamefont {K.}~\bibnamefont
  {Macieszczak}}, \bibinfo {author} {\bibfnamefont {M.~u. u. u.~u.}\
  \bibnamefont {Gu\ifmmode \mbox{\c{t}}\else \c{t}\fi{}\ifmmode~\u{a}\else
  \u{a}\fi{}}}, \bibinfo {author} {\bibfnamefont {I.}~\bibnamefont
  {Lesanovsky}}, \ and\ \bibinfo {author} {\bibfnamefont {J.~P.}\ \bibnamefont
  {Garrahan}},\ }\bibfield  {title} {\emph {\bibinfo {title} {Towards a Theory
  of Metastability in Open Quantum Dynamics},}\ }\href {\doibase
  10.1103/PhysRevLett.116.240404} {\bibfield  {journal} {\bibinfo  {journal}
  {Phys. Rev. Lett.}\ }\textbf {\bibinfo {volume} {116}},\ \bibinfo {pages}
  {240404} (\bibinfo {year} {2016})}\ \BibitemShut {NoStop}%
\bibitem [{\citenamefont {Johansson}\ \emph {et~al.}(2013)\citenamefont
  {Johansson}, \citenamefont {Nation},\ and\ \citenamefont
  {Nori}}]{qutip_JOHANSSON20131234}%
  \BibitemOpen
  \bibfield  {author} {\bibinfo {author} {\bibfnamefont {J.}~\bibnamefont
  {Johansson}}, \bibinfo {author} {\bibfnamefont {P.}~\bibnamefont {Nation}}, \
  and\ \bibinfo {author} {\bibfnamefont {F.}~\bibnamefont {Nori}},\ }\bibfield
  {title} {\emph {\bibinfo {title} {QuTiP 2: A Python framework for the
  dynamics of open quantum systems},}\ }\href {\doibase
  https://doi.org/10.1016/j.cpc.2012.11.019} {\bibfield  {journal} {\bibinfo
  {journal} {Computer Physics Communications}\ }\textbf {\bibinfo {volume}
  {184}},\ \bibinfo {pages} {1234} (\bibinfo {year} {2013})}\ \BibitemShut
  {NoStop}%
\bibitem [{\citenamefont {Mori}\ and\ \citenamefont
  {Shirai}(2020)}]{Mori_PhysRevLett.125.230604}%
  \BibitemOpen
  \bibfield  {author} {\bibinfo {author} {\bibfnamefont {T.}~\bibnamefont
  {Mori}}\ and\ \bibinfo {author} {\bibfnamefont {T.}~\bibnamefont {Shirai}},\
  }\bibfield  {title} {\emph {\bibinfo {title} {Resolving a Discrepancy between
  Liouvillian Gap and Relaxation Time in Boundary-Dissipated Quantum Many-Body
  Systems},}\ }\href {\doibase 10.1103/PhysRevLett.125.230604} {\bibfield
  {journal} {\bibinfo  {journal} {Phys. Rev. Lett.}\ }\textbf {\bibinfo
  {volume} {125}},\ \bibinfo {pages} {230604} (\bibinfo {year} {2020})}\
  \BibitemShut {NoStop}%
\bibitem [{\citenamefont {Yoshihara}\ \emph
  {et~al.}(2017{\natexlab{a}})\citenamefont {Yoshihara}, \citenamefont {Fuse},
  \citenamefont {Ashhab}, \citenamefont {Kakuyanagi}, \citenamefont {Saito},\
  and\ \citenamefont {Semba}}]{yoshihara_circuitQED_beyond_USC2017}%
  \BibitemOpen
  \bibfield  {author} {\bibinfo {author} {\bibfnamefont {F.}~\bibnamefont
  {Yoshihara}}, \bibinfo {author} {\bibfnamefont {T.}~\bibnamefont {Fuse}},
  \bibinfo {author} {\bibfnamefont {S.}~\bibnamefont {Ashhab}}, \bibinfo
  {author} {\bibfnamefont {K.}~\bibnamefont {Kakuyanagi}}, \bibinfo {author}
  {\bibfnamefont {S.}~\bibnamefont {Saito}}, \ and\ \bibinfo {author}
  {\bibfnamefont {K.}~\bibnamefont {Semba}},\ }\bibfield  {title} {\emph
  {\bibinfo {title} {Superconducting qubit--oscillator circuit beyond the
  ultrastrong-coupling regime},}\ }\href {\doibase 10.1038/nphys3906}
  {\bibfield  {journal} {\bibinfo  {journal} {Nature Physics}\ }\textbf
  {\bibinfo {volume} {13}},\ \bibinfo {pages} {44} (\bibinfo {year}
  {2017}{\natexlab{a}})}\ \BibitemShut {NoStop}%
\bibitem [{\citenamefont {Pilar}\ \emph {et~al.}(2020)\citenamefont {Pilar},
  \citenamefont {De~Bernardis},\ and\ \citenamefont
  {Rabl}}]{Pilar2020thermodynamicsof}%
  \BibitemOpen
  \bibfield  {author} {\bibinfo {author} {\bibfnamefont {P.}~\bibnamefont
  {Pilar}}, \bibinfo {author} {\bibfnamefont {D.}~\bibnamefont {De~Bernardis}},
  \ and\ \bibinfo {author} {\bibfnamefont {P.}~\bibnamefont {Rabl}},\
  }\bibfield  {title} {\emph {\bibinfo {title} {Thermodynamics of ultrastrongly
  coupled light-matter systems},}\ }\href {\doibase 10.22331/q-2020-09-28-335}
  {\bibfield  {journal} {\bibinfo  {journal} {{Quantum}}\ }\textbf {\bibinfo
  {volume} {4}},\ \bibinfo {pages} {335} (\bibinfo {year} {2020})}\
  \BibitemShut {NoStop}%
\bibitem [{\citenamefont {Rossatto}\ \emph {et~al.}(2017)\citenamefont
  {Rossatto}, \citenamefont {Villas-B\^oas}, \citenamefont {Sanz},\ and\
  \citenamefont {Solano}}]{Solano_PhysRevA.96.013849}%
  \BibitemOpen
  \bibfield  {author} {\bibinfo {author} {\bibfnamefont {D.~Z.}\ \bibnamefont
  {Rossatto}}, \bibinfo {author} {\bibfnamefont {C.~J.}\ \bibnamefont
  {Villas-B\^oas}}, \bibinfo {author} {\bibfnamefont {M.}~\bibnamefont {Sanz}},
  \ and\ \bibinfo {author} {\bibfnamefont {E.}~\bibnamefont {Solano}},\
  }\bibfield  {title} {\emph {\bibinfo {title} {Spectral classification of
  coupling regimes in the quantum Rabi model},}\ }\href {\doibase
  10.1103/PhysRevA.96.013849} {\bibfield  {journal} {\bibinfo  {journal} {Phys.
  Rev. A}\ }\textbf {\bibinfo {volume} {96}},\ \bibinfo {pages} {013849}
  (\bibinfo {year} {2017})}\ \BibitemShut {NoStop}%
\bibitem [{\citenamefont {Beaudoin}\ \emph {et~al.}(2011)\citenamefont
  {Beaudoin}, \citenamefont {Gambetta},\ and\ \citenamefont
  {Blais}}]{Blais_dissipationUSC_first_PhysRevA.84.043832}%
  \BibitemOpen
  \bibfield  {author} {\bibinfo {author} {\bibfnamefont {F.}~\bibnamefont
  {Beaudoin}}, \bibinfo {author} {\bibfnamefont {J.~M.}\ \bibnamefont
  {Gambetta}}, \ and\ \bibinfo {author} {\bibfnamefont {A.}~\bibnamefont
  {Blais}},\ }\bibfield  {title} {\emph {\bibinfo {title} {Dissipation and
  ultrastrong coupling in circuit QED},}\ }\href {\doibase
  10.1103/PhysRevA.84.043832} {\bibfield  {journal} {\bibinfo  {journal} {Phys.
  Rev. A}\ }\textbf {\bibinfo {volume} {84}},\ \bibinfo {pages} {043832}
  (\bibinfo {year} {2011})}\ \BibitemShut {NoStop}%
\bibitem [{\citenamefont {Irish}(2007)}]{Irish_PhysRevLett.99.173601}%
  \BibitemOpen
  \bibfield  {author} {\bibinfo {author} {\bibfnamefont {E.~K.}\ \bibnamefont
  {Irish}},\ }\bibfield  {title} {\emph {\bibinfo {title} {Generalized
  Rotating-Wave Approximation for Arbitrarily Large Coupling},}\ }\href
  {\doibase 10.1103/PhysRevLett.99.173601} {\bibfield  {journal} {\bibinfo
  {journal} {Phys. Rev. Lett.}\ }\textbf {\bibinfo {volume} {99}},\ \bibinfo
  {pages} {173601} (\bibinfo {year} {2007})}\ \BibitemShut {NoStop}%
\bibitem [{\citenamefont {Claude
  Cohen-Tannoudji}(1998)}]{Cohen_AtomPhoton_book:91199254}%
  \BibitemOpen
  \bibfield  {author} {\bibinfo {author} {\bibfnamefont {G.~G.}\ \bibnamefont
  {Claude Cohen-Tannoudji}, \bibfnamefont {Jacques Dupont-Roc}},\ }\href
  {libgen.li/file.php?md5=5f232a3f42019873c2d00033e102b2d2} {\emph {\bibinfo
  {title} {Atom-photon interactions: basic processes and applications}}},\
  \bibinfo {edition} {wiley}\ ed.,\ Wiley Science Paperback Series\ (\bibinfo
  {publisher} {Wiley-VCH},\ \bibinfo {year} {1998})\BibitemShut {NoStop}%
\bibitem [{\citenamefont {Cahill}\ and\ \citenamefont
  {Glauber}(1969)}]{Glauber_PhysRev.177.1857}%
  \BibitemOpen
  \bibfield  {author} {\bibinfo {author} {\bibfnamefont {K.~E.}\ \bibnamefont
  {Cahill}}\ and\ \bibinfo {author} {\bibfnamefont {R.~J.}\ \bibnamefont
  {Glauber}},\ }\bibfield  {title} {\emph {\bibinfo {title} {Ordered Expansions
  in Boson Amplitude Operators},}\ }\href {\doibase 10.1103/PhysRev.177.1857}
  {\bibfield  {journal} {\bibinfo  {journal} {Phys. Rev.}\ }\textbf {\bibinfo
  {volume} {177}},\ \bibinfo {pages} {1857} (\bibinfo {year} {1969})}\
  \BibitemShut {NoStop}%
\bibitem [{\citenamefont {Ashhab}\ and\ \citenamefont
  {Nori}(2010)}]{ashhab_nori_PhysRevA.81.042311}%
  \BibitemOpen
  \bibfield  {author} {\bibinfo {author} {\bibfnamefont {S.}~\bibnamefont
  {Ashhab}}\ and\ \bibinfo {author} {\bibfnamefont {F.}~\bibnamefont {Nori}},\
  }\bibfield  {title} {\emph {\bibinfo {title} {Qubit-oscillator systems in the
  ultrastrong-coupling regime and their potential for preparing nonclassical
  states},}\ }\href {\doibase 10.1103/PhysRevA.81.042311} {\bibfield  {journal}
  {\bibinfo  {journal} {Phys. Rev. A}\ }\textbf {\bibinfo {volume} {81}},\
  \bibinfo {pages} {042311} (\bibinfo {year} {2010})}\ \BibitemShut {NoStop}%
\bibitem [{\citenamefont {Garziano}\ \emph {et~al.}(2015)\citenamefont
  {Garziano}, \citenamefont {Stassi}, \citenamefont {Macr\`{\i}}, \citenamefont
  {Kockum}, \citenamefont {Savasta},\ and\ \citenamefont
  {Nori}}]{Savasta_2015_multiPhotonRabi_PhysRevA.92.063830}%
  \BibitemOpen
  \bibfield  {author} {\bibinfo {author} {\bibfnamefont {L.}~\bibnamefont
  {Garziano}}, \bibinfo {author} {\bibfnamefont {R.}~\bibnamefont {Stassi}},
  \bibinfo {author} {\bibfnamefont {V.}~\bibnamefont {Macr\`{\i}}}, \bibinfo
  {author} {\bibfnamefont {A.~F.}\ \bibnamefont {Kockum}}, \bibinfo {author}
  {\bibfnamefont {S.}~\bibnamefont {Savasta}}, \ and\ \bibinfo {author}
  {\bibfnamefont {F.}~\bibnamefont {Nori}},\ }\bibfield  {title} {\emph
  {\bibinfo {title} {Multiphoton quantum Rabi oscillations in ultrastrong
  cavity QED},}\ }\href {\doibase 10.1103/PhysRevA.92.063830} {\bibfield
  {journal} {\bibinfo  {journal} {Phys. Rev. A}\ }\textbf {\bibinfo {volume}
  {92}},\ \bibinfo {pages} {063830} (\bibinfo {year} {2015})}\ \BibitemShut
  {NoStop}%
\bibitem [{\citenamefont {Ma}\ and\ \citenamefont
  {Law}(2015)}]{Ken_2015_PhysRevA.92.023842}%
  \BibitemOpen
  \bibfield  {author} {\bibinfo {author} {\bibfnamefont {K.~K.~W.}\
  \bibnamefont {Ma}}\ and\ \bibinfo {author} {\bibfnamefont {C.~K.}\
  \bibnamefont {Law}},\ }\bibfield  {title} {\emph {\bibinfo {title}
  {Three-photon resonance and adiabatic passage in the large-detuning Rabi
  model},}\ }\href {\doibase 10.1103/PhysRevA.92.023842} {\bibfield  {journal}
  {\bibinfo  {journal} {Phys. Rev. A}\ }\textbf {\bibinfo {volume} {92}},\
  \bibinfo {pages} {023842} (\bibinfo {year} {2015})}\ \BibitemShut {NoStop}%
\bibitem [{\citenamefont {Niemczyk}\ \emph {et~al.}(2010)\citenamefont
  {Niemczyk}, \citenamefont {Deppe}, \citenamefont {Huebl}, \citenamefont
  {Menzel}, \citenamefont {Hocke}, \citenamefont {Schwarz}, \citenamefont
  {Garcia-Ripoll}, \citenamefont {Zueco}, \citenamefont {H{\"u}mmer},
  \citenamefont {Solano}, \citenamefont {Marx},\ and\ \citenamefont
  {Gross}}]{Gross_first_circuitUSC_2010}%
  \BibitemOpen
  \bibfield  {author} {\bibinfo {author} {\bibfnamefont {T.}~\bibnamefont
  {Niemczyk}}, \bibinfo {author} {\bibfnamefont {F.}~\bibnamefont {Deppe}},
  \bibinfo {author} {\bibfnamefont {H.}~\bibnamefont {Huebl}}, \bibinfo
  {author} {\bibfnamefont {E.~P.}\ \bibnamefont {Menzel}}, \bibinfo {author}
  {\bibfnamefont {F.}~\bibnamefont {Hocke}}, \bibinfo {author} {\bibfnamefont
  {M.~J.}\ \bibnamefont {Schwarz}}, \bibinfo {author} {\bibfnamefont {J.~J.}\
  \bibnamefont {Garcia-Ripoll}}, \bibinfo {author} {\bibfnamefont
  {D.}~\bibnamefont {Zueco}}, \bibinfo {author} {\bibfnamefont
  {T.}~\bibnamefont {H{\"u}mmer}}, \bibinfo {author} {\bibfnamefont
  {E.}~\bibnamefont {Solano}}, \bibinfo {author} {\bibfnamefont
  {A.}~\bibnamefont {Marx}}, \ and\ \bibinfo {author} {\bibfnamefont
  {R.}~\bibnamefont {Gross}},\ }\bibfield  {title} {\emph {\bibinfo {title}
  {Circuit quantum electrodynamics in the ultrastrong-coupling regime},}\
  }\href {\doibase 10.1038/nphys1730} {\bibfield  {journal} {\bibinfo
  {journal} {Nature Physics}\ }\textbf {\bibinfo {volume} {6}},\ \bibinfo
  {pages} {772} (\bibinfo {year} {2010})}\ \BibitemShut {NoStop}%
\bibitem [{\citenamefont {Yoshihara}\ \emph
  {et~al.}(2017{\natexlab{b}})\citenamefont {Yoshihara}, \citenamefont {Fuse},
  \citenamefont {Ashhab}, \citenamefont {Kakuyanagi}, \citenamefont {Saito},\
  and\ \citenamefont {Semba}}]{Yoshihara_PhysRevA.95.053824}%
  \BibitemOpen
  \bibfield  {author} {\bibinfo {author} {\bibfnamefont {F.}~\bibnamefont
  {Yoshihara}}, \bibinfo {author} {\bibfnamefont {T.}~\bibnamefont {Fuse}},
  \bibinfo {author} {\bibfnamefont {S.}~\bibnamefont {Ashhab}}, \bibinfo
  {author} {\bibfnamefont {K.}~\bibnamefont {Kakuyanagi}}, \bibinfo {author}
  {\bibfnamefont {S.}~\bibnamefont {Saito}}, \ and\ \bibinfo {author}
  {\bibfnamefont {K.}~\bibnamefont {Semba}},\ }\bibfield  {title} {\emph
  {\bibinfo {title} {Characteristic spectra of circuit quantum electrodynamics
  systems from the ultrastrong- to the deep-strong-coupling regime},}\ }\href
  {\doibase 10.1103/PhysRevA.95.053824} {\bibfield  {journal} {\bibinfo
  {journal} {Phys. Rev. A}\ }\textbf {\bibinfo {volume} {95}},\ \bibinfo
  {pages} {053824} (\bibinfo {year} {2017}{\natexlab{b}})}\ \BibitemShut
  {NoStop}%
\bibitem [{\citenamefont {Wang}\ \emph {et~al.}(2023)\citenamefont {Wang},
  \citenamefont {Ridolfo}, \citenamefont {Li}, \citenamefont {Savasta},
  \citenamefont {Nori}, \citenamefont {Nakamura},\ and\ \citenamefont
  {You}}]{you_natcom_2023Asymm_Rabi_exp}%
  \BibitemOpen
  \bibfield  {author} {\bibinfo {author} {\bibfnamefont {S.-P.}\ \bibnamefont
  {Wang}}, \bibinfo {author} {\bibfnamefont {A.}~\bibnamefont {Ridolfo}},
  \bibinfo {author} {\bibfnamefont {T.}~\bibnamefont {Li}}, \bibinfo {author}
  {\bibfnamefont {S.}~\bibnamefont {Savasta}}, \bibinfo {author} {\bibfnamefont
  {F.}~\bibnamefont {Nori}}, \bibinfo {author} {\bibfnamefont {Y.}~\bibnamefont
  {Nakamura}}, \ and\ \bibinfo {author} {\bibfnamefont {J.~Q.}\ \bibnamefont
  {You}},\ }\bibfield  {title} {\emph {\bibinfo {title} {Probing the symmetry
  breaking of a light--matter system by an ancillary qubit},}\ }\href {\doibase
  10.1038/s41467-023-40097-0} {\bibfield  {journal} {\bibinfo  {journal}
  {Nature Communications}\ }\textbf {\bibinfo {volume} {14}},\ \bibinfo {pages}
  {4397} (\bibinfo {year} {2023})}\ \BibitemShut {NoStop}%
\bibitem [{\citenamefont {Holstein}\ and\ \citenamefont
  {Primakoff}(1940)}]{HP_approx_PhysRev.58.1098}%
  \BibitemOpen
  \bibfield  {author} {\bibinfo {author} {\bibfnamefont {T.}~\bibnamefont
  {Holstein}}\ and\ \bibinfo {author} {\bibfnamefont {H.}~\bibnamefont
  {Primakoff}},\ }\bibfield  {title} {\emph {\bibinfo {title} {Field Dependence
  of the Intrinsic Domain Magnetization of a Ferromagnet},}\ }\href {\doibase
  10.1103/PhysRev.58.1098} {\bibfield  {journal} {\bibinfo  {journal} {Phys.
  Rev.}\ }\textbf {\bibinfo {volume} {58}},\ \bibinfo {pages} {1098} (\bibinfo
  {year} {1940})}\ \BibitemShut {NoStop}%
\bibitem [{\citenamefont {Rabl}(2011)}]{peter_PhysRevLett.107.063601}%
  \BibitemOpen
  \bibfield  {author} {\bibinfo {author} {\bibfnamefont {P.}~\bibnamefont
  {Rabl}},\ }\bibfield  {title} {\emph {\bibinfo {title} {Photon Blockade
  Effect in Optomechanical Systems},}\ }\href {\doibase
  10.1103/PhysRevLett.107.063601} {\bibfield  {journal} {\bibinfo  {journal}
  {Phys. Rev. Lett.}\ }\textbf {\bibinfo {volume} {107}},\ \bibinfo {pages}
  {063601} (\bibinfo {year} {2011})}\ \BibitemShut {NoStop}%
\bibitem [{\citenamefont {Minoguchi}\ \emph {et~al.}(2019)\citenamefont
  {Minoguchi}, \citenamefont {Kirton},\ and\ \citenamefont
  {Rabl}}]{yuri_minoguchi2019environmentinduced}%
  \BibitemOpen
  \bibfield  {author} {\bibinfo {author} {\bibfnamefont {Y.}~\bibnamefont
  {Minoguchi}}, \bibinfo {author} {\bibfnamefont {P.}~\bibnamefont {Kirton}}, \
  and\ \bibinfo {author} {\bibfnamefont {P.}~\bibnamefont {Rabl}},\ }\emph
  {\bibinfo {title} {Environment-Induced Rabi Oscillations in the
  Optomechanical Boson-Boson Model},}\ \href@noop {} {\Eprint
  {http://arxiv.org/abs/1904.02164} {arXiv:1904.02164 [quant-ph]} (\bibinfo
  {year} {2019})}\BibitemShut {NoStop}%
\bibitem [{\citenamefont {Rabl}(2010)}]{Peter_PhysRevB.82.165320}%
  \BibitemOpen
  \bibfield  {author} {\bibinfo {author} {\bibfnamefont {P.}~\bibnamefont
  {Rabl}},\ }\bibfield  {title} {\emph {\bibinfo {title} {Cooling of mechanical
  motion with a two-level system: The high-temperature regime},}\ }\href
  {\doibase 10.1103/PhysRevB.82.165320} {\bibfield  {journal} {\bibinfo
  {journal} {Phys. Rev. B}\ }\textbf {\bibinfo {volume} {82}},\ \bibinfo
  {pages} {165320} (\bibinfo {year} {2010})}\ \BibitemShut {NoStop}%
\bibitem [{\citenamefont {De~Bernardis}\ \emph
  {et~al.}(2018{\natexlab{b}})\citenamefont {De~Bernardis}, \citenamefont
  {Pilar}, \citenamefont {Jaako}, \citenamefont {De~Liberato},\ and\
  \citenamefont {Rabl}}]{DeBernardis_PhysRevA.98.053819}%
  \BibitemOpen
  \bibfield  {author} {\bibinfo {author} {\bibfnamefont {D.}~\bibnamefont
  {De~Bernardis}}, \bibinfo {author} {\bibfnamefont {P.}~\bibnamefont {Pilar}},
  \bibinfo {author} {\bibfnamefont {T.}~\bibnamefont {Jaako}}, \bibinfo
  {author} {\bibfnamefont {S.}~\bibnamefont {De~Liberato}}, \ and\ \bibinfo
  {author} {\bibfnamefont {P.}~\bibnamefont {Rabl}},\ }\bibfield  {title}
  {\emph {\bibinfo {title} {Breakdown of gauge invariance in
  ultrastrong-coupling cavity QED},}\ }\href {\doibase
  10.1103/PhysRevA.98.053819} {\bibfield  {journal} {\bibinfo  {journal} {Phys.
  Rev. A}\ }\textbf {\bibinfo {volume} {98}},\ \bibinfo {pages} {053819}
  (\bibinfo {year} {2018}{\natexlab{b}})}\ \BibitemShut {NoStop}%
\bibitem [{\citenamefont {Di~Stefano}\ \emph {et~al.}(2019)\citenamefont
  {Di~Stefano}, \citenamefont {Settineri}, \citenamefont {Macr{\`\i}},
  \citenamefont {Garziano}, \citenamefont {Stassi}, \citenamefont {Savasta},\
  and\ \citenamefont {Nori}}]{Savasta_NatPhys_2019}%
  \BibitemOpen
  \bibfield  {author} {\bibinfo {author} {\bibfnamefont {O.}~\bibnamefont
  {Di~Stefano}}, \bibinfo {author} {\bibfnamefont {A.}~\bibnamefont
  {Settineri}}, \bibinfo {author} {\bibfnamefont {V.}~\bibnamefont
  {Macr{\`\i}}}, \bibinfo {author} {\bibfnamefont {L.}~\bibnamefont
  {Garziano}}, \bibinfo {author} {\bibfnamefont {R.}~\bibnamefont {Stassi}},
  \bibinfo {author} {\bibfnamefont {S.}~\bibnamefont {Savasta}}, \ and\
  \bibinfo {author} {\bibfnamefont {F.}~\bibnamefont {Nori}},\ }\bibfield
  {title} {\emph {\bibinfo {title} {Resolution of gauge ambiguities in
  ultrastrong-coupling cavity quantum electrodynamics},}\ }\href {\doibase
  10.1038/s41567-019-0534-4} {\bibfield  {journal} {\bibinfo  {journal} {Nature
  Physics}\ }\textbf {\bibinfo {volume} {15}},\ \bibinfo {pages} {803}
  (\bibinfo {year} {2019})}\ \BibitemShut {NoStop}%
\bibitem [{\citenamefont {Breuer}\ \emph {et~al.}(2002)\citenamefont {Breuer},
  \citenamefont {Petruccione},\ and\ \citenamefont
  {Petruccione}}]{petruccione2002theory}%
  \BibitemOpen
  \bibfield  {author} {\bibinfo {author} {\bibfnamefont {H.}~\bibnamefont
  {Breuer}}, \bibinfo {author} {\bibfnamefont {F.}~\bibnamefont {Petruccione}},
  \ and\ \bibinfo {author} {\bibfnamefont {S.}~\bibnamefont {Petruccione}},\
  }\href {https://books.google.it/books?id=0Yx5VzaMYm8C} {\emph {\bibinfo
  {title} {The Theory of Open Quantum Systems}}}\ (\bibinfo  {publisher}
  {Oxford University Press},\ \bibinfo {year} {2002})\BibitemShut {NoStop}%
\bibitem [{\citenamefont
  {García-Ripoll}(2022)}]{garcia_ripoll_book_10261_285945}%
  \BibitemOpen
  \bibfield  {author} {\bibinfo {author} {\bibfnamefont {J.~J.}\ \bibnamefont
  {García-Ripoll}},\ }\href {\doibase 10.1017/9781316779460} {\emph {\bibinfo
  {title} {Quantum Information and Quantum Optics with Superconducting
  Circuits}}}\ (\bibinfo {year} {2022})\BibitemShut {NoStop}%
\bibitem [{\citenamefont {Dalibard}\ \emph {et~al.}(1982)\citenamefont
  {Dalibard}, \citenamefont {Dupont-Roc},\ and\ \citenamefont
  {Cohen-Tannoudji}}]{dalibard:jpa-00209544}%
  \BibitemOpen
  \bibfield  {author} {\bibinfo {author} {\bibfnamefont {J.}~\bibnamefont
  {Dalibard}}, \bibinfo {author} {\bibfnamefont {J.}~\bibnamefont
  {Dupont-Roc}}, \ and\ \bibinfo {author} {\bibfnamefont {C.}~\bibnamefont
  {Cohen-Tannoudji}},\ }\bibfield  {title} {\emph {\bibinfo {title} {{Vacuum
  fluctuations and radiation reaction : identification of their respective
  contributions}},}\ }\href {\doibase 10.1051/jphys:0198200430110161700}
  {\bibfield  {journal} {\bibinfo  {journal} {{Journal de Physique}}\ }\textbf
  {\bibinfo {volume} {43}},\ \bibinfo {pages} {1617} (\bibinfo {year} {1982})}\
  \BibitemShut {NoStop}%
\bibitem [{\citenamefont {Clerk}\ \emph {et~al.}(2010)\citenamefont {Clerk},
  \citenamefont {Devoret}, \citenamefont {Girvin}, \citenamefont {Marquardt},\
  and\ \citenamefont {Schoelkopf}}]{Clerck_RevModPhys.82.1155}%
  \BibitemOpen
  \bibfield  {author} {\bibinfo {author} {\bibfnamefont {A.~A.}\ \bibnamefont
  {Clerk}}, \bibinfo {author} {\bibfnamefont {M.~H.}\ \bibnamefont {Devoret}},
  \bibinfo {author} {\bibfnamefont {S.~M.}\ \bibnamefont {Girvin}}, \bibinfo
  {author} {\bibfnamefont {F.}~\bibnamefont {Marquardt}}, \ and\ \bibinfo
  {author} {\bibfnamefont {R.~J.}\ \bibnamefont {Schoelkopf}},\ }\bibfield
  {title} {\emph {\bibinfo {title} {Introduction to quantum noise, measurement,
  and amplification},}\ }\href {\doibase 10.1103/RevModPhys.82.1155} {\bibfield
   {journal} {\bibinfo  {journal} {Rev. Mod. Phys.}\ }\textbf {\bibinfo
  {volume} {82}},\ \bibinfo {pages} {1155} (\bibinfo {year} {2010})}\
  \BibitemShut {NoStop}%
\end{thebibliography}%
\end{document}